\documentclass[b5paper,10pt]{book}

\usepackage{phdthesis}
\usepackage[
  paperwidth=17cm
  ,paperheight=24cm
  ,inner=2.6cm
  ,outer=2.3cm
  ,top=2.3cm
  ,bottom=2.3cm
]{geometry}

\usepackage{graphicx,times}
\usepackage[hyphens]{url}
\usepackage{tabularx,amsmath}

\usepackage{multirow}
\usepackage{booktabs}
\usepackage{amssymb}
\usepackage[square,comma,numbers,sort&compress,sectionbib]{natbib}
\usepackage{wrapfig}
\usepackage{enumerate}
\usepackage{supertabular}
\usepackage{multicol}
\usepackage{setspace}
\usepackage{import}
\usepackage{booktabs} 
\usepackage{graphicx}
\usepackage{url}
\usepackage{times}
\usepackage{microtype}
\usepackage{tabularx,amsmath}
\usepackage{amssymb}
\usepackage{multirow}
\usepackage{color}
\usepackage[usenames,dvipsnames,table]{xcolor}
\usepackage[normalem]{ulem}
\usepackage[english]{babel}
\usepackage[noend]{algpseudocode}

\usepackage{algorithm}
\usepackage{paralist}
\usepackage{flushend}
\usepackage[inline]{enumitem}
\usepackage{setspace}
\usepackage{textcomp}
\usepackage{listings}
\usepackage{keyval}
\usepackage{rotating}
\usepackage{acronym}

\usepackage{lscape}
\usepackage{pdfpages}

\renewenvironment{table*}[1][]{\begin{sidewaystable}[htbp]}{\end{sidewaystable}\ignorespacesafterend}
\renewenvironment{figure*}[1][]{\begin{sidewaysfigure}[htbp]}{\end{sidewaysfigure}\ignorespacesafterend}

\usepackage{pgf, tikz}
\usetikzlibrary{arrows,automata,shapes}
\usetikzlibrary{positioning}

\newcommand{\tabincell}[2]{\begin{tabular}{@{}#1@{}}#2\end{tabular}}

\usepackage{subcaption}

\usepackage{bibentry}
\nobibliography*

\usepackage[backref=page]{hyperref}
\hypersetup{%
pdfborder = {0 0 0},
pdftitle = {Supporting Search Engines with~Knowledge~and~Context},
pdfsubject = {PhD Thesis},
pdfkeywords = {add, some, keywords},
pdfauthor = {Nikos Voskarides}
}
\usepackage{bookmark}

\renewcommand*{\backref}[1]{} 
\renewcommand*{\backrefalt}[4]{%
\ifcase #1
\or (Cited on page~#2.)  %
\else 
(Cited on pages~#2.)  
\fi
}

\let\svthefootnote\thefootnote
\newcommand\blankfootnote[1]{%
  \let\thefootnote\relax\footnotetext{#1}%
  \let\thefootnote\svthefootnote%
}
\let\svfootnote\footnote
\renewcommand\footnote[2][?]{%
  \if\relax#1\relax%
    \blankfootnote{#2}%
  \else%
    \if?#1\svfootnote{#2}\else\svfootnote[#1]{#2}\fi%
  \fi
}

\begin{document}

\includepdf[]{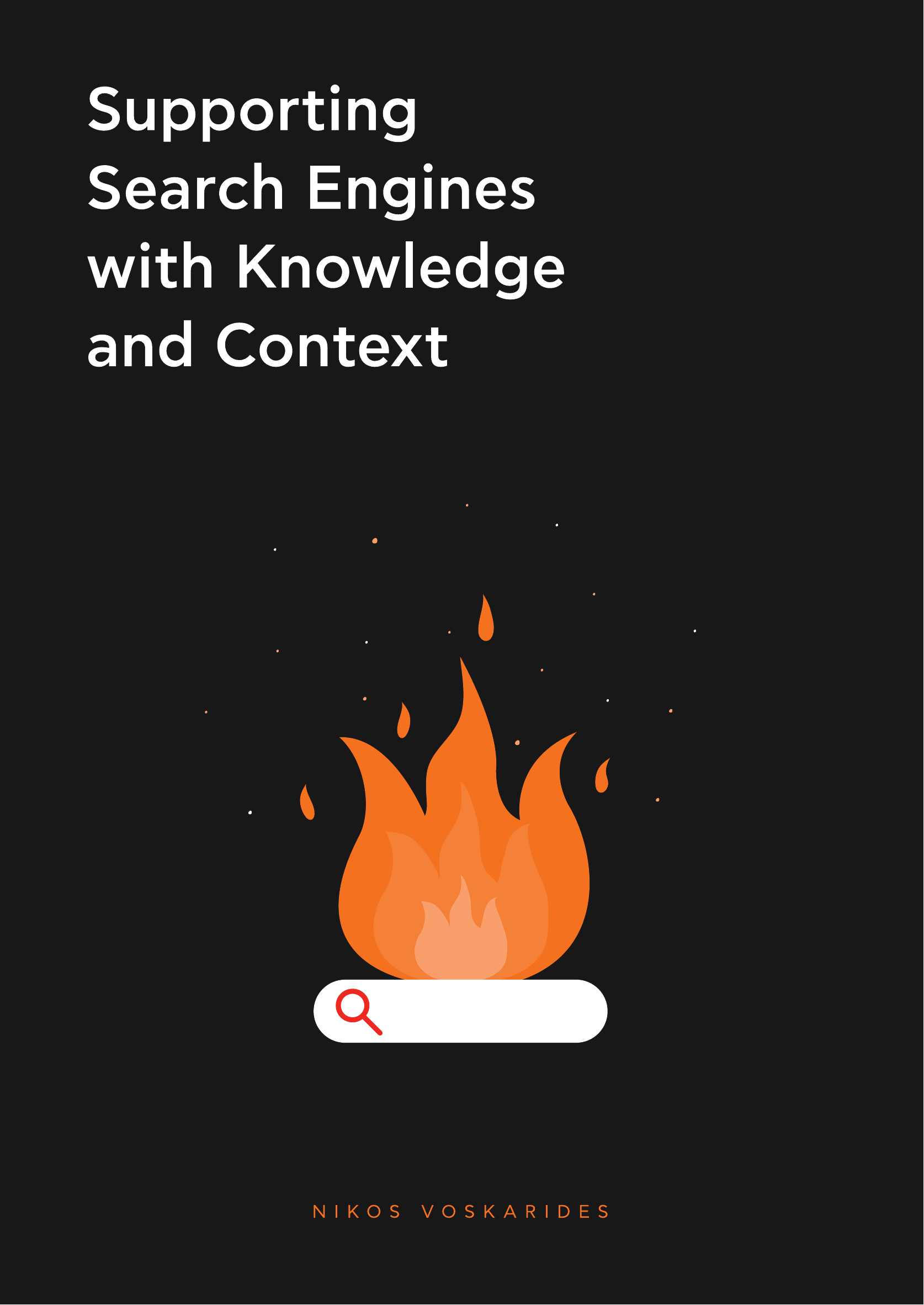}

\frontmatter

{\pagestyle{empty}
\newcommand{\printtitle}{%
{\Huge\bf Supporting Search Engines with~Knowledge~and~Context \\[0.8cm]
}}

\begin{titlepage}
\par\vskip 2cm
\begin{center}
\printtitle
\vfill
{\LARGE\bf Nikos Voskarides}
\vskip 2cm
\end{center}
\end{titlepage}

\mbox{}\newpage
\setcounter{page}{1}

\clearpage
\par\vskip 2cm
\begin{center}
\printtitle
\par\vspace {4cm}
{\large \sc Academisch Proefschrift}
\par\vspace {1cm}
{\large ter verkrijging van de graad van doctor aan de \\
Universiteit van Amsterdam\\
op gezag van de Rector Magnificus\\
prof.\ dr.\ ir.\ K.I.J. Maex\\
ten overstaan van een door het College voor Promoties ingestelde \\
commissie, in het openbaar te verdedigen in \\
de Agnietenkapel\\
op vrijdag 5 februari 2021, te 16:00 uur \\ } %
\par\vspace {1cm} {\large door}
\par \vspace {1cm}
{\Large Nikos Voskarides}
\par\vspace {1cm}
{\large geboren te Lefkosia} %
\end{center}

\clearpage
\noindent%
\textbf{Promotiecommissie} \\\\
\begin{tabular}{@{}l l l}
Promotor: 
& prof.\ dr.\ Maarten de Rijke & Universiteit van Amsterdam \\  %
\\
Copromotor: 
& dr.\ Edgar J.\ Meij & Bloomberg \\
\\
Overige leden: 
& prof.\ dr.\ Paul Groth & Universiteit van Amsterdam \\
& prof.\ dr.\ Evangelos Kanoulas & Universiteit van Amsterdam \\
& dr. Maarten Marx & Universiteit van Amsterdam \\
& prof.\ dr. Simone Teufel & University of Cambridge \\
& dr. Suzan Verberne & Universiteit Leiden \\
\end{tabular}

\bigskip\noindent%
Faculteit der Natuurwetenschappen, Wiskunde en Informatica\\

\vfill

\noindent
The research was supported by the Netherlands Organisation for Scientific Research (NWO) under project number CI-14-25.  \\
\bigskip

\noindent
Copyright \copyright~2021 Nikos Voskarides, Amsterdam, The Netherlands\\
Cover by Evros Voskarides\\
Printed by Ipskamp Printing, Amsterdam\\
\\
ISBN: 978-94-6421-207-5\\

\clearpage
}

{\pagestyle{empty}

{
\begin{center}

\noindent
\textbf{Acknowledgements} \\ \vskip .5cm
\end{center}
}

\noindent 
Maarten, your supervision has been transformative. Thank you for caring and for pushing me to become better.

\medskip \noindent 
Edgar, you have been a source of energy. Thank you for believing in me. 

\medskip \noindent 
ILPS is a fantastic research group, mainly because of Maarten’s vision and consistency. 

\begin{itemize}
\setlength\itemsep{0em}

\item[] Mano, thank you for sharing your enthusiasm and for making sure I’d find \\my way.

\item[] Petra, thank you for going above and beyond for the group.

\item[] Ridho, for being my knowledge graph buddy.

\item[] Katya, for having an alternative viewpoint.

\item[] Marzieh, for reminding me of my roots.

\item[] Anne, Daan, for encouraging me.

\item[] Vangeli, for being unconventional.

\item[] Harrie, for being a good listener.

\item[] Ana, for speaking up.

\item[] Ke, for sharing your passion.

\item[] Rolf, for being punctual.

\item[] Dan, for introducing me to the Guqin. 

\item[] Christophe, for setting the bar high.

\item[] Sabrina, for seeing my work through a different lens.

\item[] Bob, Chang, David, Hamid, Hosein, Ilya, Isaac, Maartje, Marlies, Mostafa, Pengjie, Tom, Wouter, Zhaochun, Ziming, and all the other ILPSers I worked with over the years, thank you for the fun memories.

\end{itemize}

\medskip \noindent 
I spent most of my PhD studies in Amsterdam.

\begin{itemize}
\setlength\itemsep{0em}
\item[] Giorgo, your help has been unmeasurable; thank you for sharing the burdens and multiplying the joys. 
\item[] Mario, thank you for not compromising.
\item[] You both have been there for me in happy and rough times.%
\item[] Achillea, Eleni, Fee, Gianni, Ioanna, Kyriaco, Pari, Sofia, Stathi, you brought me joy every time I was around you. Thank you.
\end{itemize}

\medskip \noindent 
During my PhD studies I did two internships, both of which were incredibly rewarding experiences. 

\begin{itemize}
\setlength\itemsep{0em}
\item[] In the summer of 2017, I interned at Bloomberg in London. 
Thank you Edgar, Ridho, Abhinav, Anju, Malvina, Miles, Diego for being great hosts.
Nicoletta, Thiago, thank you for looking out for me.
Iakove, Dora, Leo, Marilena, for all the fun times.

\item[] In the summer of 2018, I interned at Amazon in Barcelona.
Roi, Hugo, Lluis, Vassili, Marc, Jordi, thank you for making my stay memorable.
\end{itemize}

\medskip \noindent
Thank you Maarten, Paul, Simone, Suzan and Vangeli for agreeing to serve in my committee.
Giorgo, Katya, thank you for being my paranymphs. 
Harrie, thank you for translating my thesis summary to Dutch.

\medskip \medskip  \noindent 
Next, my family.

\begin{itemize}
\setlength\itemsep{0em}
\item[] Papa, Mamma, your unconditional love and the example you have set have taught me the importance of knowledge and context.
Thank you for enduring my absence all these years.
\item[] 
This thesis is dedicated to you. 
\item[] Christofore, Evro, thank you for the (silent) support.
\item[] Pappou, for the inspiration.
\end{itemize}

\medskip \noindent
Iris, thank you. You are my only home.

\medskip\medskip
\hfill Nikos Voskarides\\
\mbox{}\hfill Lefkosia, January 2021\\

\clearpage
}
\tableofcontents

\acrodef{SERP}{search engine result page}
\acrodef{IR}{information retrieval}
\acrodef{NFCM}{neural fact contextualization method}
\acrodef{QuReTeC}{query resolution by term classification}
\acrodef{KG}{knowledge graph}

\mainmatter

\chapter{Introduction}
\label{chapter:introduction}
Search engines leverage large repositories of knowledge to improve information access~\cite{manning2008introduction,nickel2015review,reinanda-2019-knowledge}.
These repositories may store unstructured knowledge such as textual documents or social media posts, or structured knowledge such as attributes of and relationships between real-world objects and topics.

In order to effectively leverage knowledge, search engines should account for context, i.e., additional information about the user and the query~\cite{allan_challenges_2003,shen_context-sensitive_2005,white_predicting_2009,bennett_modeling_2012}.
In this thesis, we study how to support search engines in leveraging knowledge while accounting for different types of context:
\begin{enumerate*}[label=(\arabic*)]
  \item context that the search engine proactively provides to enrich search results (e.g., information on tourist attractions when searching for a city),
  \item context that stems from the interactions between the user and the search engine in a conversational search session, or
  \item context provided by the user to specify a broad query.
\end{enumerate*}

\begin{figure}[t]
\captionsetup[subfigure]{aboveskip=-1pt,belowskip=-1pt}
    \centering
       \begin{subfigure}{.4\linewidth}
  \centering
  \includegraphics[width=\textwidth]{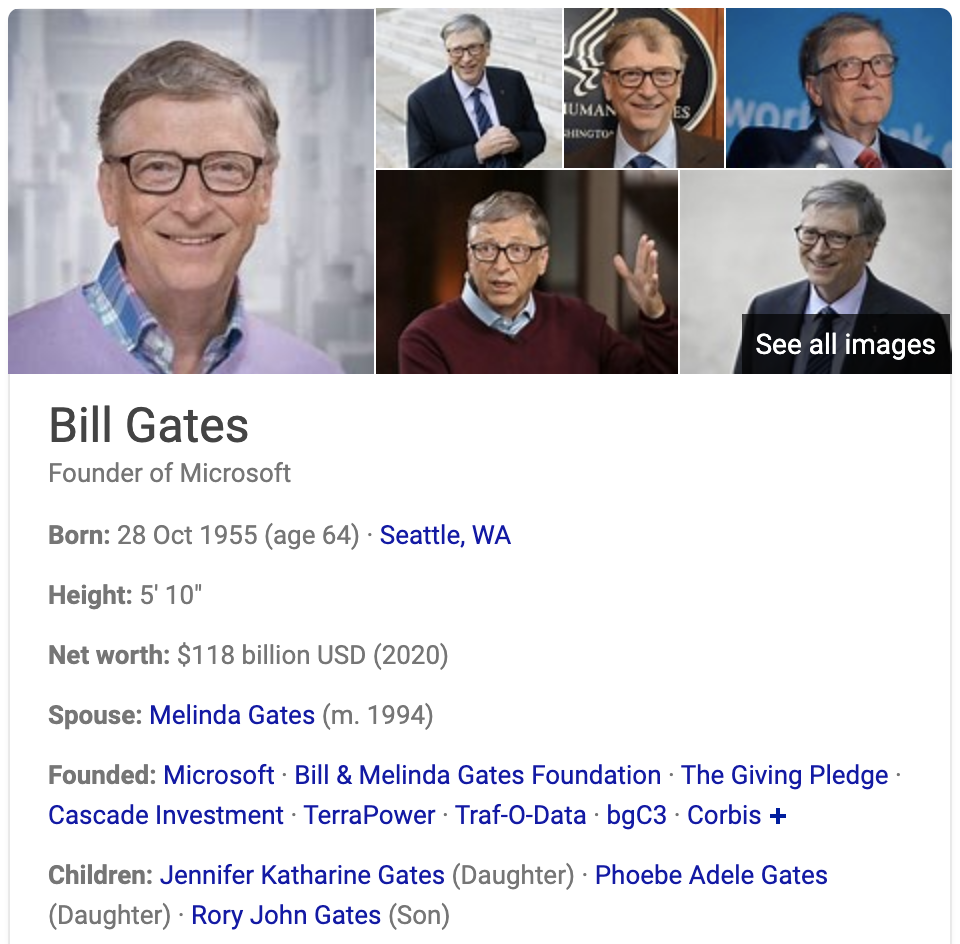}  
  \label{fig:gates1}
\end{subfigure}
~\begin{subfigure}{.4\linewidth}
  \centering
  \includegraphics[width=\textwidth]{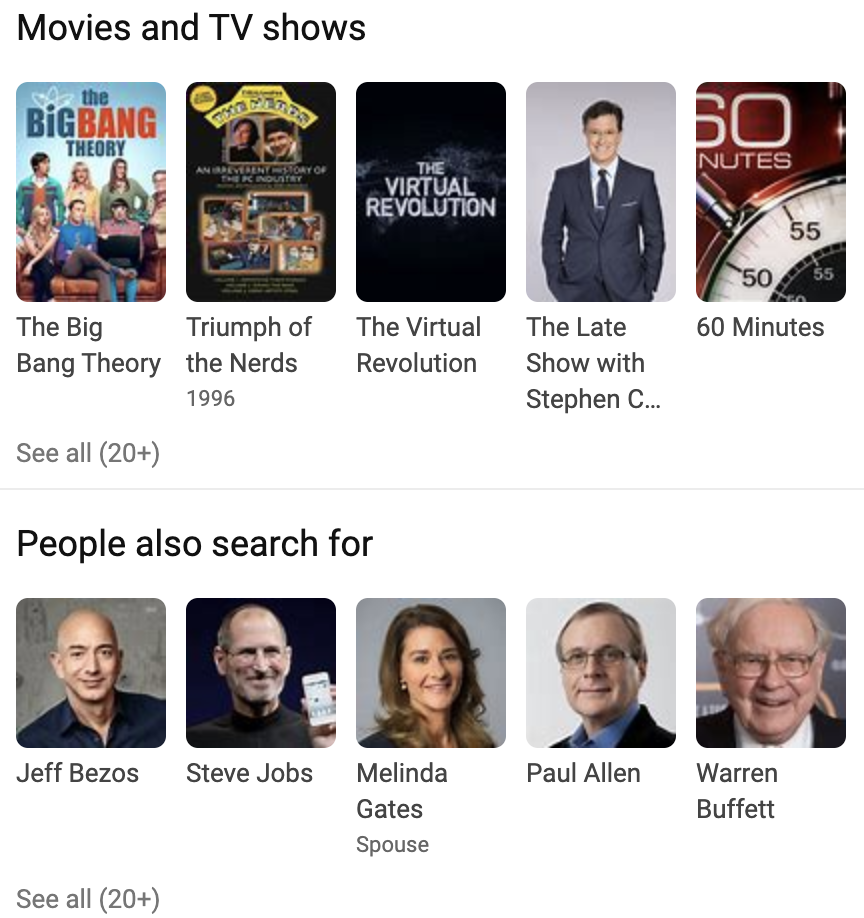}  
  \label{fig:gates2}
\end{subfigure}
\caption{Part of a SERP KG panel in response to the query ``Bill Gates'' (split in two parts).
}
\label{fig:kg_panel_gates}
\end{figure}

\Acp{SERP} present information that is meant to be relevant to the user's query~\cite{hochstotter_what_2009,bailey_evaluating_2010,smith_knowledge-context_2019}.
Apart from the traditional ``ten blue links'', modern \acp{SERP} are increasingly being enriched with additional context that often comes from structured knowledge sources to enhance the user experience~\cite{dalvi2009web}. %
\Acp{KG}, which store world knowledge in the form of facts, are the most prominent structured knowledge source for search engines~\cite{lin2012active,bota2016playing,reinanda-2019-knowledge}.
This is natural since the majority of queries issued to search engines contain entities stored in \acp{KG}~\cite{guo2009named}.
\acp{KG} are used to support different components of modern \acp{SERP}, such as direct answers to user queries and \ac{KG} panels~\cite{yih2015semantic}, which present facts about the entity identified in the query and other, related entities to support exploratory search (see Figure~\ref{fig:kg_panel_gates})~\cite{blanco2013entity,miliaraki2015selena,bota2016playing,hasibi2017dynamic}.
A challenge that arises when presenting such structured knowledge in a \ac{SERP} is that it is stored in a formal form, not directly suitable for presentation to the user.
Tackling this challenge is the focus of the first research theme of this thesis, where we study how to make structured knowledge more accessible to the search engine user.

Users interact with the search results presented to them in multiple ways and they provide signals that may be used by search engines to improve the user experience, for instance by continuously learning better ranking functions~\cite{joachims_optimizing_2002,chuklin_click_2015,hofmann_balancing_2013,hofmann_online_2016,jagerman_model_2019}.
Recent advances in natural language processing and deep learning have enabled the wide-spread use of interactive systems in real-world applications~\cite{gao_neural_2018}, which, in turn, has fueled a resurgence of research in conversational search~\cite{culpepper_research_2018,cast2019,aliannejadi_asking_2019,zamani_generating_2020,rosset_leading_2020}.
In conversational search, the user interacts with the search engine during relatively short sessions to gather knowledge over large unstructured knowledge repositories~\cite{belkin1980anomalous,Croft:1987:IRN:35053.35054}.
A prominent challenge in conversational search is that the search engine has to keep track of the evolving context during the conversation so as to enable more natural interactions.
Addressing this challenge is the focus of the second research theme of this thesis, where we study how to identify relevant context from the conversation history in order to improve interactive knowledge gathering.

Search engines facilitate knowledge gathering for different types of users.
A large portion of research in information retrieval has focused on how to answer information needs of users in web search~\cite{mcbryan_genvl_1994,brin_anatomy_1998,zeng2004learning,mohan2011web}.
In contrast to web search, in professional search, users express their information needs in a different way and aim to access and explore domain-specific knowledge~\cite{kim_automatic_2011,russell-rose_information_2018,verberne_first_2019}.
Writers are a type of professional users who heavily rely on search engines~\cite{sylvester_digital_2009,hagen_how_2016}.
For instance, writers in the scientific domain use search engines to find relevant references to include in their articles~\cite{he_context-aware_2010,bhagavatula_content-based_2018}.
Another prominent example of professional search engine users are writers in the news domain~\cite{diakopoulos_automating_2019,cucchiarelli_topic_2019}.
Such writers create narratives around specific events and use search engines to support them in this process~\cite{kirkpatrick_putting_2015,sauer_audiovisual_2017,clark_creative_2018}.
In the third research theme of this thesis, we study how to support writers explore unstructured knowledge about past events given an incomplete narrative that specifies a main event and a context.

\section{Research Outline and Questions}
\label{section:introduction:rqs}

This thesis focuses on three research themes aimed at supporting search engines with knowledge and context: 
\begin{enumerate*}[label=(\arabic*)]
	\item making structured knowledge more accessible to the user by describing and contextualizing KG facts (Chapters~\ref{chapter:acl2015}, \ref{chapter:ecir2017} and~\ref{chapter:sigir2018}), 
	\item improving interactive knowledge gathering by identifying relevant context in conversational search (Chapter~\ref{chapter:sigir2020}), and 
	\item supporting knowledge exploration for narrative creation by retrieving event-focused news articles in context (Chapter~\ref{chapter:chapter6}).
\end{enumerate*}

Below, we describe the main research questions for each chapter. In each chapter we describe more fine-grained subquestions that we ask to answer each main research question.

\subsection{Making structured knowledge more accessible to the user}
SERPs often include structured knowledge for queries that mention real-world entities in the form of KG facts.
Facts are stored in KGs in a formal form (e.g., $\langle$Bill Gates, founderOf, Microsoft$\rangle$).
When presenting a KG fact to the user, however, it is more natural to use human-readable descriptions that verbalize and contextualize the fact~\cite{gkatzia2016natural}.
For instance, a possible description of the KG fact $\langle$Bill Gates, founderOf, Microsoft$\rangle$ is: \emph{Bill Gates is an American business magnate and the principal founder of Microsoft Corporation.}
In our first study (Chapter~\ref{chapter:acl2015}), we cast the problem of finding such descriptions as a retrieval task:
\acrodef{rq:retrieve-explanations}[\ref{rq:retrieve-explanations}]{Given a KG fact and a text corpus, can we retrieve textual descriptions of the fact from the text corpus?}
\begin{enumerate}[label=\textbf{RQ\arabic*},ref={RQ\arabic*},resume]
\item \acl{rq:retrieve-explanations}\label{rq:retrieve-explanations}
\end{enumerate}
We propose a method that first extracts and enriches candidate sentences that may be referring to the entities of the fact from a text corpus, and then ranks those sentences.
Our results show that we can reliably retrieve sentences that accurately describe a given fact, under the condition that a relevant sentence exists in the underlying text corpus. 

However, it is likely that this condition does not hold in cases where a given fact is not explicitly described in the text corpus at hand. This limits the applicability of our proposed method in real-world scenarios.
In order to address this limitation, in our second study (Chapter~\ref{chapter:ecir2017}), we consider a text generation task:
\acrodef{rq:generate-explanations}[\ref{rq:generate-explanations}]{Given a KG fact, can we automatically generate a textual description of the fact in the absence of an existing description?}
\begin{enumerate}[label=\textbf{RQ\arabic*},ref={RQ\arabic*},resume]
\item \acl{rq:generate-explanations}\label{rq:generate-explanations}
\end{enumerate}
We propose to first create sentence templates for each relationship in the KG using existing fact descriptions.
Then, given a KG fact that expresses a specific relationship, we select a relevant template and fill it using additional information from the KG (other facts), if needed.
We find that our method can generate contextually rich descriptions and is robust against KG incompleteness.

KG fact descriptions often contain mentions of other, related facts that provide additional context and thus increase the user's understanding of the fact as a whole (e.g., \emph{Bill Gates founded Microsoft with Paul Allen}).
Given the large size of KGs, many facts could potentially be relevant to the fact of interest, thus we need to automate the task of finding those other facts.
This is the focus of our next study (Chapter~\ref{chapter:sigir2018}):
\acrodef{rq:contextualize-facts}[\ref{rq:contextualize-facts}]{Can we contextualize a KG query fact by retrieving other, related KG facts?}
\begin{enumerate}[label=\textbf{RQ\arabic*},ref={RQ\arabic*},resume]
\item \acl{rq:contextualize-facts}\label{rq:contextualize-facts}
\end{enumerate}
We propose a method that first enumerates other candidate facts in the neighborhood of the query fact and then ranks those facts with respect to their relevance to the query fact.
We propose the \acfi{NFCM}, a neural ranking model that combines automatically learned and hand-crafted features.
In addition, we propose to use a distant supervision method to automatically gather training data for \ac{NFCM}.
We find that \ac{NFCM} outperforms several baseline methods and that distant supervision is effective for this task.

\subsection{Improving interactive knowledge gathering}
The ultimate goal of conversational AI is interactive knowledge gathering~\cite{gao_neural_2018}.
Search engines can play a crucial role towards achieving that goal.
An interactive search engine should support conversational search, where a user aims to interactively find information stored in large unstructured knowledge repositories~\cite{culpepper_research_2018}.

In our next study (Chapter~\ref{chapter:sigir2020}), we focus on multi-turn passage retrieval as an instance of conversational search~\cite{cast2019}.
Here, the query at the current turn may be underspecified. 
Thus, we need to identify relevant context from the conversation history to arrive at a better expression of the query.
We answer the following research question:
\acrodef{rq:query-res}[\ref{rq:query-res}]{Can we use query resolution to identify relevant context and thereby improve retrieval in conversational search?}
\begin{enumerate}[label=\textbf{RQ\arabic*},ref={RQ\arabic*},resume]
\item \acl{rq:query-res}\label{rq:query-res}
\end{enumerate}
Here, \emph{query resolution} refers to the task of adding missing context from the conversation history to the current turn query, if needed.
We propose to model query resolution as a term classification task.
We design \acfi{QuReTeC}, a neural query resolution model based on bidirectional transformers.
Since obtaining human-curated training data specifically for query resolution may be cumbersome, we propose a distant supervision method that automatically generates supervision data for QuReTeC using query-passage relevance pairs.
We find that when integrating QuReTeC in a multi-stage ranking architecture we can significantly outperform baseline models.
In addition, we find that the distant supervision method we propose can substantially reduce the amount of human-curated training data required to train \ac{QuReTeC}.

\subsection{Supporting knowledge exploration for narrative creation}
Writers such as journalists often use search engines to find relevant material to include in event-oriented narratives~\cite{huurnink_search_2010,popescu_proceedings_2017,diakopoulos_automating_2019}.
Such material can provide background knowledge on the event itself or connections to other events that can help writers generate new angles on the narrative and thus better engage the reader~\cite{kirkpatrick_putting_2015,clark_creative_2018}. 
Previous work has focused on exploring knowledge for narrative creation from different sources, such as social media~\cite{diakopoulos_finding_2012,zubiaga_curating_2013,cucchiarelli_topic_2019}, or from sources with a more narrow scope, such as political speeches~\cite{maclaughlin_context-based_2020}.

In our next study (Chapter~\ref{chapter:chapter6}), we focus on supporting knowledge exploration from a corpus of event-centric news articles for narrative creation.
More specifically, we study a real-world scenario where the writer has already generated an incomplete narrative that specifies a main event and a context, and aims to retrieve relevant news articles that discuss other events from the past.
We answer the following research question:
\acrodef{rq:narrativecreation}[\ref{rq:narrativecreation}]{Can we support knowledge exploration for event-centric narrative creation by performing news article retrieval in context?}
\begin{enumerate}[label=\textbf{RQ\arabic*},ref={RQ\arabic*},resume]
\item \acl{rq:narrativecreation}\label{rq:narrativecreation}
\end{enumerate}
We formally define this task and propose a retrieval dataset construction procedure that relies on existing news articles to simulate incomplete narratives and relevant articles.
We conduct experiments on two datasets derived from this procedure and find that state-of-the-art lexical and semantic rankers are not sufficient for this task.
We find that combining those rankers with one that ranks articles by reverse chronological order outperforms those rankers alone.
We also perform an in-depth quantitative and qualitative analysis of the results along different dimensions to acquire insights into the characteristics of this task.

\section{Main Contributions}
\label{section:introduction:contributions}

In this section, we summarize the main contributions of this thesis.

\paragraph{Theoretical contributions}
\begin{enumerate}
	\item We formalize the task of retrieving \acl{KG} fact descriptions stored in a text corpus (Chapter~\ref{chapter:acl2015}).
	\item We formalize the task of generating \acl{KG} fact descriptions (Chapter~\ref{chapter:ecir2017}).
	\item We formalize the task of \acl{KG} fact contextualization (Chapter~\ref{chapter:sigir2018}).
	\item We formulate the task of query resolution for conversational search as term classification (Chapter~\ref{chapter:sigir2020}).
	\item We formalize the task of news article retrieval in context for event-centric narrative creation (Chapter~\ref{chapter:chapter6}).

\end{enumerate}

\paragraph{Algorithmic contributions}
\begin{enumerate}[resume]
	\item A learning to rank method that combines a rich set of features for retrieving \acl{KG} fact descriptions (Chapter~\ref{chapter:acl2015}).
	\item A method for generating \acl{KG} fact descriptions by template construction and filling (Chapter~\ref{chapter:ecir2017}).
	\item \Acf{NFCM}, a method for contextualizing \acl{KG} facts, and a distant supervision method for gathering training data automatically (Chapter~\ref{chapter:sigir2018}).
	\item \Acf{QuReTeC}, a method for query resolution for multi-turn passage ranking, and a distant supervision method for gathering training data automatically (Chapter~\ref{chapter:sigir2020}).
	\item A retrieval dataset construction procedure for the task of news article retrieval in context for event-centric narrative creation (Chapter~\ref{chapter:chapter6}).

\end{enumerate}

\paragraph{Empirical contributions}

\begin{enumerate}[resume]
	\item Retrieving \acl{KG} fact descriptions (Chapter~\ref{chapter:acl2015})
		\begin{enumerate}
			\item Empirical comparison of our proposed learning to rank model and other sentence retrieval methods.
			\item Empirical comparison of relationship-dependent models against an independent model.
			\item Analysis of how different feature types contribute to the performance of our model and an error analysis of common errors made by our model.
		\end{enumerate}
	\item Generating \acl{KG} fact descriptions (Chapter~\ref{chapter:ecir2017})
		\begin{enumerate}
			\item Empirical comparison of different methods by automatic and manual evaluation.
			\item Analysis of specific cases where our method succeeds or fails.
		\end{enumerate}
		
	\item Contextualizing \acl{KG} facts (Chapter~\ref{chapter:sigir2018})
		\begin{enumerate}
			\item Empirical comparison of \ac{NFCM} and heuristic baselines.
			\item We show that learning ranking functions using distant supervision is beneficial.
			\item Analysis of the effect of handcrafted and automatically learned features on retrieval effectiveness. 
		\end{enumerate}
		
	\item Query resolution for conversational search (Chapter~\ref{chapter:sigir2020})
		\begin{enumerate}
			\item Empirical comparison of \ac{QuReTeC} and multiple baselines of different nature. 
			\item We show that distant supervision can substantially reduce the amount of gold standard training data needed to train \ac{QuReTeC}.
			\item Qualitative analysis of specific cases where our method succeeds or fails.
		\end{enumerate}
		
	\item News article retrieval in context (Chapter~\ref{chapter:chapter6})
		\begin{enumerate}
			\item Empirical comparison of state-of-the-art lexical rankers on this task. 
			\item We show that a combination of lexical and semantic rankers with one that ranks articles by reverse chronological order outperforms those rankers alone.
			\item An in-depth quantitative and qualitative analysis of the performance of the rankers under comparison among different dimensions.
		\end{enumerate}

\end{enumerate}

\paragraph{Resources}

\begin{enumerate}[resume]
	\item A manually annotated dataset for \acl{KG} fact description retrieval.
	\item An automatically extracted dataset for \acl{KG} fact description generation.
	\item A manually annotated dataset for \acl{KG} fact contextualization.
	\item An open source implementation of \ac{QuReTeC}.
	\item An automatically extracted dataset for news article retrieval in context.
\end{enumerate}

\section{Thesis Overview}
\label{section:introduction:overview}

The thesis is organized in three parts. %

In the first part we study how to make KG facts more accessible to users in search applications. Specifically, given a specific KG fact, we study how to retrieve textual descriptions of the fact (Chapter~\ref{chapter:acl2015}), how to generate a textual description of the fact in the absence of an existing description (Chapter~\ref{chapter:ecir2017}), and how to retrieve other KG facts to contextualize the fact (Chapter~\ref{chapter:sigir2018}).

In the second part we study how to improve interactive knowledge gathering by performing query resolution for multi-turn passage retrieval (Chapter~\ref{chapter:sigir2020}).

In the third part we study how to support narrative creation by performing news article retrieval in context (Chapter~\ref{chapter:chapter6}).

In Chapter~\ref{chapter:conclusions} we conclude the thesis and discuss directions for future work.

\section{Origins}
\label{section:introduction:origins}

Below we list which publication is the origin of each chapter.

\medskip\noindent%
\textbf{Chapter~\ref{chapter:acl2015}} is based on the conference paper: \bibentry{voskarides-learning-2015}~\cite{voskarides-learning-2015}. 

NV designed the method and ran the experiments. EM helped with algorithmic design. All authors conributed to the text, NV did most of the writing.

\medskip\noindent%
\textbf{Chapter~\ref{chapter:ecir2017}} is based on the conference paper: \bibentry{voskarides-generating-2017}~\cite{voskarides-generating-2017}.

NV designed the method and ran the experiments. All authors contributed to the text, NV did most of the writing.

\medskip\noindent%
\textbf{Chapter~\ref{chapter:sigir2018}} is based on the conference paper: \bibentry{voskarides-weakly-supervised-2018} \cite{voskarides-weakly-supervised-2018}.

NV designed the method and ran the experiments. EM, RR contributed to the experimental design. AK helped with the infrastructure. All authors contributed to the text, NV did most of the writing.

\medskip\noindent%
\textbf{Chapter~\ref{chapter:sigir2020}} is based on the conference paper: \bibentry{voskarides-2020-query} \cite{voskarides-2020-query}.

NV designed the method and ran the experiments. DL contributed to the experimental design and ran baseline models. All authors contributed to the text, NV did most of the writing.

\medskip\noindent
\textbf{Chapter~\ref{chapter:chapter6}} is based on the conference paper: \bibentry{voskarides-2020-event}~\cite{voskarides-2020-event}.

NV designed the method and ran the experiments. All authors contributed to the text, NV did most of the writing.

\medskip\noindent%
The thesis also indirectly benefited from insights gained from the following publications: 

\begin{itemize}

	\item \bibentry{voskarides-query-dependent-2014} \cite{voskarides-query-dependent-2014}.

	\item \bibentry{voskarides2019ilps} \cite{voskarides2019ilps}.

	\item \bibentry{sidiropoulos2020knowledge} \cite{sidiropoulos2020knowledge}.

	\item \bibentry{sarvi-2020-comparison} \cite{sarvi-2020-comparison}.

	\item \bibentry{krasakis-analysing-2020} \cite{krasakis-analysing-2020}.

\end{itemize}

\part{Making Structured Knowledge more Accessible to the User}

\graphicspath{{02-acl2015/Pictures/}}

\chapter{Retrieving Knowledge Graph Fact Descriptions}
\label{chapter:acl2015}

\footnote[]{This chapter was published as~\citep{voskarides-learning-2015}.}

In the first part of this thesis, we study how to make structured knowledge more accessible to the user.
In this chapter, we aim to answer \textbf{\ref{rq:retrieve-explanations}}: \acl{rq:retrieve-explanations}

\Acf{KG} facts express entity relationships in a formal form.
In the scope of this chapter we use the term ``explaining entity relationships'' as an alias for ``retrieving KG fact descriptions''.
\section{Introduction}
\label{intro-rq}
\label{intro} 

Knowledge graphs are a powerful tool for supporting a large spectrum of search applications including ranking, recommendation, exploratory search, and web search~\cite{Dong:2014:KVW:2623330.2623623}. A knowledge graph aggregates information around entities across multiple content sources and links these entities together, while at the same time providing entity-specific properties (such as age or employer) and types (such as actor or movie).

\begin{figure}[t]
    \centering
    \includegraphics[width=0.6\textwidth]{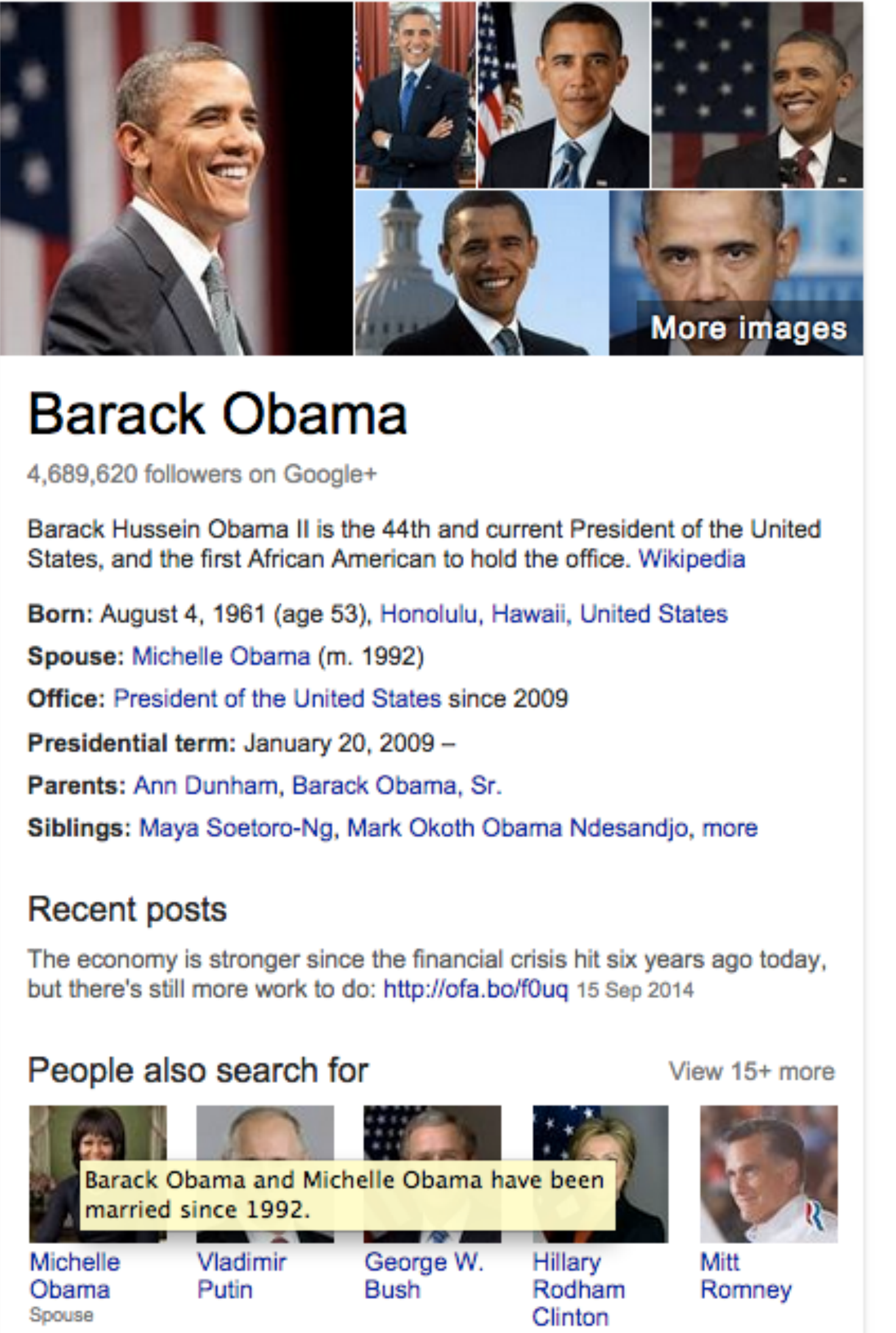}
    \caption{Part of Google's search result page for the query ``barack obama''. When hovering over the related entity ``Michelle Obama'', an explanation of the relationship between her and ``Barack Obama'' is shown.} 
    \label{fig:obama-kg}
\end{figure}

Although there is a growing interest in automatically constructing knowledge graphs, e.g., from unstructured web data~\cite{WestonBYU13,Craven200069,W12-3000}, the problem of providing evidence on why two entities are related in a knowledge graph remains largely unaddressed.
Extracting and presenting evidence for linking two entities, however, is an important aspect of knowledge graphs, as it can enforce trust between the user and a search engine, which in turn can improve long-term user engagement, e.g., in the context of related entity recommendation~\cite{blanco2013entity}.
Although knowledge graphs exist that provide this functionality to a certain degree (e.g., when hovering over Google's suggested entities, see Figure~\ref{fig:obama-kg}), to the best of our knowledge there is no previously published research on methods for entity relationship explanation. 

In this chapter we propose a method for explaining the relationship between two entities, which we evaluate on a newly constructed annotated dataset that we make publicly available.
In particular, we consider the task of explaining relationships between pairs of Wikipedia entities. We aim to infer a human-readable description for an entity pair given a relationship between the two entities. Since Wikipedia does not explicitly define relationships between entities we use a knowledge graph to obtain these relations. We cast our task as a sentence ranking problem: we automatically extract sentences from a corpus and rank them according to how well they describe a given relationship between a pair of entities.
For ranking purposes, we extract a rich set of features and use learning to rank to effectively combine them. Our feature set includes both traditional information retrieval and natural language processing features that we augment with entity-dependent features. These features leverage information from the structure of the knowledge graph. On top of this, we use features that capture the presence  in a sentence of the relationship of interest.
For our evaluation we focus on ``people'' entities and we use a large, manually annotated dataset of sentences.

We break down \textbf{\ref{rq:retrieve-explanations}} to three research sub-questions.
First, we ask what the effectiveness of state-of-the-art sentence retrieval models is for explaining a relationship between two entities (\textbf{RQ1.1}). Second, we consider whether we can improve over sentence retrieval models by casting the task in a learning to rank framework (\textbf{RQ1.2}). Third, we examine whether we can further improve performance by using relationship-dependent models instead of a relationship-independent one (\textbf{RQ1.3}). 
We complement these research questions with an error and feature analysis.

Our main contributions are a robust and effective method for explaining entity relationships, detailed insights into the performance of our method and features, and a manually annotated dataset.

\section{Related Work} %
\label{related-work} 
\label{rw-sentence-retrieval}

We combine ideas from sentence retrieval, learning to rank, and question answering to address the task of explaining relationships between entities.

Previous work that is closest to the task we address in this chapter is that of \citet{blanco2010finding} and \citet{fang2011rex}. First, \citet{blanco2010finding} focus on finding and ranking sentences that explain the relationship between an entity and a query. Our work is different in that we want to explain the relationship between two entities, rather than a query. \citet{fang2011rex} explore the generation of a ranked list of knowledge base relationships for an entity pair. Instead, we try to select sentences that describe a particular relationship, assuming that this is given.

Our approach builds on sentence retrieval, where one retrieves sentences rather than documents that answer an information need. Document retrieval models such as tf-idf, BM25, and language modeling~\cite{baeza1999modern} have been extended to tackle sentence retrieval. Three of the most successful sentence retrieval methods are TFISF~\cite{allan2003retrieval}, which is a variant of the vector space model with tf-idf weighting, language modeling with local context~\cite{murdock2006aspects,fernandez2011extending}, and a recursive version of TFISF that accounts for local context~\cite{doko2013recursive}. TFISF is very competitive compared to document retrieval models tuned specifically for sentence retrieval (e.g., BM25 and language modeling~\cite{losada2008study}) and we therefore include it as a baseline.

Sentences that are suitable for explaining relationships can have attributes that are important for ranking but cannot be captured by term-based retrieval models. One way to combine a wide range of ranking features is learning to rank (LTR). Recent years have witnessed a rapid increase in the work on learning to rank, and it has proven to be a very successful method for combining large numbers of ranking features, for web search, but also other information retrieval applications~\cite{burges2011learning,surdeanu2011learning,agarwal2012learning}. We use learning to rank and represent each sentence with a set of features that aim to capture different dimensions of the sentence. 

Question answering (QA) is the task of providing direct and concise answers to questions formed in natural language~\cite{hirschman2001natural}. QA can be regarded as a similar task to ours, assuming that the combination of entity pair and relationship form the ``question'' and that the ``answer'' is the sentence describing the relationship of interest. Even though we do not follow the QA paradigm in this chapter, some of the features we use are inspired by QA systems. In addition, we employ learning to rank to combine the devised features, which has recently been successfully applied for QA~\cite{surdeanu2011learning,agarwal2012learning}.

\section{Problem Statement}

We address the problem of explaining relationships between pairs of entities in a knowledge graph. We operationalize the problem as a problem of ranking sentences from documents in a corpus that is related to the knowledge graph. 
More specifically, given two entities $e_i$ and $e_j$ that form an entity pair $\langle e_i, e_j \rangle$, and a relation $r$ between them, the task is to extract a set of candidate sentences $S_{ij}=\{s_{ij_1}, \ldots, s_{ij_k}\}$ that refer to $\langle e_i, e_j \rangle$ and to impose a ranking on the sentences in $S_{ij}$. The relation $r$ has the general form $\langle \mathit{type}(e_i), \mathit{terms}(r), \mathit{type}(e_j)\rangle$, where $\mathit{type}(e)$ is the type of the entity $e$ (e.g., \texttt{Person} or \texttt{Actor}) and $\mathit{terms}(r)$ are the terms of the relation (e.g., \texttt{CoCastsWith} or \texttt{IsSpouseOf}). 

We are left with two specific tasks: (1)~extracting candidate sentences $S_{ij}$, and (2)~ranking $S_{ij}$,
where the goal is to have sentences that provide a perfect explanation of the relationship at the top position of the ranking. The next section describes our methods for both tasks.

\section{Explaining Entity Relationships}
\label{method}

We follow a two-step approach for automatically explaining relationships between
entity pairs. First, in Section~\ref{sentences-extraction}, we extract and enrich sentences that refer to an entity
pair $\langle e_i, e_j \rangle$ from a corpus in order to construct a set of candidate sentences. Second, in Section~\ref{sentences-ranking}, we extract a rich
set of features describing the entities' relationship $r$ and use supervised machine learning in order to rank the sentences in $S_{ij}$ according to how well they describe the relationship $r$. 

\subsection{Extracting candidate sentences}
\label{sentences-extraction}

To create a set of candidate sentences for a given entity pair and relationship, we require a corpus of documents that is pertinent to the entities at hand. Although any kind of document collection can be used, we focus on Wikipedia in this chapter, as it provides good coverage for the majority of entities in our knowledge graph.

First, we extract surface forms for the given entities: the title of
the entity's Wikipedia article (e.g., ``Barack Obama''), the titles of all
redirect pages linking to that article (e.g., ``Obama''), and all anchor text
associated with hyperlinks to the article within Wikipedia (e.g., ``president obama'').
We then split all Wikipedia articles into sentences and consider a sentence as
a candidate if (i)~the sentence is part of either entities' Wikipedia article
and contains a surface form of, or a link to, the other entity; or (ii)~the sentence contains
surface forms of, or links to, both entities in the entity pair.

Next, we apply two sentence enrichment steps for (i)~making sentences self-contained and readable outside the context of the source document and (ii)~linking the sentences to entities. For (i), we replace pronouns in candidate sentences with the title of the entity.
We apply a simple heuristic for the people entities, inspired by~\cite{wu2010open}:\footnote{We experimented with the Stanford co-reference resolution
system~\cite{lee2011stanford} and Apache OpenNLP and found that they were not able to consistently achieve the level of effectiveness that we require.} we count the frequency of the terms ``he'' and ``she'' in the article for determining the gender of the entity, and we replace the first appearance of ``he'' or ``she'' in each sentence with the entity's title. We skip this step if any surface form of the entity occurs in the sentence.

For (ii), we apply entity linking to provide links from the sentence to additional entities~\cite{milne2008learning}.
This need arises from the fact that not every sentence in an article contains explicit links to the entities it mentions, as Wikipedia guidelines only allow one link to another article in the article's text.\footnote{\url{{http://en.Wikipedia.org/wiki/Wikipedia:Manual_of_Style/Linking}}}
The algorithm takes a sentence as input and iterates over n-grams that are not yet
linked to an entity. If an n-gram matches a surface form of an entity, we
establish a link between the n-gram and the entity. We restrict our search
space to entities that are linked from within the source article of the
sentence and from within articles to which the source article links. This way,
our entity linking method achieves high precision as almost no disambiguation is
necessary.

As an example, consider the sentence ``He gave critically acclaimed performances in the crime thriller Seven\ldots'' on the Wikipedia page for Brad Pitt. After applying our enrichment steps, we obtain ``\texttt{Brad\_Pitt} gave critically acclaimed performances in the crime thriller \texttt{Seven}\ldots'', making the sentence human readable and link to the entities \texttt{Brad\_Pitt} and \texttt{Seven\_(1995\_film)}.

\subsection{Ranking sentences}
\label{sentences-ranking}
\label{features}

After extracting candidate sentences, we rank them by how well they describe the relationship of interest $r$ between entities $e_i$ and $e_j$. 
There are many signals beyond simple term statistics that can indicate relevance. Automatically constructing a ranking model using supervised machine learning techniques is therefore an obvious choice. 
For ranking we use learning to rank (LTR) and represent each sentence with a rich set of features.
Tables~\ref{tbl:features1} and~\ref{tbl:features2} list the features we use. Below we provide a brief description of the more complex ones.

\begin{table}[t]
\caption{Text and entity features used for sentence ranking.}
\label{tbl:features1}

\begin{tabularx}{\textwidth}{rlX}
\toprule
\# & Name & Gloss \\
\midrule
\multicolumn{2}{l}{\em Text features} & \\
1 & Sentence length & Length of $s$ in words \\
2 & Sum of $idf$ & Sum of IDF of terms of $s$ in Wikipedia \\
3 & Average $idf$ & Average IDF of terms of $s$ in Wikipedia\\
4 & Sentence density & Lexical density of $s$, see Equation~\ref{eq:density}~\cite{lee2001siteq} \\
5--8 & POS fractions & Fraction of verbs, nouns, adjectives, others in $s$ \cite{mintz2009distant} \\
\midrule
\multicolumn{2}{l}{\em Entity features} & \\
9 & \#entities & Total number of entities in $s$ \\
10 & Link to $e_i$ & Whether $s$ contains a link to the entity $e_i$ \\
11 & Link to $e_j$ & Whether $s$ contains a link to the entity $e_j$ \\
12 & Links to $e_i$ and $e_j$ & Whether $s$ contains links to both entities $e_i$ and $e_j$ \\
13 & Entity first & Is $e_i$ or $e_j$ the first entity in the sentence?\\
14 & Spread of $e_i$, $e_j$ & Distance between the last match of $e_i$ and $e_j$ in $s$~\cite{blanco2010finding} \\
15--22 & POS fractions left/right &  Fraction of verbs, nouns, adjectives, others to the left/right window of $e_i$ and $e_j$ in $s$ \cite{mintz2009distant} \\
23--25 & \#entities left/right/between & Number of entities to the left/right or between entities $e_i$ and $e_j$ in $s$ \\
26 & common links $e_i$, $e_j$ & Whether $s$ contains any common link of $e_i$ and $e_j$\\
27 & \#common links & The number of common links of $e_i$ and $e_j$ in $s$\\
28 & Score common links $e_i$, $e_j$ & Sum of the scores of the common links of $e_i$ and $e_j$ in $s$ \\
29--30 & \#common links prev/next & The number of common links of $e_i$ and $e_j$ in previous/next sentence of $s$ \\
\bottomrule
\end{tabularx}
\end{table}

\begin{table}[t]
\caption{Relationship and source features used for sentence ranking.}
\label{tbl:features2}
\begin{tabularx}{\textwidth}{rlX}
\toprule
\# & Name & Gloss \\
\midrule
\multicolumn{2}{l}{\em Relationship features} & \\
31 & Match $\mathit{terms}(r)$? & Whether $s$ contains any term in $\mathit{terms}(r)$ \\
32 & Match $\mathit{wordnet}(r)$? & Whether $s$ contains any phrase in $\mathit{wordnet}(r)$\\
33 & Match $\mathit{word2vec}(r)$? & Whether $s$ contains any phrase in $\mathit{word2vec}(r)$\\
34--36 & or's & Boolean OR of feature 31 and one or both of features 32 and 33\\
37--38 & {or(31, 32, 33)} prev/next & Boolean OR of features 31, 32, 33 for the previous/next sentence of $s$ \\
39 & Average $\mathit{word2vec}(r)$ & Average cosine similarity of phrases in $\mathit{word2vec}(r)$ that are matched in $s$\\
40 & Maximum $\mathit{word2vec}(r)$ & Maximum cosine similarity of phrases in $\mathit{word2vec}(r)$ that are matched in $s$\\
41 & Sum $\mathit{word2vec}(r)$ & Sum of cosine similarity of phrases in $\mathit{word2vec}(r)$ that are matched in $s$\\
42 & Score LC & Lucene score of $s$ with $\mathit{titles}(e_i,e_j)$, $\mathit{terms}(r)$, $\mathit{wordnet}(r)$, $\mathit{word2vec}(r)$ as query\\
43 & Score R-TFISF & R-TFISF score of $s$ with queries constructed as above\\ 
\midrule
\multicolumn{2}{l}{\em Source features} & \\
44 & Sentence position & Position of $s$ in document from which it originates \\
45 & From $e_i$ or $e_j$? & Does $s$ originate from the Wikipedia article of $e_i$ or $e_j$?\\
46 & \#($e_i$ or $e_j$) & Number of occurrences of $e_i$ or $e_j$ in document from which $s$ originates, inspired by document smoothing for sentence retrieval~\cite{murdock2005translation}\\
\bottomrule
\end{tabularx}
\end{table}

\paragraph{Text features}
This feature type regards the importance of the sentence $s$ at the term level. 
We compute the density of $s$ (feature 4) as:
\begin{align}
	\label{eq:density}
	density(s) = \frac{1}{K \cdot (K+1)} 
	\sum_{j=1}^n \frac{idf(t_j) \cdot idf(t_{j+1})}{distance(t_j,t_{j+1})^2},
\end{align}
where $K$ is the number of keyword terms in $s$ and $distance(t_j, t_{j+1})$ is the number of non-keyword terms between keyword terms $t_j$ and $t_{j+1}$. We treat stop words and numbers in $s$ as non-keywords and the remaining terms as keywords. Features 5--8 capture the distribution of part-of-speech tags in the sentence.

\paragraph{Entity features}
These features partly build on~\cite{tsagkias:linking:2011,meij2012adding} and describe the entities and are dependent on the knowledge graph. 
Whether $e_i$ or $e_j$ is the first appearing entity in a sentence might be an indicator of importance (feature 13). The spread of $e_i$ and $e_j$ in the sentence (feature 14) might be an indicator of their centrality in the sentence~\cite{blanco2010finding}.
Features 15--22 capture the distribution of part-of-speech tags in the sentence  in a window of four words around $e_i$ or $e_j$ in $s$~\cite{mintz2009distant}, complemented by the number of entities between, to the left of, and to the right of the entity pair (features 23--25).

We assume that two articles that have many common articles that point to them are strongly related~\cite{witten2008effective}. We hypothesize that, if a sentence contains common inlinks from $e_i$ and $e_j$, the sentence might contain important information about their relationship. Hence, we add whether the sentence contains a common link (feature 26) and the number of common links (feature 27) as features.
We score a common link $l$ between $e_i$ and $e_j$ using:
\begin{align}
	score(l, e_i, e_j) = sim(l, e_i) \cdot sim(l, e_j),
	\label{eq:simCommonLinks}
\end{align}
where $\mathit{sim}(\cdot, \cdot)$ is defined as the similarity between two Wikipedia articles, computed using a variant of Normalized Google Distance~\cite{witten2008effective}.
Feature 28 then measures the sum of the scores of the common links.

Previous research shows that using surrounding sentences is beneficial for sentence retrieval~\cite{doko2013recursive}. We therefore consider the number of common links in the previous and next sentence (features 29--30).

\paragraph{Relationship features}
\label{relation-features}
Feature 31 indicates whether any of the relationship-specific terms occurs in the sentence. Only matching the terms in the relationship may have low coverage since terms such as ``spouse'' may have many synonyms and/or highly related terms, e.g., ``husband'' or ``married''. 
Therefore, we use WordNet to find synonym phrases of $r$ (feature 32); we refer to this method as $\mathit{wordnet}(r)$.

Alternatively, we use word embeddings to find such similar phrases~\cite{mikolov2013distributed}. Such embeddings take a text corpus as input and learn vector representations of words and phrases consisting of real numbers.
Given the set $V_r$ consisting of the vector representations of all the relationship terms and the set $V$ which consists of the vector representations of all the candidate phrases in the data, we calculate the distance between a candidate phrase represented by a vector $\mathbf{v}_i \in V$ and the vectors in  $V_r$ as:
\begin{align}
	\mathit{distance}(\mathbf{v}_i, V) = \cos\left(\mathbf{v}_i, \sum_{\mathbf{v}_j \in V_r}\mathbf{v}_j\right),
	\label{eq:distance}
\end{align}
where $\sum_{\mathbf{v}_j \in V_r}\mathbf{v}_j$ is the element-wise sum of the vectors in $V_r$ and the distance between two vectors $\mathbf{v}_1$ and $\mathbf{v}_2$ is measured using cosine similarity.
The candidate phrases in $V$ are then ranked using Equation~\ref{eq:distance} and the top-$m$ phrases are selected, resulting in features 33, 39, 40, and 41; we refer to the ranked set of phrases that are selected using this procedure as $\mathit{word2vec}(r)$.

In addition, we employ state-of-the-art retrieval functions and include the scores for queries that are constructed using the entities $e_i$ and $e_j$, the relation $r$, $\mathit{wordnet}(r)$, and $\mathit{word2vec}(r)$. We use the titles of the entity articles $\mathit{titles}(e)$ to represent the entities in the query and two ranking functions, Recursive TFISF (R-TFISF) and LC,\footnote{In preliminary experiments R-TFISF and LC were the best performing among a pool of sentence retrieval methods.} (features 42--43). 
TFISF is a sentence retrieval model that determines the level of relevance of a sentence $s$ given a query $q$ as:
\begin{align}
	R(s,q) = \sum_{t \in q} \log(\mathit{tf}_{t,q} + 1) \cdot \log(\mathit{tf}_{t,s} + 1) \cdot \log\left(\frac{n+1}{0.5+\mathit{sf}_t}\right),
	\label{eq:tfisf}
\end{align}
where $\mathit{tf}_{t,q}$ and $\mathit{tf}_{t,s}$ are the number of occurrences of term $t$ in the query $q$ and the sentence $s$ respectively, $\mathit{sf}_t$ is the number of sentences in which $t$ appears, and $n$ is the number of sentences in the collection.
R-TFISF is an improved extension of the TFISF method~\cite{doko2013recursive}, which incorporates context from neighboring sentences in the ranking function:
\begin{align}
\label{eq:rtfisf}
	R_c(s,q) = (1-\mu) R(s,q) + 
	  \mu [R_c(s_{prev}(s),q) + R_c(s_{next}(s),q)],\nonumber
\end{align}
where $\mu$ is a free parameter and $s_{prev}(s)$ and $s_{next}(s)$ indicate functions to retrieve the previous and next sentence, respectively. We use a maximum of three recursive calls. 

\paragraph{Source features} Here, we refer to features that are dependent on the source document of the sentences. We have three such features.

\section{Experimental Setup}
\label{experimental-setup}
\label{expsetup-annotations}

In this section we describe the dataset, 
manual annotations, learning to rank algorithm, and evaluation metrics that we use to answer our
research questions.

\subsection{Dataset}

\label{expsetup-entitypairs}
We draw entities and their relationships from a proprietary knowledge graph that is created from Wikipedia, Freebase, IMDB, and other sources, and that is used by the Yahoo web search engine. We focus on ``people'' entities and relationships between them.\footnote{Note that, except for the co-reference resolution step described in Section~\ref{sentences-extraction}, our method does not depend on this restriction.} For our experiments we need to select a manageable set of entities, which we obtain as follows.
We consider a year of query logs from a large commercial search engine, count the number of times a user clicks on a Wikipedia article of an entity in the results page and perform stratified sampling of entities according to this distribution. As we are bounded by limited resources for our manual assessments, we sample 1476 entity pairs that together with nine unique relationship types form our experimental dataset. 

We use an English Wikipedia dump dated July 8, 2013, containing approximately 4M articles, of which 50638 belong to ``people'' entities that are also in our knowledge graph. We extract sentences using the approach described in Section~\ref{sentences-extraction}, resulting in 36823 candidate sentences for our entities. 
On average we have {24.94} sentences per entity pair (maximum {423} and minimum {0}). Because of the large variance, it is not feasible to obtain exhaustive annotations for all sentences. We rank the sentences using R-TFISF and keep the top-10 sentences per entity pair for annotation. 
This results in a total of 5689 sentences.

Five human annotators provided relevance judgments, manually judging sentences based on how well they describe the relationship for an entity pair, for which we use a five-level graded relevance scale (perfect, excellent, good, fair, bad).\footnote{\url{https://github.com/nickvosk/acl2015-dataset-learning-to-explain-entity-relationships}}
Of all relevance grades 8.1\% is perfect, 15.69\% excellent, 19.98\% good, 8.05\% fair, and 48.15\% bad. Out of 1476 entity pairs, 1093 have at least one sentence annotated as fair. As is common in information retrieval evaluation, we discard entity pairs that have only ``bad'' sentences. 
We examine the difficulty of the task for human annotators by measuring inter-annotator agreement on a subset of 105 sentences that are judged by 3 annotators.
Fleiss' kappa is $k=0.449$, which is considered to be moderate agreement.

\subsection{Machine learning}
For ranking sentences we use a Random Forest (RF) classifier~\cite{ML:2001:breiman}.\footnote{In preliminary experiments, we contrasted RF with
gradient boosted regression trees and LambdaMART and found that RF consistently
outperformed other methods.} We set the number of iterations to 300 and the
sampling rate to 0.3. Experiments with varying these two parameters did not
show any significant differences. We also tried several feature normalization
methods, none of them being able to significantly outperform the runs without
feature normalization.

We obtain POS tags using the Stanford part-of-speech tagger and filter out a standard list of 33 English stopwords. For the word embeddings we use $\mathit{word2vec}$ and train our model on all text in Wikipedia using negative sampling and the continuous bag of words architecture. We set the size of the phrase vectors to 500 and $m = 30$.

\subsection{Evaluation metrics} 

We employ two main evaluation metrics in our experiments, NDCG~\cite{jarvelin2002cumulated} and ERR~\cite{chapelle2009expected}. The former measures the total accumulated gain from the top of the ranking that is discounted at lower ranks and is normalized by the ideal cumulative gain. The latter models user behavior and measures the expected reciprocal rank at which a user will stop her search.
We consider these ranking-based graded evaluation metrics at two cut-off points: position 1, corresponding to showing a single sentence to a user, and 10, which accounts for users who might look at more results. We report on NDCG@1, NDCG@10, ERR@1, ERR@10, and 
Exc@1, which indicates whether we have an ``excellent'' or ``perfect'' sentence at the top of the ranking. Likewise, Per@1 indicates whether we have a ``perfect'' sentence at the top of the ranking (not all entity pairs have an excellent or a perfect sentence).

We perform 5-fold cross validation and test for statistical significance using a paired two-tailed t-test.
We depict a significant difference in performance for $p<0.01$ with $^\blacktriangle$ (gain) and $^\blacktriangledown$ (loss) and for $p<0.05$ with $^\vartriangle$ (gain) and $^\triangledown$ (loss). Boldface indicates the best score for a metric. 

\section{Results and Analysis}
\label{results-discussion} 

\begin{table}[t]
   \caption{\label{tab:baselines} Results for five baseline variants. See text for their description and significant differences. }
\centering

  \begin{tabular}{l cccc}
  \toprule
   Baseline &  NDCG@1 & NDCG@10 & ERR@1 & ERR@10\\ 
   \midrule
  B1 &  0.7508 & 0.8961 & 0.3577 &	0.4531\\ 
  B2 & 0.7511 & 0.8958 & 0.3584	& 0.4530\\
  B3 & 0.7595 & 0.8997 & 0.3696 &	0.4600\\
  B4 & 0.7767 & 0.9070 & 0.3774	& 0.4672\\
  B5 & \bf 0.7801 & \bf 0.9093 &  \bf 0.3787	& \bf 0.4682\\
  \bottomrule
  \end{tabular}
  
\end{table}

We compare the performance of typical document retrieval models and state-of-the-art sentence retrieval models in order to answer \textbf{RQ1.1}.
We consider five sentence retrieval models: Lucene ranking (LC), language modeling with Dirichlet smoothing (LM), BM25, TFISF, and
Recursive TF-ISF (R-TFISF).
We follow related work and set $\mu=0.1$ for R-TFISF, $k = 1$ and $b = 0$ for BM25 and $\mu=250$ for LM \cite{fernandez2011extending}.

In our experiments, a query $q$ is constructed using various combinations of surface forms of the two entities $e_i$ and $e_j$ and the relationship $r$. Each entity in the entity pair can be represented by its title, the titles of any redirect pages pointing to the entity's article, the n-grams used as anchors in Wikipedia to link to the article of the entity, or the union of them all. The relationship $r$ can be represented by the terms in the relationship, synonyms in $\mathit{wordnet}(r)$, or by phrases in $\mathit{word2vec}(r)$.

First, we fix the way we represent $r$. Baseline B1 does not include any representation of $r$ in the query. B2 includes the relationship terms of $r$, while B3 includes the relationship terms of $r$ and the synonyms in $\mathit{wordnet}(r)$. B4 includes the terms of $r$ and the phrases in $\mathit{word2vec}(r)$, and B5 includes the relationship terms of $r$, the synonyms in $\mathit{wordnet}(r)$ and the phrases in $\mathit{word2vec}(r)$.
Combining these variations with the entity representations, we find that all combinations that use the titles as representation and R-TFISF as the retrieval function outperform all other combinations.%
This can be explained by the fact that titles are least ambiguous, thus reducing the possibility of accidentally referring to other entities. BM25 and LC perform almost as well as R-TFISF, with only insignificant differences in performance.

Table~\ref{tab:baselines} shows the best performing combination of each baseline, i.e., varying the representation of $r$ and using titles and R-TFISF. B4 and B5 are the best performing baselines, suggesting that $\mathit{word2vec}(r)$ and $\mathit{wordnet}(r)$ are beneficial. B5 significantly outperforms all  baselines except B4. 

We also experiment with a supervised combination of the baseline rankers using LTR. Here, we consider each baseline ranker as a separate feature and train a ranking model. The trained model is not able to outperform the best individual baseline, however.

\subsection{Learning to rank sentences}

Next, we provide the results of our method using the features described in Section~\ref{features}, exploring whether our learning to rank (LTR) approach improves over sentence retrieval models (\textbf{RQ1.2}).
We compare an LTR model using the features in Tables~\ref{tbl:features1} and~\ref{tbl:features2} against the best baseline (B5).
\begin{table*}[th]
\caption{Results for the best baseline (B5) and the learning to rank method (LTR).}
\label{tab:fullVSbaselines}    
\centering
  \begin{tabular}{lrrllllllll}
  \toprule
   Has one & \# pairs & \# sentences & Method &   NDCG@1 & NDCG@10 & ERR@1 &  ERR@10 &  Exc@1 &  Per@1\\ \midrule
      fair & 1093 & 4435 & B5 & 0.7801& 0.9093 & 0.3787 & 0.4682 & -- & -- \\
       &  & & LTR & \bf 0.8489$^\blacktriangle$ & \bf 0.9375$^\blacktriangle$ &  \bf  0.4242$^\blacktriangle$ &   \bf 0.4980$^\blacktriangle$  & -- & -- \\ 
  \midrule
      good & 1038 & 4285 &  B5 & 0.7742 & 0.9078 & 0.3958 & 0.4894  & -- & -- \\
      &   & &LTR &  \bf 0.8486$^\blacktriangle$ &   \bf 0.9374$^\blacktriangle$ &   \bf 0.4438$^\blacktriangle$ &   \bf 0.5208$^\blacktriangle$  & -- & -- \\ 
  \midrule
      excellent & 752 & 3387  & B5 & 0.7455 & 0.8999 & 0.4858 & 0.5981 & 0.7314 & --\\
       &  & & LTR & \bf  0.8372$^\blacktriangle$ &  \bf  0.9340$^\blacktriangle$ &  \bf 0.5500$^\blacktriangle$  &  \bf 0.6391$^\blacktriangle$ &  \bf 0.8298$^\blacktriangle$ & --\\
  \midrule
     perfect & 339 & 1687 & B5 & 0.7082 & 0.8805 & 0.6639 & 0.7878 & 0.7729 & 0.6136\\
    &  & &  LTR & \bf 0.8150$^\blacktriangle$ &  \bf 0.9245$^\blacktriangle$  &  \bf 0.7640$^\blacktriangle$  &  \bf 0.8518$^\blacktriangle$ &  \bf 0.8909$^\blacktriangle$ &  \bf 0.7227$^\blacktriangle$ \\ 
  \bottomrule
  \end{tabular}
\end{table*}

Table~\ref{tab:fullVSbaselines} shows the results. Each group in the table contains the results for the entity pairs that have at least one candidate sentence of that relevance grade for B5 and LTR.

We find that LTR significantly outperforms B5 by a large margin. 
The absolute performance difference between LTR and B5 becomes larger for all metrics as we move from ``fair'' to ``perfect,'' which shows that LTR is more robust than the baseline for entity pairs that have at least one high quality candidate sentence.
LTR ranks the best possible sentence at the top of the ranking for $\sim$83\% of the cases for entity pairs that contain an ``excellent'' sentence and for $\sim$72\% of the cases for entity pairs that contain a ``perfect'' sentence. 

Note that, as indicated in Section~\ref{expsetup-entitypairs}, we discard entity pairs that have only ``bad'' sentences in our experiments. For the sake of completeness, we report on the results for all entity pairs in our dataset---including those without any relevant sentences---in Table~\ref{tab:fullVSbaselinesBad}. %
\begin{table*}[th]
\caption{Results for the best baseline (B5) and the learning to rank method (LTR), using all entity pairs in the dataset, including those without any relevant sentences.}
\label{tab:fullVSbaselinesBad}   
\centering
{\small
      \begin{tabular}{lrrllllllll}
      \toprule
          Has one & \# pairs & \# sentences & Method &   NDCG@1 & NDCG@10 & ERR@1 &  ERR@10 &  Exc@1 &  Per@1\\ 
       \midrule
       - & 1476 & 5689 & B5 & 0.5776 & 0.6733 & 0.2804 & 0.3467 & -- & -- \\
       & & & LTR & \bf 0.6285$^\blacktriangle$ & \bf 0.6940$^\blacktriangle$ &  \bf  0.3155$^\blacktriangle$ &   \bf 0.3694$^\blacktriangle$   & -- & -- \\ 
      \bottomrule
      \end{tabular}
  }
\end{table*}

\subsection{Relationship-dependent models}
\begin{table*}[th]
\caption{Results for relationship-dependent models. Similar relationships are grouped together.}
\label{tab:perRel}
\centering
  \begin{tabular}{p{7cm}rrcccc}
  \toprule
   Relationship &  \# pairs & \# sentences & NDCG@1 & NDCG@10 & ERR@1 &  ERR@10 \\ \midrule
   $\langle \mathit{MovieActor}, \mathit{CoCastsWith}, \mathit{MovieActor} \rangle$  & 410 & 1403 & 0.8604 & 0.9436 & 0.3809 & 0.4546\\ \hline
   $\langle \mathit{TvActor}, \mathit{CoCastsWith}, \mathit{TvActor} \rangle$ & 210 & 626 & 0.8729  & 0.9482 & 0.3271  & 0.3845 \\ \hline
  $\langle \mathit{MovieActor}, \mathit{IsDirectedBy}, \mathit{MovieDirector} \rangle$ \\ 
  $\langle \mathit{MovieDirector}, \mathit{Directs}, \mathit{MovieActor} \rangle$ &  112 & 492 & 0.8795  & 0.9396 & 0.4709 & 0.5261 \\ \hline
  $\langle \mathit{Person}, \mathit{isChildOf}, \mathit{Person} \rangle$ \\ \small{$\langle \mathit{Person}, \mathit{isParentOf}, \mathit{Person} \rangle$} &  108 & 716 & 0.8428 & 0.9081 &  0.6395   & 0.7136 \\ \hline
  $\langle \mathit{Person}, \mathit{isPartnerOf}, \mathit{Person} \rangle$ \\ \small{$\langle \mathit{Person}, \mathit{isSpouseOf}, \mathit{Person} \rangle$}&  155 & 877 & 0.8623& 0.9441 & 0.6153 & 0.6939\\ \hline
  $\langle \mathit{Athlete}, \mathit{PlaysSameSportTeamAs}, \mathit{Athlete} \rangle$  & 98 & 321  &  0.8787 &  0.9535 & 0.3350  & 0.3996\\
   \midrule
      Average results over all data  & 1093  & 4435 &\bf 0.8661 & \bf  0.9395 & \bf 0.4615 & \bf 0.5287 \\
      LTR (Table~\ref{tab:fullVSbaselines}; fair) &  &  & 0.8489 & 0.9375 & 0.4242 & 0.4980 \\ 
        
  \bottomrule
  \end{tabular}
\end{table*}
\begin{table}[th]
\caption{Results using relationship-dependent models, removing individual feature types.}
\label{tab:type-all} 
  \centering
  \begin{tabular}{lllll}
  \toprule
   Features & NDCG@1 & NDCG@10 & ERR@1 & ERR@10\\ \midrule    
 All  &   \bf 0.8661 &  \bf 0.9395 &  \bf 0.4615 & \bf 0.5287\\
  \midrule
  All$\setminus$text   & 0.8620 & 0.9372 & 0.4606 & 0.5274\\ 
  All$\setminus$source & 0.8598 & 0.9372 & 0.4582 & 0.5261\\
    All$\setminus$entity  & 0.8421$^\triangledown$ &  0.9282$^\blacktriangledown$  & 0.4497 & 0.5202$^\triangledown$\\
      All$\setminus$relation  & 0.8183$^\blacktriangledown$ & 0.9201$^\blacktriangledown$ & 0.4352$^\blacktriangledown$ & 0.5112$^\blacktriangledown$\\
  \bottomrule
  \end{tabular}
\end{table}

Relevant sentences may have different properties for different relationship types. For example, a sentence describing two entities being partners would have a different form than one describing that two entities costar in a movie. A similar idea was investigated in the context of QA for associating question and answer types~\cite{yao2013automatic}.
To answer \textbf{RQ1.3} we examine whether learning a relationship-dependent model improves over learning a single model for all types. 
We split our dataset per relationship type and train a model per type using 5-fold cross-validation within each. Table \ref{tab:perRel} shows the results.
Our method is robust across different relationships in terms of NDCG. However, we observe some variation in ERR as this metric is more sensitive to the distribution of relevant items than NDCG---the distribution over relevance grades varies per relationship type. For example, it is much more likely to find candidate sentences that have a high relevance grade for {$\langle \mathit{Person}$, $\mathit{isSpouseOf}$, $\mathit{Person} \rangle$} than for {$\langle \mathit{Athlete}$, $\mathit{PlaysSameSportTeamAs}$, $\mathit{Athlete} \rangle$} in our dataset. 
We plan to address this issue by exploring other corpora in the future.

The second-to-last row in Table~\ref{tab:perRel} shows the averaged results over the different relationship types, which is a significant improvement over LTR at $p<0.01$ for all metrics.
This method ranks the best possible sentence at the top of the ranking for $\sim$85\% of the cases for entity pairs that contain an ``excellent'' sentence ($\sim$2\% absolute improvement over LTR) and for $\sim$75\% of the cases for entity pairs that contain a ``perfect'' sentence ($\sim$3\% absolute improvement over LTR). 

\subsection{Feature type analysis}
\label{results-feature-analysis}

Next, we analyze the impact of the feature types.
Table~\ref{tab:type-all} shows how performance varies when removing one feature type at a time from the full feature set.
Relationship type features are the most important, although entity type features are important as well. This indicates that introducing features based on entities identified in the sentences and the relationship is beneficial for this task.
Furthermore, the limited dependency on the source feature type indicates that our method might be able to generalize in other domains.
Finally, text type features do contribute to retrieval effectiveness, although not significantly.
Note that calculating the sentence features is straightforward, as none of our features requires heavy linguistic analysis.

\subsection{Error analysis}
\label{results-insights}
When looking at errors made by the system, we find that some are due to the fact that entity pairs might have more than one relationship (e.g., actors that costar in movies also being partners) but the selected sentence covers only one of the relationships.\footnote{The annotators marked sentences that do not refer to the relationship of interest as ``bad'' but indicated whether they describe another relationship or not. We plan to account for such cases in future work.} For example, \texttt{Liza Minnelli} is the daughter of \texttt{Judy Garland}, but they have also costarred in a movie, which is the relationship of interest. The model ranks the sentence ``Liza Minnelli is the daughter of singer and actress Judy Garland\ldots'' at the top, while the most relevant sentence is: ``Judy Garland performed at the London Palladium with her then 18-year-old daughter Liza Minnelli in November 1964.''

Sentences that contain the relationship in which we are interested, but for which this cannot be directly inferred, are another source of error.
Consider, for example, the following sentence, which explains director \texttt{Christopher Nolan} directed actor \texttt{Christian Bale}: ``Jackman starred in the 2006 film The Prestige, directed by Christopher Nolan and costarring Christian Bale, Michael Caine, and Scarlett Johansson''.  Even though the sentence contains the relationship of interest, it focuses on actor \texttt{Hugh Jackman}. The sentence ``In 2004, after completing filming for The Machinist, Bale won the coveted role of Batman and his alter ego Bruce Wayne in Christopher Nolan's Batman Begins\ldots'', in contrast, refers to the two entities and the relationship of interest directly, resulting in a higher relevance grade. Our method, however, ranks the first sentence on top, as it contains more phrases that refer to the relationship. 

\section{Conclusions and Future Work} %
\label{conclusion}

We have presented a method for explaining relationships between knowledge graph entities with human-readable descriptions. 
We first extract and enrich sentences that refer to an entity pair and then rank the sentences according to how well they describe the relationship. For ranking, we use learning to rank with a diverse set of features.
Evaluation on a manually annotated dataset of ``people'' entities shows that our method significantly outperforms state-of-the-art sentence retrieval models for this task. Experimental results also show that using relationship-dependent models is beneficial.

In future work we aim to evaluate how our method performs on entities and relationships of any type and popularity, including tail entities and miscellaneous relationships. We also want to investigate moving beyond Wikipedia and extract candidate sentences from documents that are not related to the knowledge graph, such as web pages or news articles. Employing such documents also implies an investigation into more advanced co-reference resolution methods. 

Our analysis showed that sentences may cover different relationships between entities or different aspects of a single relationship---we aim to account for such cases in follow-up work.
Furthermore, sentences may contain unnecessary information for explaining the relation of interest between two entities. Especially when we want to show the obtained results to end users, we may need to apply further processing of the sentences to improve their quality and readability. We would like to explore sentence compression techniques to address this. 
Finally, relationships between entities have an inherit temporal nature and we aim to explore ways to explain entity relationships and their changes over time.

In this chapter, we studied the task of retrieving existing KG fact descriptions (explaining entity relationships).
In the next chapter, we study how to generate such descriptions instead of retrieving existing ones.

\chapter{Generating Knowledge Graph Fact Descriptions}
\label{chapter:ecir2017}

\footnote[]{This chapter was published as~\citep{voskarides-generating-2017}.}

In the previous chapter, we studied how to retrieve existing KG fact descriptions.
However, a scenario where a description for a KG fact does not exist in the underlying text corpus is not unlikely. 
Therefore, in this chapter, we aim to answer \textbf{\ref{rq:generate-explanations}}: \acl{rq:generate-explanations}
As in the previous chapter, we use the term ``entity relationship'' to refer to a KG fact.

\section{Introduction}

Results displayed on a modern search engine result page (SERP) are sourced from multiple, heterogeneous sources. For so-called organic results it has been known for a long time that result snippets, i.e., brief descriptions explaining the result item and its relation to the query, positively influence the user experience~\cite{tombros1998advantages}.  In this chapter, we focus on generating descriptions for results sourced from another important ingredient of modern SERPs: knowledge graphs.
Knowledge graphs (KGs) contain information about entities and their relationships.
A large and diverse set of search applications utilize KGs to improve the user experience.
For instance, web search engines try to identify KG entities in queries and augment their result pages with knowledge graph panels that provide contextual entity information~\cite{lin2012active,blanco2015fast}.
Such panels usually focus on a single entity and may include attributes of the entity and other, related entities.

Entities can be connected with more than one relationship in a KG, however.
For example, two actors might have appeared in the same film, be born in the same country and also be partners.
Recent work has focused on finding relationships between a pair of entities and ranking the relationships by a predefined relevance criterion \cite{fang2011rex}.
When using relationships in real-world search applications, with SERPs being the prime example, a crucial problem  is that they are typically represented in a formal manner that is not suitable to present to an end user.
Instead, human-readable descriptions that verbalize and provide context about entity relationships are more natural to use~\cite{gkatzia2016natural}.
They can be used, e.g., for entity recommendations~\cite{blanco2013entity} or for KG-based timeline generation~\cite{althoff2015timemachine}.

Descriptions of KG relationships themselves are usually not included in large-scale knowledge graphs and previous work on automatically generating such descriptions has either relied on hand-crafted templates~\cite{althoff2015timemachine} or on external text corpora~\cite{voskarides-learning-2015}. 
The main limitations of the former are that manually creating these templates is expensive, not generalizable, and thus it does not scale well. 
The latter approach is limited as the underlying text corpus may not contain descriptions for all certain relationship instances;
it will not produce meaningful results for instances that do not appear in the text corpus. 

We propose a method that overcomes these limitations by automatically generating descriptions of KG entity relationships.
Since there exist textual descriptions of a certain relationship for some relationship instances, we aim to use these descriptions to learn how the relationship is generally expressed in text and use this information to generate descriptions for other instances of the same relationship.
Existing relationship descriptions are usually complex and tailored to the entities they discuss.
Also, it is likely that the KG does not contain all the information included in a description.
For example, the KG might not contain any information about the second part of the following sentence: ``\textit{Catherine Zeta-Jones starred in the romantic comedy The Rebound, in which she played a 40-year-old mother of two} \ldots''.
Nevertheless, descriptions of the same relationship share patterns that are specific to that relationship.
Therefore, we first create sentence templates for a certain relationship and then, for a new relationship instance, we select appropriate templates, which we formulate as a ranking problem, and fill them with the appropriate entities to generate a description.

We propose a method that generates descriptions of entity relationships for a relationship instance given a knowledge graph and a set of relationship instances coupled with their descriptions; we evaluate this method using an automatic and manual evaluation method, and release the datasets used to the community.\footnote{\url{https://github.com/nickvosk/ecir2017-gder-dataset/}}  We show that we generate contextually rich relationship descriptions that are meant to be valid under the KG closed-world assumption. Moreover, our template-based method is naturally robust against KG incompleteness, since in the case of lack of contextual information about the relationship instance, it can still generate a basic description.

\section{Related Work}

Web search engine result pages (SERPs) can be augmented with information about the query and the documents from KGs in order to improve the user experience~\cite{lin2012active}.
Also, SERPs can be augmented with textual descriptions and/or summaries with a prominent example being snippet generation for web search~\cite{tombros1998advantages,turpin2007fast}.
Closest to our setting, relationship descriptions have been studied in the context of providing evidence for entity recommendation for web search~\cite{voskarides-learning-2015} and timeline generation for knowledge base entities~\cite{althoff2015timemachine}.
Our task, generating a description of a relationship instance given a KG, is similar to event headline generation, where the task is to generate a short sentence that summarizes a specific event.
Similar to our templates, the headline patterns constructed in~\cite{pighin2014modelling} consist of words and entity slots.
Our method differs however, since relationships are more general than events and we thus have to deal with ambiguity at generation time when selecting which template matches a relationship instance.

Our task is also similar to concept-to-text generation, where the task is to generate a textual description given a set of database records~\cite{reiter2000building}. %
In this context, our task is most closely related to~\cite{saldanha2016entity,lebret2016generating}.
\citet{saldanha2016entity} use a template-based approach for generating company descriptions from Freebase.
They construct sentence templates by replacing the entities in existing sentences by the Freebase relation of the entity to the company (e.g., $\langle company \rangle$ was founded by $\langle founder \rangle$).
They add a preprocessing step where they remove phrases from the sentence that contain entities that are not connected to the company directly.
At generation time, the authors replace the entity slots with the appropriate entities.
\citet{lebret2016generating} propose a neural model to generate the first sentence of a person's biography in Wikipedia conditioned on Wikipedia infoboxes.
Our setting is different from these papers since our generated descriptions are neither restricted to having entities that are directly connected to the subject entity in a KG nor need they be contained in a Wikipedia infobox.

\section{Problem Definition}
\label{sec:prob-def}

In this section we formally define the task of generating descriptions of entity relationships.
Table~\ref{tab:glossary} lists the main notation we use in this chapter.

\begin{table}[t]
\caption{Glossary.}
\label{tab:glossary}
\begin{tabularx}{\textwidth}{X p{10cm}}
\toprule
\bf Symbol & \bf Description
\\
\midrule
$\mathcal{K}$ 				& knowledge graph  \\
$\mathcal{E}$	 			& set of entities\\
$\mathcal{P}$ 				& set of predicates\\
$\langle s, p, o \rangle$ 	& knowledge graph triple with $s, o \in \mathcal{E}$ and $p \in \mathcal{P}$ \\
$v$ 						& word in vocabulary $\mathcal{V}$ \\
$a$ 						& sentence \\
$r_i$ 						& relationship instance of relationship $r$ \\
$T_r$ 						& set of templates $t \in T_r$ for relationship $r$\\
$R_t$						& set of relationship instances that support the template $t$\\
$X$ 						& set of pairs $\langle r_{i'}, y' \rangle$, where $y'$ is a textual description (a single sentence) \\
$C$ 						& mapping from an entity to an entity cluster \\
$K$ 						& entity dependency graph of a sentence \\
$G$ 						& compression graph \\
$P$ 						& set of paths in $G$ \\
\bottomrule
\end{tabularx}
\end{table}

\subsection{Prelimilaries}

Let $\mathcal{E}$ be a set of entities and $\mathcal{P}$ a set of predicates. 
A \emph{knowledge graph} $\mathcal{K}$ is a set of triples $\langle s, p, o \rangle$, where $s, o \in \mathcal{E}$ and $p \in \mathcal{P}$.
We follow the closed-world assumption for $\mathcal{K}$ and use Freebase as our knowledge graph~\cite{bollacker2008freebase,nickel2015review}.
A \emph{sentence} $a$ is a sequence of words $[v_1, \ldots, v_n]$, where each $v_i \in a$ is also in $\mathcal{V}$. Non-overlapping sub-sequences of $a$ might refer to a single entity $e\in\mathcal{E}$.

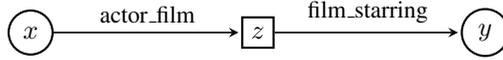
\begin{figure}[t]
     \centering
    \begin{tikzpicture}[
			align=center,
            > = stealth, %
            shorten > = 1pt, %
            auto,
            node distance = 3cm, %
            semithick %
        ]

        \tikzstyle{every state}=[
            draw = black,
            thick,
            fill = white,
            minimum size = 4mm
        ]

        \node[state] (x) {$x$};
        \node[state, rectangle] (z) [right of=x] {$z$};
        \node[state] (y) [right of=z]{$y$};

		\path[->] (x) edge node[midway, above] {\small actor\_film}  (z);
		\path[->] (z) edge node[midway, above] {\small film\_starring} (y);
    \end{tikzpicture}
        \caption{Graphical representation of the logical form of the $\mathit{starsInFilm}$ relationship.
	 Lambda variables are shown in circles and existential variables in rectangles.
        }
        \label{fig:relationship-examples}
\end{figure}

A \emph{relationship} $r$ is a logical form in $\lambda$-calculus that consists of two lambda variables ($x$ and $y$), at least one predicate, and zero or one existential variables \cite{yih2015semantic}.
Lambda variables can be substituted with Freebase entities, excluding compound value type (CVT) entities.\footnote{CVT entities are special entities in Freebase that are used to model attributes of relationships (e.g., date of marriage).}
Existential variables, on the other hand, can be substituted with Freebase entities, including CVT entities.
For example, the logical form of the relationship 
$\mathit{starsInFilm}$ is
$\lambda x . \lambda y . \exists z  . actor\_film(x,z) \wedge film\_starring(z,y)$.
Figure~\ref{fig:relationship-examples} shows the equivalent graphical representation of this relationship.

A pair $r_i=r\langle s,o\rangle$ is a \emph{relationship instance} of $r$ for entities $s, o \in \mathcal{E}$ if 
by substituting $x=s$ and $y=o$ in $r$ and by executing the resulting logical form in the knowledge graph $\mathcal{K}$ we get at least one result.
For example, $\mathit{starsInFilm}(\mathit{Brad Pitt}, \mathit{Troy})$ is a relationship instance of the $\mathit{starsInFilm}$ relationship.

\subsection{Task definition}
We assume that a relationship instance $r_i$ can be expressed with a human-readable description (such as a single sentence) that contains mentions of both $s$ and $o$ and possibly other entities which may provide contextual information for the relationship $r$ or the entities $s$ and $o$.
The task we address in this chapter is to generate such a textual description $y$ of the relationship instance $r_i$ given the KG. For this we leverage a set of pairs $X$, where each $x \in X$ is a pair of $r_{i'}$ and $y'$, and $y'$ is the description of $r_{i'}$.
We describe how we obtain this set in Section~\ref{sec:exp-setup}.

We aim to generate descriptions that are 
valid (expressing a relationship that can be found in the knowledge graph under the closed-world assumption),
natural (grammatically correct), 
and informative, i.e., not just replicating the formal relationship but providing additional contextual information where possible.

We conclude our task definition with an example.
Assume that we are given the relationship instance $\mathit{starsInFilm}(Brad Pitt, Troy)$.
A possible description of this relationship instance is the following:
``Brad Pitt
appeared in the 
American
epic
adventure
film
Troy.''
This description not only contains mentions of the entities of the relationship instance 
and a verbalization of the relationship (``appeared in''), but also mentions of other entities that provide additional context.
In particular, it contains mentions of \texttt{Troy}'s type (\texttt{Film}), its genres (\texttt{Epic, Adventure}), and its country of origin.

\section{Generating Textual Descriptions}

In this section we detail our method which consists of three main steps.
First, we enrich the description $y'$ for each pair $\langle r_{i'}, y' \rangle \in X$ with additional entities from the KG (Section~\ref{sec:corpus-enrichment}).
Second, we use $\mathcal{K}$ and the set $X$ to create a set of sentence templates $T_r$ for the relationship $r$ (Section~\ref{sec:template-creation}).
Third, given a new relationship instance, we use $T_r$ and $\mathcal{K}$ to generate a description (Section~\ref{sec:generating-description}).

\subsection{Enriching the textual descriptions}
\label{sec:corpus-enrichment}

In this step we perform entity linking to enrich the description $y'$ for each pair $\langle r_{i'}, y' \rangle \in X$ with additional entities from the KG.
This is done in order to facilitate the template creation step (Section~\ref{sec:template-creation}).
Each $y'$ is a sentence that is about an entity $e\in\mathcal{E}$ and in the context of this chapter we obtain these sentences from Wikipedia as our KG provides explicit links to Wikipedia articles.
Although Wikipedia articles already contain explicit links to other articles and thus entities, these links are quite sparse.
Therefore, we apply an algorithm for entity linking similar to~\cite{voskarides-learning-2015}.

\begin{table}[t]
\caption{Additional surface forms per entity type.}
\label{tab:surface-forms-per-type}
\centering
\begin{tabularx}{23em}{l l}
\toprule
\bf Entity type & \bf Surface form
\\
\midrule
Person & ``he'' or ``she'', person's surname
\\
Film & ``the film'' 
\\
Music album & ``the album''
\\
Music composition & ``the song'', ``the track''
\\
\bottomrule
\end{tabularx}
\smallskip
\end{table}

Since $y'$ originates from a Wikipedia article that is about a specific entity, we restrict the \emph{candidate entities} (i.e., the entities that we consider adding to enrich $y'$) to $e$ itself, the in-links and out-links of the article of $e$ in the Wikipedia structure, and the one-hop and two-hop neighbors of $e$ in the KG.
We infer the \emph{surface forms} of each entity using the Wikipedia link structure, as is common in entity linking~\cite{meij2012adding}, and we also use the aliases of each entity provided by the KG.\footnote{We tag the sentences with POS tags and ignore unigram surface forms that are verbs.}
In order to increase coverage for $e$, we enhance the set of surface forms of entity $e$ using the rules in Table~\ref{tab:surface-forms-per-type}.

We iterate over the n-grams of the sentence that are not yet linked to an entity in decreasing order of length; if the n-gram matches a surface form of a candidate entity, we \emph{link} the n-gram to the entity.
If multiple entity candidates exist for a surface form, we rank the candidate entities by the number of entity neighbors they have in the sentence and select the top-ranked entity.
Because of the very restricted set of candidate entities, the linking is usually unambiguous (with only one entity candidate per surface form).\footnote{A manual evaluation of this algorithm on a held-out, random sample of 100 sentences in our dataset revealed an average of 93\% precision and 85\% recall per sentence.}

\subsection{Creating sentence templates}
\label{sec:template-creation}
In this step, we create a set of templates $T_r$ for a relationship $r$ using the KG and the set of $\langle r_{i'}, y' \rangle$ pairs.
The templates in $T_r$ will be used in the next step to generate a novel description for the relationship instance $r_i$.

A \emph{sentence template} $t$ is a tuple $(k, l, R_t)$, where
(i)~$k = [u_1 u_2 \ldots u_n]$ is a sequence, such that $\forall u_i \in l: u_i \in \mathcal{V} \cup \mathcal{E}_t$, 
(ii)~$l$ is a logical form in $\lambda$-calculus that consists of all the lambda variables in $\mathcal{E}_t$, at least one predicate and zero or more existential variables, and
(iii) $R_t$ is a set of relationship instances that support $t$.

The procedure we follow is outlined in Algorithm \ref{alg:template-creation}.
\begin{algorithm}[t]
\centering
\caption{Template creation}
\label{alg:template-creation}
\begin{algorithmic}[1]
\Require{A set $X$, the knowledge graph $\mathcal{K}$}
\Ensure{A set of templates $T_r$}
\State $X' \gets []$
\For{$\langle r_{i'}, y' \rangle \in X$}
	\State $K \gets$ \textproc{BuildEntityDependencyGraph($y', \mathcal{K}$)}
	\State $X'.append(\langle r_{i'}, y', K\rangle)$
\EndFor
\State $C \gets$ \textproc{ClusterEntities($X'$)}
\State $G \gets$ \textproc{BuildCompressionGraph($X', C$)}
\State $P \gets$ \textproc{FindValidPaths($G$)}
\State $T_r \gets \{\}$

\For{$p \in P$}
	\State $t \gets$ \textproc{ConstructTemplate($p, G, X'$)}
	\If{$t \neq NULL$}
		\State $T_{r}.add(t)$
	\EndIf
\EndFor
\end{algorithmic}
\end{algorithm}
First, we augment each $\langle r_{i'}, y' \rangle$ pair with an entity dependency graph $K$ in order to capture dependencies between entities in a sentence (lines 1--4).
Next, we build a mapping $C$ that maps each entity in each sentence to a single cluster id (line 5).
This is done in order to facilitate the detection of useful patterns in the sentences since each sentence describes a relationship for a particular entity pair.
Then, we build a compression graph $G$ (line 6) and use it to find valid paths $P$ (line 7).
Finally, for each path $p \in P$, we construct a template $t$ and add it to the set of templates (lines 8--12).
We now describe each procedure in Algorithm~\ref{alg:template-creation}.

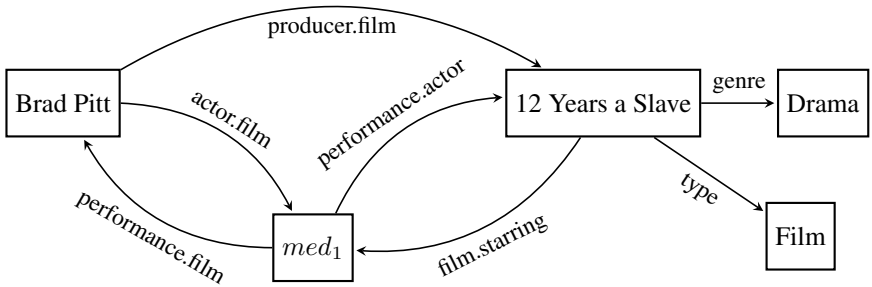
\begin{figure}[t]
	\centering
    \begin{tikzpicture}[
			align=center,
            > = stealth, %
            shorten > = 1pt, %
            auto,
            node distance = 3cm, %
            semithick %
        ]

        \tikzstyle{every state}=[
            draw = black,
            thick,
            fill = white,
        ]

        \node[state, rectangle] (x) {Brad Pitt};
        \node[state, rectangle] (z) [below right=1cm and 2cm of x] {$med_1$};  
        \node[state, rectangle] (y) [above right=1cm and 2cm of z]{12 Years a Slave};
        \node[state, rectangle] (d) [below right=1.2cm of y]{Film};
		\node[state, rectangle] (g) [right=1cm of y]{Drama};

		\path[->] (x) edge [bend left] node[midway, sloped, above] {\small actor.film}  (z);
		\path[->] (z) edge [bend left] node[midway, sloped, below] {\small performance.film} (x);
		\path[->] (y) edge [bend left] node[midway, sloped, below] {\small film.starring} (z);
		\path[->] (z) edge [bend left] node[midway, sloped, above] {\small performance.actor} (y);
		\path[->] (y) edge node[midway, sloped, below] {\small type} (d);
		\path[->] (y) edge node[midway] {\small genre} (g);
		\path[->] (x) edge [bend left] node[midway, sloped, below] {\small producer.film} (y);
    \end{tikzpicture}
	 \caption{
	 Entity dependency graph for the sentence ``Brad Pitt appeared in the drama film 12 Years a Slave''. Nodes represent entities and edge labels represent predicates ($med_1$ is a CVT entity).
}
	 \label{fig:entity-dependency-graph}
\end{figure}

\smallskip\noindent%
\textbf{\textproc{BuildEntityDependencyGraph(.)}}
In order to build the graph $K$ for a sentence $y'$, we retrieve all paths between each pair of entities mentioned in $y'$ from the KG and add them to $K$. We only consider 1-hop paths and 2-hop paths that pass through a CVT entity.
Figure~\ref{fig:entity-dependency-graph} shows the entity dependency graph for an example sentence.

\smallskip\noindent%
\textbf{\textproc{ClusterEntities(.)}}
In order to obtain $C$, we consider all $x'=\langle r_{i'}, y', K \rangle \in X'$ and map two entities in the same cluster if they share at least one incoming or outgoing edge label in their corresponding entity dependency graph $K$.
For example, in the $\mathit{starsInFilm}$ relationship, this procedure will create separate clusters for persons, films, dates and CVT entities.

\smallskip\noindent%
\textbf{\textproc{BuildCompressionGraph(.)}}
In this step, we build a compression graph $G=(V,E)$ using the sentence $y'$ of each $\langle r_{i'}, y', K \rangle \in X'$.
$V$ is a set of nodes and $E$ is a set of edges.
We follow a similar procedure to~\cite{ganesan2010opinosis}, in which
each node holds a list of $\langle sid, pid \rangle$ pairs, where $sid$ is a sentence id and $pid$ is the index of the word/entity in the sentence.
In our case a node can be a word or an entity cluster.
We map two words onto the same node if they have the same lowercase form and the same POS tag.
We map two entities on the same node if they have the same cluster id.

\smallskip\noindent%
\textbf{\textproc{FindValidPaths(.)}}
In order to find valid paths in the graph $G$, we set all the entity cluster nodes as valid start/end nodes and traverse $G$ to find a set of paths $P$ from a start to an end node.
In order to build templates that are natural we enforce the following constraints for the paths in $P$:
(i) the path must contain a verb and
(ii) the path must have been seen as a complete sentence at least once in the input sentences.
For example, given the following sentences (the corresponding cluster id per entity are listed in brackets):
\begin{itemize}

	\item $y'_1$: ``Bruce\_Willis[$c_1$] appeared in Moonrise\_Kingdom[$c_2$]''
	\item $y'_2$: ``Liam\_Neeson[$c_1$] appeared in the action[$c_3$] film[$c_4$] Taken[$c_2$]''
	\item $y'_3$: ``Brad\_Pitt[$c_1$] appeared in the drama[$c_3$] film[$c_4$] 12\_Years\_a\_Slave[$c_2$]''
\end{itemize}
we obtain the following valid paths by traversing the graph:
\begin{itemize}
	\item $p_1$: ``$c_1$ appeared in $c_2$''
	\item $p_2$: ``$c_1$ appeared in the $c_3$ $c_4$ $c_2$''
\end{itemize}

\smallskip\noindent%
\textbf{\textproc{ConstructTemplate(.)}}
Algorithm~\ref{alg:construct-template} outlines the procedure for constructing a template $t$ from a path $p$.
First, for each $\langle r_{i'}, y', K \rangle \in X'$, we check whether $y'$ is a (possibly non-continuous) subsequence $h$ of path $p$ by using the positional information of each node in $p$ from $G$.\footnote{For example, the path $p_1$ is a subsequence of $y'_2$.} 
If it is, we check whether $h$ contains links to both the subject and the object of the relationship instance $r_{i'}$.
If it does, we store the entity dependency graph and the relationship instance.
Next, if the number of instances is less than a parameter $\alpha$, we consider the template to be invalid.
Subsequently, we build the logical form $l$ by aggregating the entity dependency graphs $D_g$.
Entity nodes that were part of the path $p$ become lambda variables (nodes constructed from subject and object entities have special identifiers).
Entity nodes that were not part of the path $p$ (CVT entities) become existential variables.
We ignore edges appearing in less than $|D_g| \cdot \beta$ entity dependency graphs.
Lastly, we replace the cluster ids in $p$ with the corresponding lambda variables to obtain a sequence $k$.

\begin{figure}
	\centering
    \begin{tikzpicture}[
			align=center,
            > = stealth, %
            shorten > = 1pt, %
            auto,
            node distance = 3cm, %
            semithick %
        ]

        \tikzstyle{every state}=[
            draw = black,
            thick,
            fill = white,
        ]

        \node[state] (x) {$x_{subj}$};
        \node[state, rectangle] (z) [right=1cm and 2cm of x] {$z$};  %
        \node[state] (y) [right=1cm and 2cm of z]{$x_{obj}$};
        \node[state] (d) [below right=0.8cm of y]{$x_3$};
		\node[state] (g) [right=1cm of y]{$x_4$};

		\path[->] (x) edge [bend left] node[midway, sloped, above] {\small actor.film}  (z);
		\path[->] (z) edge [bend left] node[midway, sloped, below] {\small performance.film} (x);
		\path[->] (y) edge [bend left] node[midway, sloped, below] {\small film.starring} (z);
		\path[->] (z) edge [bend left] node[midway, sloped, above] {\small performance.actor} (y);
		\path[->] (y) edge node[midway, sloped, above] {\small type} (d);
		\path[->] (y) edge node[midway] {\small genre} (g);
    \end{tikzpicture}
	 \caption{
	 Logical form of the template constructed using $p_2$ and $y'_1, y'_2, y'_3$ (with their corresponding relationship instances).
	 $k=$``$x_{subj}$ appeared in the $x_3$ $x_4$ $x_{obj}$''. 
	 Lambda variables are shown in circles and existential variables in rectangles.
}
	 \label{fig:example-template}
\end{figure}
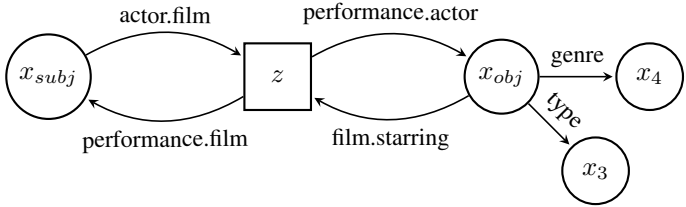

Figure~\ref{fig:example-template} shows the logical form of a template constructed using the example sentences $y'_1$, $y'_2$ and $y'_3$ and their corresponding instances in graphical form ($\beta = 0.5$).
Note that the edge ``producer.film'' has been eliminated since it only appears in one out of the three instances.

\begin{algorithm}[t]
\caption{\textproc{ConstructTemplate(.)}}
\label{alg:construct-template}
\begin{algorithmic}[1]
\Require{A path $p$, the compression graph $G$, a set $X'$, parameters $\alpha, \beta$}
\Ensure{A template $t$}
\State $D_g \gets []$  \Comment entity dependency graphs
\State $R_t \gets []$ \Comment relationship instances that support the template
\For{$\langle r_{i'}, y', K \rangle \in X'$}
	\If{\textproc{IsSubsequence($p, y', G$)}}
		\State $h \gets$ \textproc{GetSubsequence($p, y', G)$}  \Comment get the actual subsequence
		\State $\langle s, o\rangle \gets r_{i'}$ \Comment subject/object of the relationship instance
		\If{\textproc{ContainsLink($h, s$)} \textbf{and} \textproc{ContainsLink($h, o$)}}
			\State $D_g.append(K)$
			\State $R_t.append(r_{i'})$
		\EndIf
	\EndIf
\EndFor
\If{$|R_t| < \alpha$}  \Comment too few relationship instances
	\State \Return $NULL$
\EndIf
\State $l \gets$ \textproc{BuildLogicalForm($D_g, \beta$)}  \Comment aggregate the entity dependency graphs
\State $k \gets$ \textproc{ReplaceClusterIdsWithVariables($p$)}
\State $t=(k, l, R_t)$

\end{algorithmic}
\end{algorithm}

\subsection{Generating the description}
\label{sec:generating-description}
In this step we generate a novel description for a relationship instance $r_i$ using the set of templates $T_r$ and the knowledge graph $\mathcal{K}$.
This comes down to selecting the template from $T_r$ that best describes the relationship instance $r_i$ and filling it with the appropriate entities.

The procedure is as follows.
First, we rank the templates in $T_r$ for the relationship instance using a scoring function $f(r_i, t)$.
Subsequently, for each template $t=(k, l, R_t)$ we replace the subject and object lambda variables in $l$ to obtain $l' = l[x_{subj}=s, x_{obj}=o]$.
We then query the knowledge graph $\mathcal{K}$ using $l'$ and if 
at least one instantiation of $l'$ exists, we randomly pick one and replace all the entity variables in $k$ with the entity names to generate the description $y$, otherwise we proceed to the next template.
As an example, assume we are given the instance $r_i=\mathit{starsInFilm}(\mathit{Ryan\_Reynolds}$, $\mathit{Deadpool})$ and we consider the template shown in Figure~\ref{fig:example-template}.
A possible instantiation of the template for this relationship instance will result in the description ``Ryan Reynolds appeared in the comedy film Deadpool''.\footnote{Note that there might be multiple instantiations (e.g., Deadpool is also a science fiction film) and selecting the optimal  one depends on the application---we leave this for future work.}

The template scoring function $f(r_i, t)$ returns a score for a relationship instance $r_i$ and template $t$.
As we want to generate descriptions that are valid under the closed-world assumption of the KG, we promote templates that are semantically closest to the relationship instance.
For a new relationship instance $r_i$ we extract binary features for each entity in the $r_i$. 
Recall that $r_i$ has two or more entities (subject $s$, object $o$ and possibly a CVT entity $z$).
For each entity $e$ of $r_i$, we extract all triples $\langle e, p, e'\rangle$ from the KG $\mathcal{K}$.
We restrict the feature space by discriminating between entity attributes and entity relations depending on the predicate $p$ as in \cite{lin1knowledge}.
If the predicate $p$ is an attribute (e.g., ``gender''), we use the complete triple as a feature (e.g. $\langle s, \mathit{gender}, \mathit{female}\rangle$).
If the predicate $p$ is a relation (e.g., ``date\_of\_death''), we only keep the subject and the predicate of the triple as a feature (e.g., $\langle e, \mathit{person.date\_of\_death}\rangle$).
We also add a count feature for the relation predicates (e.g., $\langle s, \mathit{person.children}, 2\rangle$, i.e., a person has two children).
We denote the resulting binary vector for $r_i$ as $\mathit{vec}(r_i)$.
We obtain a vector $\mathit{vec}(t)$ for template $t$ by summing the vectors of all the instances $R_t$ of $t$.
We also compute a vector $\mathit{vec\_tfidf}(t)$ that is a TF.IDF weighted vector of $\mathit{vec}(t)$, where IDF is calculated at the template level.
Based on these ingredients, we define two scoring functions:
\begin{itemize}
	\item \textbf{Cosine}
Calculates the cosine similarity between vectors $vec(r_i)$ and $vec\_tfidf(t)$.
	\item \textbf{Supervised}
Learns a scoring function using a supervised learning to rank algorithm.
We treat $r_i$ as a ``query'' and $t$ as a ``document.''
\end{itemize}

\noindent%
We create training data for the supervised algorithm as follows.
Recall that each $r_i$ is coupled with a description $y'$.
For each $r_i$, we assign a relevance label of 3 for templates that best match $y$ (measured by the number of entities) and a relevance label of 2 for the rest of the templates that match $y$.
In order to create ``negative'' training data, we sample templates that are dissimilar to the ones that match $y$ in the following way.
First, we calculate the average vector of all the templates that match $y$ and build a distribution of templates based on the cosine distance from the average vector to each of the templates in $T_r$ (excluding the ones that match $y$).
Lastly, we sample at most the number of matching templates from the resulting distribution and assign them a relevance label of 1 (we ignore templates that have a cosine similarity to the average vector greater than $0.9$).
For the supervised model we use the following features: each element/value pair in $\mathit{vec}(r_i)$, the cosine similarity between vectors $\mathit{vec}(r_i)$ and $\mathit{vec\_tfidf}(t)$, the words in $t$, the number of entities in $t$ and the size of $R_t$. We use LambdaMART~\cite{wu2008ranking} as the learning algorithm and optimize for NDCG@1.\footnote{For this method we use 20\% of the training data as validation data. The same test data is used for all methods.}

\section{Experimental Setup}
\label{sec:exp-setup}
In this section we describe the experimental setup we designed to answer~\textbf{\ref{rq:generate-explanations}}.

\subsection{Datasets}

We use an English Wikipedia dump dated 5 February 2015 as our document corpus.
We perform sentence splitting and POS tagging using the Stanford CoreNLP toolkit. 
We use a subset of the last version of Freebase as our KG~\cite{bollacker2008freebase}:
all the triples in the people, film and music domains, as these are well-represented in Freebase.

In order to create an evaluation dataset for our task, we first need a set of KG relationships.
We rank the predicates in each domain by the number of instances and keep the 10 top-ranked predicates.
We exclude trivial predicates such as ``dateOfDeath''.
We then use the predicates to manually construct the logical forms of the relationships (see Figure~\ref{fig:relationship-examples} for an example).
Second, we need a set of $\langle r_{i'}, y'\rangle$ pairs for each relationship $r$, where $r_{i'} = r\langle s', o' \rangle$ is an instance of relationship $r$, $s'$ and $o'$ are entities and $y'$ is a description of $r_{i'}$.
To this end, for each relationship $r$, we randomly sample 12000 relationship instances from the KG.
For each relationship instance $r_{i'}$, we pick the first sentence in the Wikipedia article of the subject entity $s'$ that contains links to both $s'$ and $o'$.
If such a sentence does not exist, we proceed to the next instance.
We manually inspected a subset of the sentences selected with this heuristic and the quality of the selected sentences was relatively good.
Our final dataset contains 10 relationships and 90058 $\langle r_{i'}, y'\rangle$ instances in total and 8187 instances on average per relationship.
We randomly select 80\% of each relationship sub-dataset for training and 20\% for testing.

\subsection{Evaluation metrics}
We perform two types of evaluation: automatic and manual.
For automatic evaluation we use METEOR~\cite{lavie2007meteor}, ROUGE-L~\cite{lin2004rouge} and BLEU-4~\cite{papineni2002bleu} as metrics. 
METEOR was originally proposed in the context of machine translation but has also
been used
in a task similar to ours~\cite{saldanha2016entity}.
ROUGE is a standard metric in summarization %
and 
BLEU is widely used in machine translation and generation. %
As is common in text generation~\cite{konstas2013global}, we also employ manual evaluation. 
We ask human annotators to annotate each output sentence on three dimensions: validity under the KG closed-world assumption (0 or 1), informativeness (1--5) and grammaticality (1--5).
One human annotator (not one of the authors) annotated 11 generated sentences per relationship per system (440 sentences in total).

\subsection{Compared approaches}
We compare 4 variations of our method.
The variations differ in the way they rank templates for a given relationship instance. 
The first variation (\emph{Random}) ranks the templates randomly.
The second (\emph{Most-freq}) ranks templates by the number of relationship instances that support the template.
The third (\emph{Cosine}) ranks templates based on the cosine similarity between the vectors of the relationship instance and the template (Section \ref{sec:generating-description}).
The fourth (\emph{Supervised}) ranks templates using a learning to rank model (Section \ref{sec:generating-description}), 
for which we use LambdaMART with the default number of trees (1000).
We set $\alpha = 20$ and $\beta = 0.5$ (Section~\ref{sec:generating-description}).
We depict a significant improvement in performance over \emph{Random} with
$^\blacktriangle$ (paired two-tailed t-test, $p<0.05$).

\newcommand{\yo}{\phantom{$^\blacktriangle$}}

\section{Results}
In this section we describe our experimental results. We compare all methods discussed previously, using the automatic and manual setups, respectively.

\subsection{Automatic evaluation}

Table~\ref{tab:auto-eval} shows the automatic evaluation results.
We observe that \emph{Supervised} and \emph{Cosine} outperform \emph{Random} and \emph{Most-freq} on all metrics.
This is expected since the former two try to capture the semantic similarity between a relationship instance and a template.
Although \emph{Supervised} consistently outperforms \emph{Cosine}, the differences between \emph{Cosine} and \emph{Supervised} are not significant.

We also observe that the scores for the automatic measures are relatively low.
This is because of two reasons: (i) we generally generate much shorter sentences than the reference sentence as not all information that appears in the reference sentence is represented in the KG, and (ii) since the reference sentences are extracted automatically, some of the reference sentences describe a minor aspect of the relationship or do not discuss the relationship at all.

\begin{table}[t]
\centering
\caption{\label{tab:auto-eval}  Automatic evaluation results, averaged per relationship.
}
  \begin{tabular}{lccc}
  \toprule
  \textbf{Method} & \textbf{BLEU} & \textbf{METEOR} & \textbf{ROUGE} \\
   \midrule
  Random &  1.14\yo & 16.56\yo & 24.13\yo \\ 
  Most-freq &  0.13\yo & 13.99\yo	 & 21.96\yo \\ 
  Cosine &  1.76$^\blacktriangle$ & 17.37\yo & 25.84$^\blacktriangle$ \\ 
  Supervised & \textbf{2.14}$^\blacktriangle$ & \textbf{19.18}$^\blacktriangle$ & \textbf{26.54}$^\blacktriangle$ \\ 
  \bottomrule
  \end{tabular}
  \medskip
\end{table}

\subsection{Manual evaluation}
\begin{table}[t]
\centering
\caption{\label{tab:manual-eval} Manual evaluation results, averaged per relationship.
}
  \begin{tabular}{lccc}
  \toprule
     \textbf{Method} &    \textbf{Validity} &  \textbf{Informativeness} & \textbf{Grammaticality} \\ 
   \midrule
  Random & 0.4545\yo  &  1.98\yo & 3.67 \\ 
  Most-freq &  0.5000\yo & 1.60\yo & 3.62  \\ 
  Cosine &  0.5636$^\blacktriangle$ & 2.05\yo & \textbf{4.00}  \\ 
  Supervised &  \textbf{0.5818}$^\blacktriangle$ & \textbf{2.18}$^\blacktriangle$ & 3.90 \\ 
  \bottomrule
  \end{tabular}
  \medskip
\end{table}
Table~\ref{tab:manual-eval} shows the results for manual evaluation.
The results follow a similar trend as in the automatic evaluation; \emph{Supervised} and \emph{Cosine} outperform \emph{Random} and \emph{Most-freq} on all metrics.
\emph{Supervised} significantly outperforms \emph{Random} in terms of validity and informativeness. The differences between \emph{Cosine} and \emph{Supervised} are not significant.

\subsection{Analysis}
We have also examined specific examples and identify cases where the best performing approach (\emph{Supervised}) succeeds or fails.
In terms of validity, it succeeds in matching attributes of the relationship instance and the template.
E.g., in the context of the relationship $\mathit{parentOf}$, it correctly figures out what the genders of the entities are and the semantically valid expression of the relationship between them, often better than \emph{Cosine}, as illustrated by the following example:
\begin{description}
\item[\rm(\emph{Supervised})] ``Emperor Francis I (1708 - 1765) was the father of Emperor Leopold II'' (VALID)
\item[\rm(\emph{Cosine})] ``Emperor Francis I was the son of Emperor Leopold II'' (INVALID)
\end{description}
\emph{Supervised} benefits from training a model that combines multiple features such as the template words with attributes of the relationship instance to describe whether the relationship is still ongoing or not.
One of the main cases where \emph{Supervised} fails is in ranking a relationship instance in a temporal dimension with regards to other relationship instances, as illustrated by the following example for the $\mathit{childOf}$ relationship:
\begin{description}
\item ``Thomas Howard was the second son of Henry Howard and Frances de Vere."\\(INVALID: Thomas Howard was the \emph{first} son of Henry Howard)
\end{description}
The fact that our best performing approach (\emph{Supervised}) has a relatively low validity score (0.5818) shows that there is room for improvement in capturing the semantic similarity between a relationship instance and a template.

In terms of informativeness, \emph{Supervised} succeeds in offering contextual information about the relationship instance, such as dates, locations, occupations and film genres.
The fact that informativeness scores are relatively low is because they are dependent on validity: when a generated sentence was assigned a validity of score 0, it was also assigned an informativeness score of just 1.

Grammaticality scores are high for all the systems with no significant differences.
This is expected as the templates were generated using the same procedure for all the compared systems.
Mainly, grammaticality is harmed when some entities in the generated sentence have the wrong surface form (e.g., `Britain', `British'), which is not surprising as we do simple surface realization (deciding which surface form of the entity best fits with the generated sentence) and only use the entity names as surface forms.

\section{Conclusion}
We have addressed the problem of generating descriptions of entity relationships from KGs.
We have introduced a method that first creates sentence templates for a specific relationship, and then, for a new relationship instance, it generates a novel description by selecting the best template and filling the template slots with the appropriate entities from the KG.
We have experimented with different scoring functions for ranking templates for a relationship instance and performed an automatic and a manual evaluation.

When using information about the relationship instance and the template taken from the KG, both automatic and manual evaluation outcomes are improved.
A supervised method that uses both KG features and other template features (template words, number of entities) consistently outperforms an unsupervised method on all automatic evaluation metrics and also in terms of validity and informativeness.

As to future work, 
our error analysis showed that we need more sophisticated modeling for capturing the semantic similarity between a relationship instance and a template, especially for capturing temporal dimensions that also involve other relationship instances.
We also want to explore more sophisticated methods for selecting the correct surface form for an entity to improve grammaticality.
Finally, we aim to evaluate our method on generating descriptions for less popular KG relationships.

In this chapter, we studied the task of generating KG fact (entity relationship) descriptions.
In the next chapter, we move on to study how to contextualize KG facts using other, related KG facts.

\graphicspath{{04-sigir2018/figures/}}

\chapter{Contextualizing Knowledge Graph Facts}
\label{chapter:sigir2018}

\footnote[]{This chapter was published as~\citep{voskarides-weakly-supervised-2018}.}

In Chapter~\ref{chapter:acl2015} and~\ref{chapter:ecir2017}, we studied how retrieve and generate descriptions of \acf{KG} facts.
KG fact descriptions often contain mentions to other, related KG facts that are not trivial to find given the large size of KGs.
In this chapter, we address \textbf{\ref{rq:contextualize-facts}}: \acl{rq:contextualize-facts}

\section{Introduction}

Knowledge graphs (KGs) have become essential for applications such as search,
query understanding, recommendation and question
answering because they provide a unified view of real-world entities and the facts (i.e., relationships) that hold between them~\cite{blanco2013entity, blanco2015fast, yih2015semantic,
miliaraki2015selena}.
For example,  KGs are increasingly being used to provide direct answers to user queries~\cite{yih2015semantic},
or to construct so-called \emph{entity cards} that provide useful
information about the entity identified in the query.
Recent work~\cite{bota2016playing, hasibi2017dynamic} suggests that search
engine users find entity cards useful and engage with them when they contain
information that is relevant to their search task, for instance in the form of a
set of recommended entities and facts that are related to the
query~\cite{blanco2013entity}.
Previous work has focused on augmenting entity cards with facts that are centered around, i.e., one-hop away from, the main entity of the query~\cite{hasibi2017dynamic}. %

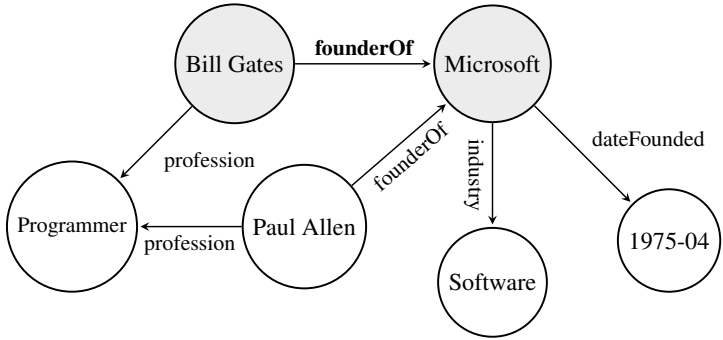
\begin{figure}[t]
	\centering
	 \resizebox{0.8\linewidth}{!}{
      \begin{tikzpicture}[
      align=center,
            > = stealth, %
            shorten > = 1pt, %
            auto,
            node distance = 3cm, %
            semithick %
        ]

        \tikzstyle{every state}=[
            draw = black,
            thick,
            fill = white,
        ]
        \node[state, fill=gray!15] (x) {Bill Gates};
        \node[state, fill=gray!15] (y) [right=2cm of x]{Microsoft};
        \node[state] (s) [below=1.5cm of y]{Software};
         \node[state] (pob) [below left=1.5cm of x] {\small Programmer};
\node[state] (pa) [right=1.5cm of pob] {Paul Allen};
        \node[state] (da) [below right=2cm of y] {1975-04};

    \path[->] (x) edge [font=\bf] node[midway] {\small founderOf} (y);
    \path[->] (y) edge node[midway, sloped, below] {\small industry} (s);
        \path[->] (y) edge node[midway] {\small dateFounded} (da);
        \path[->] (pa) edge node[midway, sloped, below] {\small founderOf} (y);
        \path[->] (x) edge node[midway] {\small profession} (pob);
        \path[->] (pa) edge node[midway] {\small profession} (pob);
    \end{tikzpicture}
    }
     \caption{ A Freebase subgraph that consists of relevant facts to the
         query fact $\mathit{founderOf}(\text{Bill Gates}, \text{Microsoft})$.
    }
	\label{fig:contextualization-graph-gates-microsoft}
\end{figure}
However, oftentimes a user is interested in KG facts that by definition involve more than one entity (e.g., ``Who founded Microsoft?'' $\longrightarrow$ ``Bill Gates'').
In such cases, we can exploit the richness of the KG by providing query-specific additional facts that increase the user's understanding of the fact as a whole, and that are not necessarily centered around only one of the entities. %
Additional relevant facts for the running example would include Bill Gates' profession, Microsoft's founding date, its main industry and its co-founder Paul Allen (see Figure~\ref{fig:contextualization-graph-gates-microsoft}).
In this case, Bill Gates' personal life %
is less relevant to the fact that he founded Microsoft.

Query-specific relevant facts can also be used in other applications to enrich the user experience.
For instance, they can be used to increase the utility of KG question answering (QA) systems that currently only return a single fact as an answer to a natural language question~\cite{yih2015semantic,FNTIR:2016:Bast}.
Beyond QA, systems that focus on automatically generating natural language from KG facts~\cite{lebret2016generating} would also benefit from query-specific relevant facts, which can make the generated text more natural and human-like.
This becomes even more important for KG facts that involve tail entities, for which natural language text might not exist for training~\cite{voskarides-generating-2017}.

In this chapter, we address the task of KG fact \emph{contextualization}, that is, given a KG fact that consists of two entities and a relation that connects them, retrieve additional facts from the KG that are relevant to that fact.
This task is analogous to ad-hoc retrieval: (i) the ``query'' is a KG fact, (ii) the ``documents'' are other facts in the KG that are in the neighborhood of the ``query''.  
We propose a \emph{neural fact contextualization method} (NFCM), a method that first generates a set of candidate facts that are part of \{1,2\}-hop paths from the entities of the main fact.
NFCM then ranks the candidate facts by how relevant they are for contextualizing the main fact.
We estimate our learning to rank model using supervised data.
The ranking model combines (i) features we automatically learn from data and (ii) those that represent the query-candidate facts with a set of hand-crafted features we devised or adjusted for this task.
Due to the size and heterogeneous nature of KGs, i.e., the large number of entities and relationship types, we turn to distant supervision to gather training data.
Using another, human-verified test collection we gauge the performance of our proposed method and compare it with several baselines.
We sum up our contributions as follows.
\begin{itemize}
\item We introduce the task of KG fact contextualization where the goal is to, given a fact that consists of two entities and a relationship that connects them, rank other facts from a KG that are relevant to that fact.
\item We propose NFCM, a method to solve KG fact contextualization using distant supervision and learning to rank. Our results show that: (i)~distant supervision is an effective means for gathering training data for this task and
(ii)~a neural learning to rank model that is trained end-to-end outperforms several baselines on a human-curated evaluation set. 
\item We provide a detailed result analysis and insights into the nature of our task.
\end{itemize}
The remainder of this chapter is organized as follows.
We first provide a definition of our task in Section~\ref{sec:problem}
and then introduce our method in Section~\ref{sec:sigir2018-method}.
We describe our experimental setup and detail our results and analyses in Sections~\ref{sec:expsetup} and~\ref{sec:sigir2018-results}, respectively.
We conclude with an overview of related work and an outlook on future directions.

\section{Problem Statement}
\label{sec:problem}

In this section we provide background definitions and formally define the task of KG fact contextualization.
\subsection{Preliminaries}
\label{sec:preliminaries}
\begin{figure}[t]
	\centering
	\resizebox{0.7\linewidth}{!}{
	\begin{tikzpicture}[
  align=center,
  > = stealth, %
  shorten > = 1pt, %
  auto,
  node distance = 3cm, %
  semithick %
  ]

  \tikzstyle{every state}=[
  draw = black,
  thick,
  fill = white,
  minimum size = 4mm
  ]

  \node[state] (x) {Barack\\Obama};
  \node[state, rectangle] (z) [right of=x] {M1};
  \node[state] (y) [right of=z]{Michelle\\Obama};
  \node[state] (d) [below right=1.6cm of z]{1992-10};
  \node[state] (l) [below right=1.3cm of x]{Hawaii};
  
  \path[->] (x) edge[below]  node[midway] {\small spouse} (z);
  \path[->] (z) edge[below]  node[midway] {\small spouse} (y);
  \path[->] (z) edge[midway]  node[midway, sloped, below] {\small marriageDate} (d);
  \path[->] (x) edge[below]  node[midway, sloped, below] {\small bornIn} (l);
\end{tikzpicture}
	}
	 \caption{
KG subgraph that consists of three facts: $\mathit{bornIn}\langle \text{Barack Obama}, \text{Hawaii}\rangle$, $\mathit{spouseOf}\langle \text{Barack Obama}, \text{Michelle Obama}\rangle$ and $\mathit{marriageDate}\langle \text{M1}, \text{1992-10}\rangle$.
M1 is a CVT entity.
Note that the third fact is an attribute of the second fact.
}
	 \label{rel:spouseOf}
\end{figure}
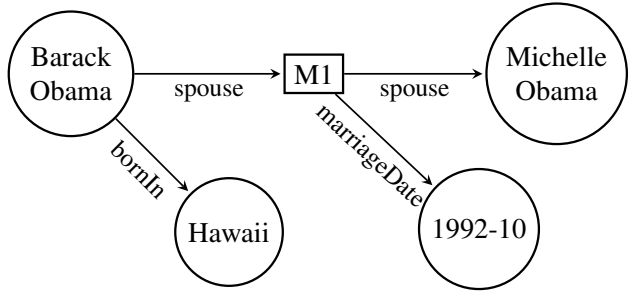

Let $E = E_n \cup E_c$ be a set of entities, where $E_n$ and $E_c$ are disjoint sets of non-CVT and CVT entities, respectively.\footnote{Compound Value Type (CVT) entities are special entities frequently used in KGs such as Freebase and Wikidata to model fact attributes. See Figure~\ref{rel:spouseOf} for an example.}
Furthermore, let $P$ be a set of predicates.
A \emph{knowledge graph} $K$ is a set of triples $\langle s, p, o \rangle$, where $s, o \in E$ and $p \in P$. By viewing
each triple in $K$ as a labelled directed edge, we can interpret $K$ as a labelled directed graph.
We use Freebase as our knowledge graph~\cite{bollacker2008freebase,nickel2015review}. 

A path in K is a non-empty sequence $\langle s_0, p_0, t_0 \rangle, \ldots ,\langle s_m, p_m, t_m \rangle$  of triples from K such that $t_i = s_{i+1}$ for each $i \in {0, m-1}$.

We define a \emph{fact} as a path in $K$ that either:
(i)~consists of 1 triple, $s_0 \in E$ and $t_0 \in E_n$ (i.e., $s_0$ may be a CVT entity), or
(ii)~consists of 2 triples, $s_0, t_1 \in E_n$ and $t_0=s_1 \in E_c$ (i.e., $t_0=s_1$ must be a CVT entity).
A fact of type (i) can be an attribute of a fact of type (ii), iff they have a common CVT entity (see Figure~\ref{rel:spouseOf} for an example).

Let $R$ be a set of relationships where a \emph{relationship} $r \in R$ is a label for a set of facts that share the same predicates but differ in at least one entity.
For example, $\mathit{spouseOf}$ is the label of the fact depicted in the top part of Figure~\ref{rel:spouseOf} and consists of two triples.
Our definition of a relationship corresponds to direct relationships between entities, i.e., one-hop paths or two-hop paths through a CVT entity.
For the remainder of this chapter, we refer to a specific fact $f$ as $r\langle s, t\rangle$, where $r \in R$ and $s, t \in E$.

\subsection{Task definition}
\label{sec:task_definition}
Given a query fact $f_q$ and a KG $K$, we aim to find a set of other, relevant facts from $K$.
Specifically, we want to enumerate and rank a set of candidate facts $F = \{f_c: f_c \subseteq K, f_c \neq f_q\}$ based on their relevance to $f_q$.
A candidate fact $f_c$ is \emph{relevant} to the query fact $f_q$ if it provides useful and contextual information.
Figure \ref{fig:contextualization-graph-gates-microsoft} shows an example part of our KG that is relevant to the query fact $\mathit{founderOf}\langle \text{Bill Gates}, \text{Microsoft} \rangle$.
Note that a candidate fact does not have to be directly connected to both entities of the query fact to be relevant, e.g., $\mathit{profession} \langle \text{Paul Allen}, \text{Programmer}\rangle$.
Similarly, a fact can be related to one or more entities in the relationship instance, e.g., $\mathit{parentOf} \langle \text{Bill Gates}, \text{Jennifer Katharine Gates} \rangle$, but not provide any context, thus being considered irrelevant.

\section{Method}
\label{sec:sigir2018-method}
In this section we describe our proposed neural fact contextualization method (NFCM) which works in two steps.
First, given a query fact $f_q$, we enumerate a set of candidate facts $F = \{f_c: f_c \subseteq K\}$ (see Section~\ref{sec:enumerate-facts}).
Second, we rank the facts in $F$ by relevance to $f_q$ to obtain a final ranked list $F'$ using a supervised learning to rank model (see Section~\ref{sec:ranking-facts}).
We describe how we use distant supervision to automatically gather the required annotations to train the supervised learning to rank model in Section \ref{sec:training-data}.

\subsection{Enumerating KG facts}
\label{sec:enumerate-facts}
\begin{algorithm}[t]
\caption{Fact enumeration for a given query fact $f_q$.}
\label{alg:fact-enumeration}
\begin{algorithmic}[1]
\Require{A query fact $f_q=r\langle s, t \rangle$}
\Ensure{A set of candidate facts $F$}
\State $F \gets \{ \} $
\For{$e \in \{s, t\}$}
    \For{$n \in$ \textproc{GetOutNeighbors($e$)} + \textproc{GetInNeighbors($e$)}}
        \State $F.addAll($\textproc{GetFacts($e, n$)}$)$
        \If{\textproc{IsClassOrType($n$)}}
            \State continue
        \EndIf
        \For{$n_2 \in$ \textproc{GetOutNeighbors($n$)}}
            \State $F.addAll($\textproc{GetFacts($n, n_2$)}$)$
        \EndFor
        \For{$n_2 \in$ \textproc{GetInNeighbors($n$)}}
            \State $F.addAll($\textproc{GetFacts($n_2, n$)}$)$
        \EndFor
    \EndFor
\EndFor
\State \Return $F$
\end{algorithmic}
\end{algorithm}
In this section we describe how we obtain the set of candidate facts $F$ from $K$ given a query fact $f_q=r \langle s,t \rangle$.
Because of the large size of real-world KGs---which can easily contain upwards of 50 million entities and 3 billion facts~\cite{pellissier2016freebase}---
it is computationally infeasible to add all possible facts of $K$ in $F$.
Therefore, we limit $F$ to the set of facts that are in the broader neighborhood of the two entities $s$ and $t$.  %
Intuitively, facts that are further away from the two entities of the query fact are less likely to be relevant. %

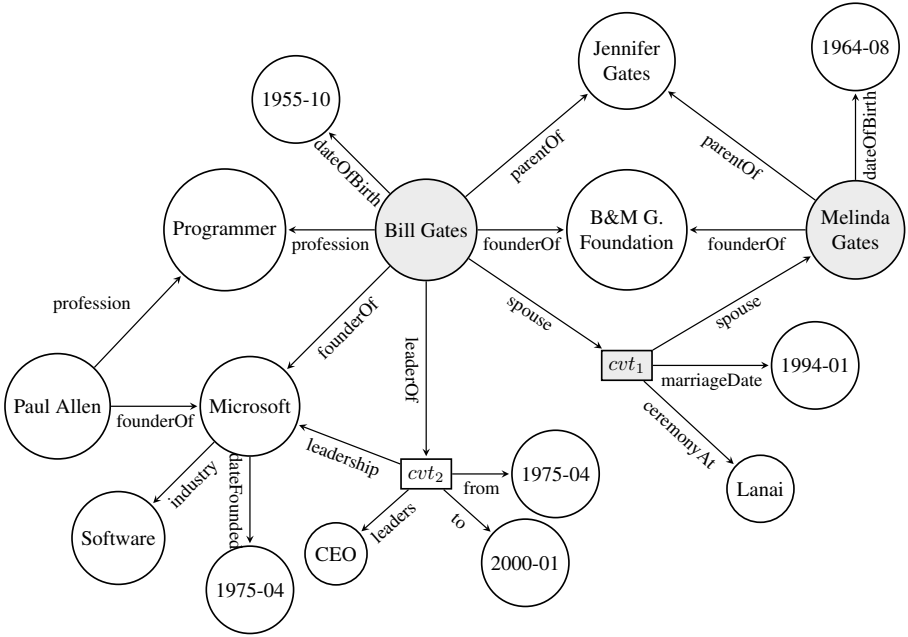
\begin{figure}[t]
	\centering
    \resizebox{\linewidth}{!}{
    	\begin{tikzpicture}[
      align=center,
            > = stealth, %
            shorten > = 1pt, %
            auto,
            node distance = 3cm, %
            semithick %
        ]

        \tikzstyle{every state}=[
            draw = black,
            thick,
            fill = white,
            minimum size = 5mm
        ]

        \node[state, fill=gray!15] (x) {Bill Gates};
                \node[state] (bmgf) [right=1.5cm of x] {B\&M G.\\Foundation};
    \node[state, rectangle, fill=gray!15] (med1)  [below=1cm of bmgf] {$cvt_1$};
    \node[state, fill=gray!15] (y) [right=2cm of bmgf]{Melinda\\Gates};
    \node[state] (m) [below left=2.5cm of x]{Microsoft};
        \node[state] (s) [below left=1.5cm of m]{Software};
        \node[state] (pa) [left=1.5cm of m] {Paul Allen};
        \node[state] (prog) [left=1.5cm of x] {Programmer};
    \node[state] (daf) [below=1.5cm of m] {1975-04};
    \node[state] (dobb) [above left=1.5cm of x] {1955-10};
    \node[state] (dobm) [above=1.5cm of y] {1964-08};
        \node[state] (lanai) [below right=2cm of med1] {Lanai};
        \node[state] (med1from) [right=2cm of med1] {1994-01};

    \node[state] (jkg) [above=1cm of bmgf] {Jennifer\\Gates};

        \node[state, rectangle] (med2)  [below=3cm of x] {$cvt_2$};
        \node[state] (chair)  [below left=1cm of med2] {CEO};
        \node[state] (chairfrom)  [right=1cm of med2] {1975-04};
        \node[state] (chairto)  [below right=1cm of med2] {2000-01};

        \path[->] (x) edge node[midway, sloped, below] {\small spouse} (med1);
        \path[->] (med1) edge node[midway, sloped, below] {\small spouse} (y);
    \path[->] (x) edge node[midway, sloped, below] {\small founderOf} (m);
    \path[->] (m) edge node[midway, sloped, below] {\small industry} (s);
        
        \path[->] (pa) edge node[midway, sloped, below] {\small founderOf} (m);

        \path[->] (x) edge node[midway] {\small profession} (prog);
        \path[->] (pa) edge node[midway] {\small profession} (prog);
        
        \path[->] (m) edge node[midway, sloped, below] {\small dateFounded} (daf);
        \path[->] (x) edge node[midway, sloped, below] {\small dateOfBirth} (dobb);
        \path[->] (y) edge node[midway, sloped, below] {\small dateOfBirth} (dobm);
        
        \path[->] (med1) edge node[midway, sloped, below] {\small ceremonyAt} (lanai);
        \path[->] (med1) edge node[midway, sloped, below] {\small marriageDate} (med1from);

                  \path[->] (x) edge node[midway, sloped, below] {\small parentOf} (jkg);
         \path[->] (y) edge node[midway, sloped, below] {\small parentOf} (jkg);
         
         \path[->] (x) edge node[midway, sloped, below] {\small founderOf} (bmgf);
         \path[->] (y) edge node[midway, sloped, below] {\small founderOf} (bmgf);

         \path[->] (x) edge node[midway, sloped, below] {\small leaderOf} (med2);
         \path[->] (med2) edge node[midway, sloped, below] {\small leadership} (m);
         \path[->] (med2) edge node[midway, sloped, below] {\small leaders} (chair);         
         \path[->] (med2) edge node[midway, sloped, below] {\small from} (chairfrom);
         \path[->] (med2) edge node[midway, sloped, below] {\small to} (chairto);

    \end{tikzpicture}
        }
     \caption{Graph with a subset of the facts that are enumerated for the query fact $\mathit{spouseOf}(\text{Bill Gates}, \text{Melinda Gates})$. The entities of the query fact are shaded. %
     }
	 \label{fig:candidate-graph-example}
 \end{figure}

The procedure we follow is outlined in Algorithm~\ref{alg:fact-enumeration}.
This algorithm enumerates the candidate facts for $f_q = r \langle s, t \rangle $ that are at most 2 hops away from either $s$ or $t$.
Three exceptions are made to this rule: (i)~CVT entities are not counted as hops, 
(ii)~we do not include $f_q$ in $F$ as it is trivial,
and (iii)~to reduce the search space, we do not expand intermediate neighbors that
represent an entity class or a type (e.g., ``actor'') as these can have
millions of neighbors.
Figure~\ref{fig:candidate-graph-example} shows an example graph with a subset of the
facts that we enumerate for the query fact $\mathit{spouseOf}\langle \text{Bill
Gates}, \text{Melinda Gates} \rangle$ using Algorithm~\ref{alg:fact-enumeration}.

\subsection{Fact ranking}
\label{sec:ranking-facts}

Next, we describe how we rank the set of enumerated candidate facts $F$ 
with respect to their relevance to the
query fact $f_q=r\langle s, t \rangle$.
The overall methodology is as follows. For each candidate fact $f_c \in F$, we create a
pair $(f_q, f_c)$---an analog to a query-document
pair---and score it using a function $u: (f_q, f_c) \to [0,1] \in R$ (higher values indicate higher relevance).
We then obtain a ranked list of facts $F'$ by sorting the facts in $F$ based on their score.

We begin by describing the training procedure we follow and continue with the network architecture we use for learning our scoring function $u$.

\paragraph{Learning procedure}
\label{sec:learning-procedure}
We train a network that learns the scoring function $u(f_q, f_c)$ end-to-end in mini-batches using stochastic gradient descent (we define the network architecture below).
We optimize the model parameters using Adam~\cite{kingma2014adam}.
During training we minimize a pairwise loss to learn the function $u$, while during inference we use the learned function $u$ to score a query-candidate fact pair ($f_q$, $f_c$).
This paradigm has been shown to outperform pointwise learning methods in ranking tasks, while keeping inference efficient~\cite{dehghani2017neural}.
Each batch $B$ consists of query-candidate fact pairs ($f_q$, $f_c$) of a single query fact $f_q$.
For constructing $B$ for a query fact $f_q$, we use all pairs ($f_q$, $f_c$) that are labeled as relevant and sample $k$ pairs ($f_q$, $f_c$) that are labeled as irrelevant.  
During training, we minimize the mean pairwise squared error between all pairs of ($f_q$, $f_c$) in $B \times B$:
\begin{align}
	L(B, \theta) = \frac{1}{|B|} \sum_{\langle x_1, x_2 \rangle \in B \times B } ([l(x_1) - l(x_2)] - [u(x_1) - u(x_2)] )^2,
\end{align}
where $x_1=(f_q, f_{c_1})$ and $x_2=(f_q, f_{c_2})$ are query-candidate fact pairs in the set $B \times B$,  $l(x) \in \{0, 1\}$ is the relevance label of a query-candidate fact pair $x$, $|B|$ is the batch size, and $\theta$ are the parameters of the model which we define below.

\begin{figure}[t]
	\centering
	\includegraphics[scale=0.6]{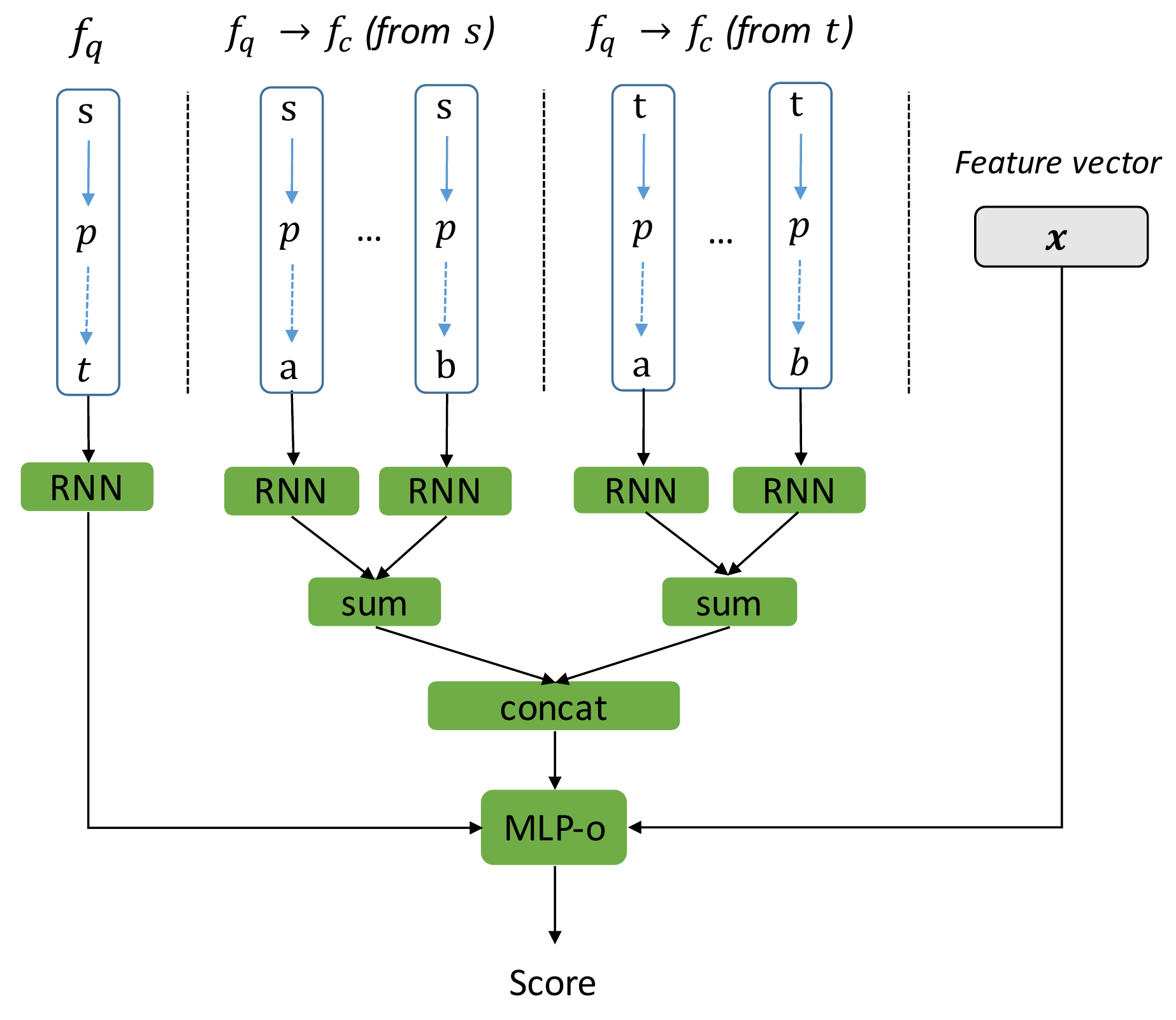}
	 \caption{
Network architecture that learns a scoring function $u(f_q, f_c)$.
Given a query fact $f_q=\mathit{r} \langle \text{s}, \text{t}\rangle$ and a candidate fact $f_c=\mathit{r'} \langle \text{a}, \text{b}\rangle$ it outputs a score $u(f_q, f_c)$.
``$f_q \rightarrow f_c$ (from $e$)'' is a label for the paths that start from an entity $e$ of the query fact (either $s$ or $t$) and end at an entity $e'$ of the candidate fact $f_c$.
Note that $p$ is a variable in this figure, i.e., it might refer to different predicates.
}
	 \label{fig:architecture}
\end{figure}

\paragraph{Network architecture}
Figure~\ref{fig:architecture} shows the network architecture we designed for learning the scoring function $u(f_q, f_c)$.
We encode the query fact $f_q$ in a vector $\boldsymbol{v_q}$ using an RNN. %
As we will explain further in that section, we do not model the entities in the facts independently due to the large number of entities; instead, we model each entity as an aggregation of its types.
Therefore, instead of modeling the candidate fact $f_c$ in isolation and losing per-entity information, we first enumerate all the paths up to two hops away from both the entities of the query fact $f_q$ ($s$ and  $t$) to all the entities of the candidate fact $f_c$ ($s'$ and $t'$).
Let $A_s$ denote the set of paths from $s$ to all the entities of $f_c$.
Let $A_t$ denote the set of paths from $t$ to all the entities of $f_c$.
For each $A \in \{A_s, A_t\}$, we first encode all the paths in $A$ using an RNN, and then combine the resulting encoded paths using the procedure described later in this section.
We denote the vectors obtained from the above procedure for $A_s$ and $A_t$ as $\boldsymbol{v}_{as}$ and $\boldsymbol{v}_{at}$, respectively.
Then we obtain a vector $\boldsymbol{v}_{a}=[\boldsymbol{v}_{as}, \boldsymbol{v}_{at}]$, where $[\cdot,\cdot]$ denotes the concatenation operation (middle part of Figure~\ref{fig:architecture}).
Note that we use the same RNN parameters for all the above operations.
To further inform the scoring function, we design a set of hand-crafted features $\boldsymbol{x}$ (right-most part of Figure~\ref{fig:architecture}).
We detail the hand-crafted features later in this section.

Finally, $\text{MLP-o}([\boldsymbol{v}_q, \boldsymbol{v}_{a}, \boldsymbol{x}])$ is a multi-layer perceptron with $\alpha$ hidden layers of dimension $\beta$ and one output layer that outputs $u(f_q, f_c)$.
We use a ReLU activation function in the hidden layers and a sigmoid activation function in the output layer.
We vary the number of layers to capture non-linear interactions between the features in $\boldsymbol{v}_q$, $\boldsymbol{v}_a$, and $\boldsymbol{x}$.

The remainder of this section describes how we encode a single fact, how we combine the representations of a set of facts, and, finally, the hand-crafted features.

\subsubsection{Encoding a single fact}
Recall from Section~\ref{sec:preliminaries} that a fact $f$ is a path in the KG. 
In order to model paths we turn to neural representation learning.
More specifically, since paths are sequential by nature we employ recurrent neural networks (RNNs) to encode them in a single vector~\cite{guu2015traversing,das2017chains}. 
This type of modeling has proven successful in predicting missing links in KGs~\cite{das2017chains}.
One restriction that we have in modeling such paths is the very large number of entities ($\sim1.5$ million entities in our dataset) and, since learning an embedding for such large numbers of entities requires prohibitively large amounts of memory and data, we represent each entity using an aggregation of its types~\cite{das2017chains}.
Formally, let $\boldsymbol{W}_z$ denote a $|Z| \times d_z$ matrix, where each row is an embedding of an entity type $z$, $|Z|$ is the number of entity types in our dataset and $d_z$ is the entity type embedding dimension.
Let $\boldsymbol{W}_p$ denote a $|P| \times d_p$ matrix, where each row is an embedding of a predicate $p$, $|P|$ is the number of predicates in our dataset, and $d_p$ is the predicate embedding dimension.
In order to model \emph{inverse} predicates in paths (e.g., $\text{Microsoft} \rightarrow \mathit{founderOf}^{-1} \rightarrow \text{Paul Allen}$), we also define a $|P| \times d_p$ matrix $\boldsymbol{W}_{p_i}$, which corresponds to embeddings of the inverse of each predicate~\cite{guu2015traversing}.

The procedure we follow for modeling a fact $f$ is as follows.
For simplicity in the notation, in this Section we denote a path as a sequence of alternate entities and predicates $[s_0, p_0, \ldots t_m]$, instead of a sequence of triples as defined in Section~\ref{sec:preliminaries}.
For each entity $e \in f$, we first retrieve the types of $e$ in $K$.
From these, we only keep the 7 most frequent types in $K$, which we denote as $Z_e$~\cite{das2017chains}.
We then project each $z \in Z_e$ to its corresponding type embedding $\boldsymbol{w}_z \in \boldsymbol{W}_z$ and perform element-wise sum on these embeddings to obtain an embedding $w_e$ for entity $e$. 
We project each predicate $p \in f$ to its corresponding embedding $\boldsymbol{w}_p$ ($\boldsymbol{w}_p \in \boldsymbol{W}_{p_i}$ if $p$ is inverse, $\boldsymbol{w}_p \in \boldsymbol{W}_{p}$ otherwise).

The resulting projected sequence $X_f =[\boldsymbol{w}_{s_0}, \boldsymbol{w}_{p_0} , \ldots, \boldsymbol{w}_{t_m}]$ is passed to a uni-directional recurrent neural network (RNN).
The RNN has a sequence of hidden states $[\boldsymbol{h}_1, \boldsymbol{h}_2, \ldots, \boldsymbol{h}_n]$, where $\boldsymbol{h}_i=tanh(\boldsymbol{W_{hh}} \boldsymbol{h}_{i-1} + \boldsymbol{W_{xh}} \boldsymbol{x}_i)$, and $\boldsymbol{W_{hh}}$ and $\boldsymbol{W_{xh}}$ are the parameters of the RNN.
The RNN is initialized with zero state values.
We use the last state of the RNN $\boldsymbol{h}_n$ as the representation of the fact $f$. 

\subsubsection{Combining a set of facts}
We obtain the representation of the set of encoded facts using element-wise summation of the encoded facts (vectors).
We leave more elaborate methods for combining facts such as attention mechanisms~\cite{bahdanau2014neural,das2017chains} for future work.

\subsubsection{Hand-crafted features}

\begin{table}[t]
\caption{Notation
}
\label{tab:notation}

\begin{tabularx}{\linewidth}{l p{3.2cm} p{8.5cm}}
\toprule
\bf Name & \bf Description & \bf Definition \\
\midrule
$\mathit{NumTriples}$ & Number of triples in $K$ &  $|\{\langle s, p, t\rangle: \langle s, p, t\rangle \in K\} |$  \\
$\mathit{TriplesPred}(p)$ & Set of triples that have predicate $p$ & $\{ \langle s, p', t \rangle :  \langle s, p', t \rangle \in K , p' = p \}$\\

$\mathit{TriplesEnt}(e)$ & Set of triples that have entity $e$ & $\{ \langle s, p, t \rangle :  \langle s, p, t \rangle \in K, s = e \vee t = e \}$ \\

$\mathit{TriplesSubj}(e)$ & Set of triples that have entity $e$ as subject & $\{ \langle s, p, t \rangle :  \langle s, p, t \rangle \in K, s = e\} $\\

$\mathit{TriplesObj}(e)$ & Set of triples that have entity $e$ as object & $\{ \langle s, p, t \rangle :  \langle s, p, t \rangle \in K, t = e\} $\\
$\mathit{UniqEnt}(T)$ & The unique set of entities in a set of triples $T$ & $ \bigcup \{ \{s, t\}: \langle s, p, t \rangle \in T \} $
\\

$\mathit{Types}(e)$ & The set of types of entity $e$ & $\{ z:  \langle e, \mathit{type}, z \rangle \in K\} $ \\

$\mathit{Entities}(f)$  & The set of entities of fact $f$ & $  \bigcup \{ \{s, t\}: \forall \langle s, p, t \rangle \in f \}  $ \\
$\mathit{Preds}(f)$  & The set of predicates of fact $f$ & $\{ p:  \langle s, p, t \rangle  \in f \} $\\ 
\bottomrule

\end{tabularx}
\end{table}

Here, we detail the hand-crafted features $\boldsymbol{x}$ we designed or adjusted for this task.
Table~\ref{tab:notation} lists the notation we use.
We generate features based on feature templates that are divided into three groups: (i)~those that give us a sense of
\emph{importance} of a fact, (ii)~those that give us a sense of \emph{relevance} of $(f_q, f_c)$,
and (iii)~a set of miscellaneous features.
Note that we use log-computations to avoid underflows.

\paragraph{(i) Fact importance}
This group of feature templates give us a sense on how important a fact $f$ is when taking statistics of the knowledge graph $K$ into account at a global level.
Note that we calculate these features for both facts $f_q$ and $f_c$.
The first of these feature templates measures \emph{normalized predicate frequency} of each predicate $p$ that participates in fact $f$ (we also include the minimum, maximum and average value for each fact as metafeatures~\citep{borisov-2016-using}).
This is defined as the ratio of the size of the set of triples that have predicate $p$ in the KG to the total number of triples:
\begin{align}
  \mathit{PredFreq}(p) = \frac{|\mathit{TriplesPred}(p)|}{\mathit{NumTriples}}.
  \label{eq:pred-freq}
\end{align}
The second feature template is the \emph{normalized entity frequency} for each entity $e$ that participates in fact $f$ (we also include the minimum, maximum and average value for each fact as metafeatures).
This is defined as the ratio of the number of triples in which $e$ occurs in the KG over the number of triples in the KG:
\begin{align}
  \mathit{EntFreq}(e) = \frac{|\mathit{TriplesEnt}(e)|}{\mathit{NumTriples}}.
  \label{eq:ent-freq}
\end{align}
The final feature template in this feature group is \emph{path informativeness}, proposed by~\citet{pirro2015explaining}, which we apply for both $f_q$ and $f_c$ (recall from Section~\ref{sec:preliminaries} that a fact $f$ is a path in the KG).
This feature is analog to TF.IDF and aims to estimate the importance of predicates for an entity.
The informativeness of a path $\pi$ is defined as follows~\cite{pirro2015explaining}:
\begin{align}
  I(\pi) = \frac{1}{2 |\pi|} \sum_{\langle s, p, t\rangle \in \pi} \mathit{PFITF}_{out}(p, s, K) + \mathit{PFITF}_{in}(p, t, K),
  \label{eq:pathInformativeness}
\end{align}
where:
\begin{align}
  \mathit{PFITF}_{x}(p, e, K) = \mathit{PF}_{x}(p, e) * \mathit{ITF}(p), x \in \{in, out\},
    \nonumber
\end{align}
where $\mathit{ITF}(p)$ is the inverse triple frequency of predicate $p$:
\begin{align}
  \mathit{ITF}(p) = \log \frac{\mathit{NumTriples}}{|\mathit{TriplesPred}(p)|},
    \nonumber
\end{align}
$\mathit{PF}_{out}(p, e)$ is the outgoing predicate frequency of $e$ when $p$ is the predicate:
\begin{align}
  \mathit{PF}_{out}(p, e) = \frac{|\mathit{TriplesSubj}(e) \cap \mathit{TriplesPred}(p)|}{|\mathit{TriplesSubj}(e)|},
    \nonumber
\end{align}
and $\mathit{PF}_{in}(p, e)$ is the incoming predicate frequency of $e$ when $p$ is the predicate:
\begin{align}
  \mathit{PF}_{in}(p, e) = \frac{|\mathit{TriplesObj}(e) \cap \mathit{TriplesPred}(p)|}{|\mathit{TriplesObj}(e)|}.
    \nonumber
\end{align}

\paragraph{(ii) Relevance}
This group of feature templates gives us signal on the relevance of a candidate fact $f_c$ w.r.t.\ the query fact $f_q$.
The first of these feature templates measures \emph{entity similarity} for each pair $(e_1, e_2) \in  \mathit{Entities}(f_q) \times \mathit{Entities}(f_c)$ (we also include the minimum, maximum and average entity similarity as metafeatures).
We measure entity similarity using type-based Jaccard similarity:
\begin{align}
  \mathit{EntTypeSim}(e_1, e_2) = \mathit{JaccardSim}(\mathit{Types}(e_1), \mathit{Types}(e_2)).
  \label{eq:entity-sim-type-jaccard}
\end{align}

\noindent
The next feature template in the \emph{relevance} category is \emph{entity distance},
which allows us to reason about the distance of two entities
$(e_1, e_2) \in  \mathit{Entities}(f_q) \times \mathit{Entities}(f_c)$ (we also include the minimum, maximum and average entity distance as metafeatures).
This feature is defined as the length of the shortest path between $e_1$ and $e_2$ in $K$.
The intuition is that we can get a signal for the relevance of $f_c$ by measuring how ``close'' the entities in $f_c$ are to the entities of $f_q$ in the KG.

The next set of features measure \emph{predicate similarity} between every pair
of predicates $(p_1, p_2) \in \mathit{Preds}(f_q) \times \mathit{Preds}(f_c)$ (we also include the minimum, maximum and average predicate similarity as metafeatures).
The intuition is that if $f_c$ has predicates that are highly similar to the predicates in $f_q$, then $f_c$ might be relevant to $f_q$.
We measure predicate similarity in two ways.
First, by measuring the co-occurrence of entities that participate in the
predicates $p_1$ and $p_2$:
\begin{align}
  &\mathit{PredCooccSim}(p_1, p_2) = {} 
    \label{eq:pred-sim-ent-cooccur}\\
  &\mathit{JaccardSim}(\mathit{UniqEnt}(\mathit{TriplesPred}(p_1)), \mathit{UniqEnt}(\mathit{TriplesPred}(p_2))).
  \nonumber
\end{align}
For instance, $\mathit{PredCooccSim}(p_1, p_2)$ would be high for $p1 = \mathit{starredIn}$ and $p2 = \mathit{directedBy}$.

Second, by measuring the jaccard similarity of the set of predicates in $f_q$ with the set of predicates in $f_c$~\cite{pirro2015explaining}:
\begin{align}
  &\mathit{SetPredicatesJaccardSim}(f_q, f_c) = {} 
    \label{eq:path-sim-jaccard} \\
  &\mathit{JaccardSim}(\mathit{Preds}(f_q), \mathit{Preds}(f_c)).
  \nonumber
\end{align}
Finally, we add a binary feature %
that captures whether $f_q$ and $f_c$ have the same CVT entity, i.e., $f_c$ is an attribute of $f_q$.

\paragraph{(iii) Miscellaneous}
This set of features includes whether $f_q$ has a CVT entity (same for $f_c$).
We also include whether an entity is a date (for all entities of $f_q$ and $f_c$).
Finally, we include the concatenation of the predicates of $f_q$ as a feature using one-hot encoding.

\section{Experimental Setup}
\label{sec:expsetup}

In this section we describe the setup of our experiments that aim to answer~\textbf{\ref{rq:contextualize-facts}}, which we break down to the following research sub-questions:

\begin{description}[nosep]%

\item[\textbf{RQ3.1}] How does NFCM perform compared to a set of heuristic baselines on a crowdsourced dataset?

\item[\textbf{RQ3.2}] How does NFCM perform compared to a scoring function that scores candidate facts w.r.t. a query fact using the relevance labels gathered from distant supervision on a crowdsourced dataset?

\item[\textbf{RQ3.3}] Does NFCM benefit from both the handcrafted features and the automatically learned features? 

\item[\textbf{RQ3.4}] What is the per-relationship performance of NFCM?
How does the number of instances per relationship affect the ranking performance? 
\end{description}

\subsection{Knowledge graph}
\label{sec:datasets-kg}
We use the latest edition of Freebase as our knowledge graph~\cite{bollacker2008freebase}.
We include Freebase relations from the following set of domains:
\textit{People, Film, Music, Award, Government, Business, Organization, Education}.
Following previous work~\cite{mintz2009distant}, we exclude triples that have an equivalent reversed triple.

\subsection{Dataset}
\label{sec:dataset}
Our dataset consists of query facts, candidate facts, and a relevance label for each query-candidate fact pair.
In order to construct our evaluation dataset we need to start with a set of relationships.
Given that most of our domains are people-centric, we obtain this set by extracting all relationships from Freebase that have an entity of type \emph{Person} as one of the entities.
In the end, we are left with 65 unique relationships in total (see Table~\ref{tab:relationships} for example relationships).
\begin{table}[t]
\centering
\caption{Examples of relationships used in this work.}
\label{tab:relationships}
\begin{tabularx}{22em}{l l}
\toprule
\bf Domain & \bf Relationship
\\
\midrule
People & $\mathit{spouseOf}(\mathit{person}, \mathit{person})$  %
\\
& $\mathit{parentOf}(\mathit{person}, \mathit{person})$  %
\\
& $\mathit{educatedAt}(\mathit{person}, \mathit{organization})$  %
\\
\midrule
Business & $\mathit{founderOf}(\mathit{person}, \mathit{organization})$  %
\\
& $\mathit{boardMemberOf}(\mathit{person}, \mathit{organization})$ %
\\

& $\mathit{leaderOf}(\mathit{person}, \mathit{organization})$ %
\\

\midrule
Film & $\mathit{starredIn}(\mathit{person}, \mathit{film})$  %
\\
& $\mathit{directorOf}(\mathit{person}, \mathit{film})$ %
\\
 & $\mathit{producerOf}(\mathit{person}, \mathit{film})$ %
 \\
\bottomrule
\end{tabularx}
\end{table}
We then proceed to gather our set of query facts.
For each relationship, we sample at most 2,000 query facts, provided that they have at least one relevant fact after applying the procedure described in Section~\ref{sec:training-data}.
In total, the dataset contains 62,044 query facts (954.52 on average per relationship).
After gathering query facts for each relation, we enumerate candidate facts for each query fact using the procedure described in Section~\ref{sec:enumerate-facts}.
Finally, we randomly split the dataset per relationship (70\% of the query facts for training, 10\% for validation, 20\% for testing).
Table~\ref{tab:dataset-stats} shows statistics of the resulting dataset.

\begin{table}[t]
\centering
\caption{Statistics of the dataset gathered using distant supervision (see Section~\ref{sec:training-data}).}
\label{tab:dataset-stats}
\begin{tabularx}{27em}{l l l l l l}
\toprule
\bf Part & \bf \# query facts & \multicolumn{4}{c}{\bf \# candidate facts}
\\
& &  average & median & max. & min.
\\
\midrule
Training & 44,632 & 1,420 & 741 & 9,937 & 2 
\\
Validation & \phantom{0}4,983 & 1,424 & 749  & 9,796 & 3
\\
Test & 12,429 &  1,427 & 771 &  9,924 & 3
\\
\bottomrule
\end{tabularx}
\end{table}

Note that we train and tune the fact ranking models with the training and validation sets in Table~\ref{tab:dataset-stats} respectively, using the automatically gathered relevance labels (see Section~\ref{sec:training-data}).
The test set was only used for preliminary experiments (not reported) and for constructing our manually curated evaluation dataset (see Section \ref{sec:evaluation}).
We describe how we automatically gather noisy relevance labels for our dataset in the next section.

\subsection{Gathering noisy relevance labels}
\label{sec:training-data}

Gathering relevance labels for our task is challenging due to the size and heterogeneous nature of KGs, i.e., having a large number of facts and relationship types.
Therefore, we turn to distant supervision~\cite{mintz2009distant} to gather relevance labels at scale.
We choose to get a supervision signal from Wikipedia for the following reasons:
(i) it has a high overlap of entities with the KG we use, and 
(ii) facts that are in KGs are usually expressed in Wikipedia articles alongside other, related facts.
We filter Wikipedia to select articles whose main entity is in Freebase, and the entity type corresponds to one of the domains listed in Section~\ref{sec:datasets-kg}.
This results in a set of 1,743,191 Wikipedia articles.

The procedure we follow for gathering relevance labels given a query fact $f_q$ and its set of candidate facts $F$ is as follows.
For a query fact $f_q=r \langle s, t\rangle$, we focus on the Wikipedia article of entity $s$.
First, as Wikipedia style guidelines dictate that only the first mention of another entity should be linked, we augment the articles with additional entity links using an entity linking method proposed in~\cite{voskarides-generating-2017}.
Next, we retain only segments of the Wikipedia article that contain references to $t$.
Here, a segment refers to the sentence that has a reference to $t$ and also one sentence before and one after the sentence.
For each such extracted segment, we assume that it expresses the fact $f_q$, which is a common assumption in gathering noisy training data for relation extraction~\cite{mintz2009distant}.
From the segments, we then collect a set of other entities, $O$, that occur in the same sentence that mentions $t$: for computational efficiency, we enforce $|O| \le{} 20$.
Then, we extract facts for all possible pairs of entities $\langle e_1, e_2 \rangle \in \{O \cup \{s, t\}\} \times \{O \cup \{s, t\}\}$.
If there is a single fact $f_c$ in $K$ that connects $e_1$ and $e_2$, we deem $f_c$ relevant for $f_q$.
However, if there are multiple facts connecting $e_1$ and $e_2$ in $K$, the mention of the fact in the specific segment is ambiguous and thus we do not deem any of these facts as relevant~\cite{sorokin2017context}.
The rest of the facts in $F$ are deemed irrelevant for $f_q$.

The distribution of relevant/non-relevant labels in the distantly supervised data is heavily skewed: out of 87,998,956 facts in total, only 225,032 are deemed to be relevant (0.26\%).
This is expected since the candidate fact enumeration step can generate thousands of facts for a certain query fact (see Section~\ref{sec:enumerate-facts}).

As a sanity check, we evaluate the performance of our approach to collect distant supervision data by sampling 5 query facts for each relation in our dataset.
For these query facts, we perform manual annotations on the extracted candidate facts that were deemed as relevant by the distant supervision procedure.
We obtain an overall precision of 76\% when comparing the relevance labels of the distant supervision against our manual annotations.
This demonstrates the potential of our distant supervision strategy for creating training data.

\subsection{Manually curated evaluation dataset}
\label{sec:evaluation}
In order to evaluate the performance of NFCM on the KG fact contextualization task, we perform crowdsourcing to collect a human-curated evaluation dataset.
The procedure we use to construct this evaluation dataset is as follows.
First, for each of the 65 relationships we consider, we sample five query facts of the relationship from the test set (see Section~\ref{sec:dataset}).
Since fact enumeration for a query fact can yield hundreds or thousands of facts (Section~\ref{sec:enumerate-facts}), it is infeasible to consider all the candidate facts for manual annotation.
Therefore, we only include a candidate fact in the set of facts to be annotated if:
(i) the candidate fact was deemed relevant by the automatic data gathering procedure (Section~\ref{sec:training-data}), or
(ii) the candidate fact matches a fact pattern that is built using relevant facts that appear in at least 10\% of the query facts of a certain relationship. An example fact pattern is $\mathit{parentOf} \langle ?, ?\rangle$, which would match the fact $\mathit{parentOf}\langle \text{Bill Gates}, \text{Jennifer Gates} \rangle$. 

We use the CrowdFlower platform, and ask the annotators to judge a candidate fact w.r.t.\ its relevance to a query fact.
We provide the annotators with the following scenario (details omitted for brevity):
\begin{quote}
\textit{We are given a specific real-world fact, e.g.,  ``Bill Gates is the founder of Microsoft'', which we call the query fact.
We are interested in writing a description of the query fact (a sentence or a small paragraph).
The purpose of this assessment task is to identify other facts that could be included in a description of the query fact. Note that even though all facts presented for assessment will be accurate, not all will be relevant or equally important to the description of the main fact.}
\end{quote}
We ask the annotators to assess the relevance of a candidate fact in a 3-graded scale:
\begin{itemize}
	\item \emph{very relevant}: I would include the candidate fact in the description of the query fact; the candidate fact provides additional context to the query fact.
	\item \emph{somewhat relevant}: I would include the candidate fact in the description of the query fact, but only if there is space. 
	\item \emph{irrelevant}: I would not include the candidate fact in the description of the query fact.
\end{itemize}
Alongside each query-candidate fact pair, we provide a set of extra facts that could possibly be used to decide on the relevance of a candidate fact.
These include facts that connect the entities in the query fact with the entities in the candidate fact.
For example, if we present the annotators with the query fact 
$\mathit{spouseOf} \langle \text{Bill Gates}$, Melinda Gates$\rangle$
and the candidate fact 
$\mathit{parentOf} \langle \text{Melinda Gates}$, Jennifer Gates$\rangle$
we also show the fact 
$\mathit{parentOf} \langle \text{Bill Gates}$, Jennifer Gates$\rangle$.

Each query-candidate fact pair is annotated by three annotators.
We use majority voting to obtain the gold labels, breaking ties arbitrarily. 
The annotators get a payment of 0.03 dollars per query-candidate fact pair.

By following the crowdsourcing procedure described above, we obtain 28,281 fact judgments for 2,275 query facts (65 relations, 5 query facts each).
Table~\ref{tab:cf-label-distribution} details the distribution of the relevance labels.
One interesting observation is that facts that are attributes of other facts (see Section \ref{sec:preliminaries}) tend to have relatively more relevant judgments than the ones that are not.
This is expected since some of them are attributes of the query fact (e.g., date of marriage for a \textit{spouseOf} query fact).
Finally, Fleiss' kappa is $\kappa$ = 0.4307, which is considered moderate agreement.
Note that all the results reported in Section~\ref{sec:sigir2018-results} are on the manually curated dataset described here.
\begin{table}[t]
\centering
\caption{Relevance label distribution of the crowdsourced evaluation dataset.
}
\label{tab:cf-label-distribution}
\begin{tabularx}{27em}{l@{} c c}
\toprule
\bf Relevance & \bf \mbox{}\hspace*{-.35cm}Non-attribute facts (\%) & \bf Attribute facts (\%) \\
\midrule
Irrelevant
&60.86
&34.34
\\
Somewhat relevant
&34.49
&57.81
\\
Very relevant
&\phantom{0}4.63
&\phantom{0}7.84
\\
\bottomrule
\end{tabularx}
\end{table}
  
\paragraph{Evaluation metrics}
We use the following standard retrieval evaluation metrics: MAP, NDCG@5, NDCG@10 and MRR. In the case of MAP and MRR, which expect binary labels, we consider ``very relevant'' and``somewhat relevant'' as ``relevant".
We report on statistical significance with a paired two-tailed t-test.
\subsection{Heuristic baselines}
\label{sec:baselines}
To the best of our knowledge, there is no previously published method that addresses the task introduced in this chapter.
Therefore, we devise a set of intuitive baselines that are used to showcase that our task is not trivial.
We derive them by combining features we introduced in Section~\ref{sec:ranking-facts}.
We define these heuristic functions below:
\begin{itemize}
\item \textit{Fact informativeness (FI).}
Informativeness of the candidate fact $f_c$ \cite[Eq. \ref{eq:pathInformativeness}]{pirro2015explaining}. This baseline is independent of $f_q$.
\item \textit{Average predicate similarity (APS).}
Average predicate similarity of all pairs of predicates $(p_1, p_2) \in \mathit{Preds}(f_q) \times \mathit{Preds}(f_c)$ (Eq. ~\ref{eq:pred-sim-ent-cooccur}). 
The intuition here is that $f_c$ might be relevant to $f_q$ if it contains predicates that are similar to the predicates of $f_q$.
\item \textit{Average entity similarity (AES).}
Average entity similarity of all pairs of entities in $(e_1, e_2) \in \mathit{Entities}(f_q) \times \mathit{Entities}(f_c)$  (Eq. ~\ref{eq:entity-sim-type-jaccard}). 
The assumption here is that $f_c$ might be relevant to $f_q$ if it contains entities that are similar to the entities of $f_q$.
\end{itemize}

\subsection{Implementation details}
The models described in Section~\ref{sec:ranking-facts} are implemented in TensorFlow v.1.4.1~\cite{abadi2016tensorflow}.
Table~\ref{tab:hyperparams} lists the hyperparameters of NFCM.
We tune the variable hyper-parameters of this table on the validation set and optimize for NDCG@5.

\begin{table}[h]
\centering
\caption{Hyperparameters of NFCM, tuned on the validation set.}
\label{tab:hyperparams}
\begin{tabularx}{28em}{p{6cm} l}
\toprule
\bf Description & \bf Value(s)\\
\midrule
\# negative samples $k$ during training & [1, 10, 100] \\
Learning rate & [0.01, 0.001, 0.0001] \\
$d_z$: entity type embedding dimension & [64, 128, 256] \\
$d_p$: Predicate embedding dimension & [64, 128, 256] \\
RNN cell size & [64, 128, 256] \\
RNN cell dropout & [0.0, 0.2] \\
$\alpha$: \# hidden layers of MLP-o & [0, 1, 2] \\
$\beta$: \# dimension of MLP-o hidden layers& [50, 100] \\
L2 regularization factor for MLP-o kernel & [0.0, 0.1, 0.2] \\
\bottomrule
\end{tabularx}
\end{table}

\section{Results and Discussion}
\label{sec:sigir2018-results}
In this section we discuss and analyze the results of our evaluation, answering the research questions listed in Section~\ref{sec:expsetup}.

In our first experiment, we compare NFCM to a set of heuristic baselines we derived to answer \textbf{RQ3.1}.
Table~\ref{tab:results-baselines} shows the results.
We observe that NFCM significantly outperforms the heuristic baselines by a large margin.
We have also experimented with linear combinations of the above heuristics but the performance does not improve over the individual ones and therefore we omit those results.
We conclude that the task we define in this chapter is not trivial to solve and simple heuristic functions are not sufficient.
\begin{table}[t]
\centering
\caption{Comparison between NFCM and the heuristic baselines. Significance is tested between NFCM and AES, the best performing baseline. We depict a significant improvement of NFCM over AES for $p<0.05$ as $^\blacktriangle$.}
\label{tab:results-baselines}
\begin{tabularx}{27em}{p{1.8cm} X X X X}
\toprule
\bf Method & MAP & NDCG@5 &  NDCG@10  &  MRR \\
\midrule
FI
&0.1222
&0.0978
&0.1149
&0.1928
\\
APS
&0.2147
&0.2175
&0.2354
&0.3760
\\
AES
& 0.2950
& 0.3284
& 0.3391
& 0.5214
\\
\midrule
NFCM
& \bf 0.4874$^\blacktriangle$
& \bf 0.5110$^\blacktriangle$
& \bf 0.5289$^\blacktriangle$
& \bf 0.7749$^\blacktriangle$
\\
\bottomrule
\end{tabularx}
\end{table}

In our second experiment we compare NFCM with distant supervision and aim to answer \textbf{RQ3.2}. That is, how does NFCM perform compared to DistSup, a scoring function that scores candidate facts w.r.t. a query fact using the relevance labels gathered from distant supervision.
The aim of this experiment is to investigate whether it is beneficial to learn ranking functions based on the signal gathered from distant supervision, and to see if we can improve performance over the latter.
Table~\ref{tab:results-distsup} shows the results.
We observe that NFCM significantly outperforms DistSup on MAP, NDCG@5, and NDCG@10 and conclude that learning ranking functions (and in particular NFCM) based on the signal gathered from distant supervision is beneficial for this task.
We also observe that NFCM performs significantly worse than DistSup on MRR.
One possible reason for this is that NFCM returns facts that are indeed relevant but were not selected for annotation and thus assumed not relevant, since the data annotation procedure is biased towards DistSup (see Section ~\ref{sec:evaluation}).
We aim to validate this hypothesis by conducting an additional user study in future work.
Nevertheless, having an automatic method for KG fact contextualization trained with distant supervision becomes increasingly important for tail entities for which we might only have information in the KG itself and not in external text corpora or other sources.
\begin{table}[t]
\centering
\caption{Comparison between NFCM and the distant supervision baseline. We depict a significant improvement of NFCM over DistSup as $^\blacktriangle$ and a significant decrease as $^\blacktriangledown$ ($p<0.05$).}
\label{tab:results-distsup}
\begin{tabularx}{27em}{p{1.8cm} X X X X X}
\toprule
\bf Method & MAP & NDCG@5 &  NDCG@10  &  MRR \\
\midrule
DistSup
&0.2831
&0.4489
&0.3983
&\bf 0.8256
\\
NFCM
& \bf 0.4874$^\blacktriangle$
& \bf 0.5110$^\blacktriangle$
& \bf 0.5289$^\blacktriangle$
& 0.7749$^\blacktriangledown$
\\
\bottomrule
\end{tabularx}
\end{table}

\begin{table}[t]
\centering
\caption{Comparison between the full NFCM model and its variations. Significance is tested between NFCM and its best variation (LF). We depict a significant improvement of NFCM over LF for $p<0.05$ as $^\blacktriangle$.}
\label{tab:results-constituent}
\begin{tabularx}{28em}{p{1.8cm} X X X X}
\toprule
\bf Method & MAP & NDCG@5 &  NDCG@10 &  MRR \\
\midrule
HF
&0.4620
&0.4753
&0.4989
&0.7180
\\
LF
& 0.4676
& 0.4993
& 0.5134
& 0.7647
\\
\midrule
NFCM
& \bf 0.4874$^\blacktriangle$
& \bf 0.5110
& \bf 0.5289$^\blacktriangle$
& \bf 0.7749
\\
\bottomrule
\end{tabularx}
\end{table}

\begin{figure}[t]
	\centering
	\includegraphics[width=.8\linewidth]{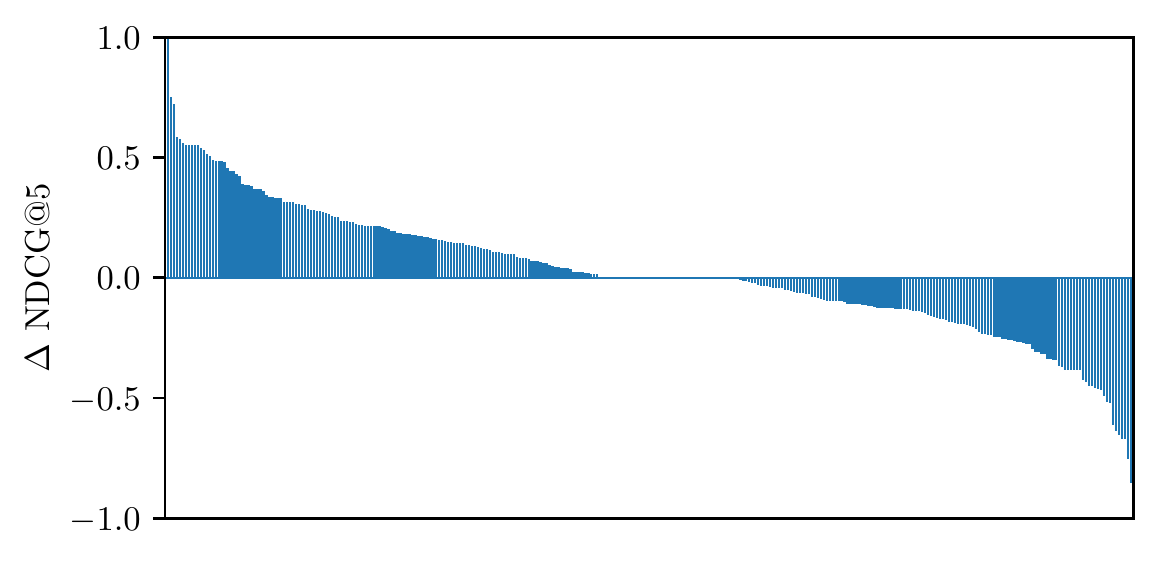}
	 \caption{
	 Per query fact differences in NDCG@5 between the variation of NFCM that only uses the learned features (LF) and the best-performing variation of NFCM that only uses the hand-crafted features (HF). A positive value indicates that LF performs better than HF on a query fact and vice versa.
	 }
	 \label{fig:LF_minus_HF}
\end{figure}

In order to answer \textbf{RQ3.3}, that is, whether NFCM benefits from both the hand-crafted features and the learned features, we perform an ablation study.
Specifically, we test the following variations of NFCM that only modify the final layer of the architecture (see Section~\ref{sec:ranking-facts}):
\begin{enumerate}[label=(\roman*)]
\item LF: Keeps the learned features ($\boldsymbol{v_q}$ and $\boldsymbol{v_{a}}$), and ignores the hand-crafted features $\boldsymbol{x}$.
\item HF: Keeps the hand-crafted features ($\boldsymbol{x}$) and ignores the learned features ($\boldsymbol{v_q}$ and $\boldsymbol{v_{a}}$).
\end{enumerate}
We tune the parameters of LF and HF on the validation set.
Table~\ref{tab:results-constituent} shows the results.
First, we observe that NFCM outperforms HF by a large margin. Also, NFCM outperforms LF on all metrics (significantly so for MAP and NDCG@10) which means that by combining HF and LF we are able to obtain more relevant results at lower positions of the ranking.  
We aim to explore more sophisticated ways of combining LF and HF in future work.  %
In order to verify whether LF and HF have complementary signals, we plot the per-query differences in NDCG@5 for LF and HF in Figure~\ref{fig:LF_minus_HF}.
We observe that the performance of LF and HF varies across query facts, confirming the hypotheses that LF and HF yield complementary signals.

\begin{figure}[t]
	\centering
	\includegraphics[width=.8\linewidth]{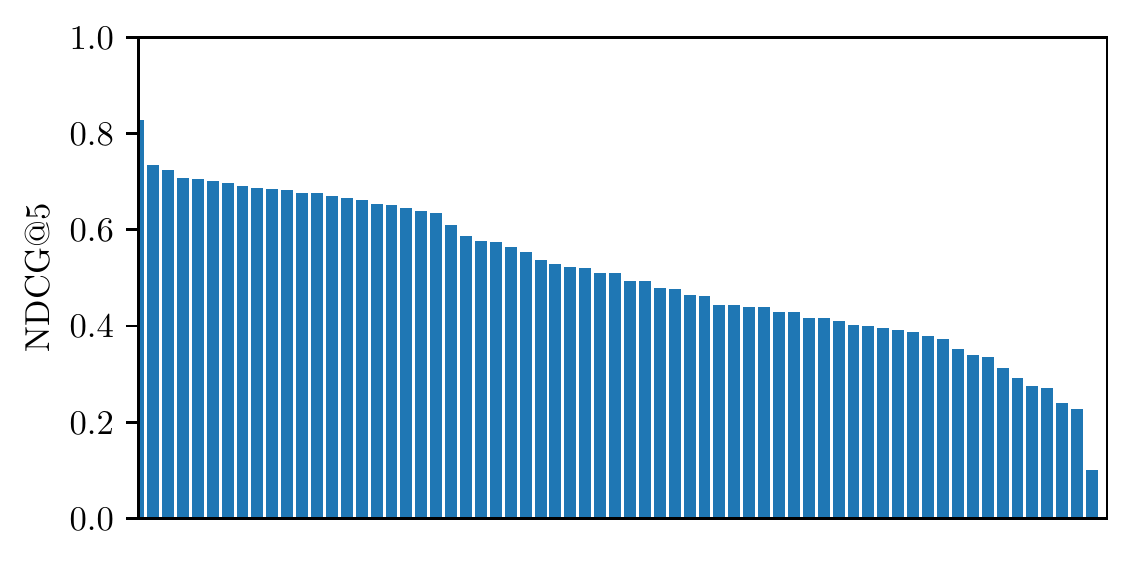}
	 \caption{
	 NDCG@5 for NFCM per relationship.
	 }
	 \label{fig:NFCM_per_relationship}
\end{figure}

In order to answer \textbf{RQ3.4}, we conduct a performance analysis per relationship.
Figure~\ref{fig:NFCM_per_relationship} shows the per-relationship NDCG@5 performance of NFCM -- query fact scores are averaged per relationship.
The relationship for which NFCM performs best is $\mathit{profession}$, which has a NDCG@5 score of 0.8275.
The relationship for which NFCM performs worst at is $\mathit{awardNominated}$, which has a NDCG@5 score of 0.1.
Further analysis showed that $\mathit{awardNominated}$ has a very large number of candidate facts on average, which might explain the poor performance on that relationship.

\begin{figure}[t]
	\centering
	\includegraphics[width=0.8\linewidth]{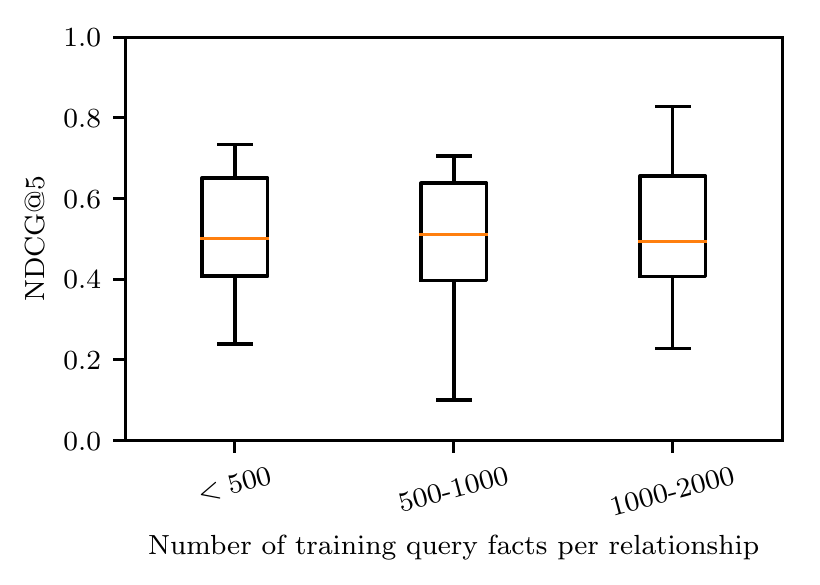}
	 \caption{
	 Box plot that shows NDCG@5 per number of training query facts of each relationship (binned).
	 Each box shows the median score with an orange line and the upper and lower quartiles (maximum and lower values shown outside each box).
	 }
	 \label{fig:per_relationship_numqueries}
\end{figure}
Furthermore, we investigate how the number of queries we have in the training set for each relationship affects the ranking performance.
Figure~\ref{fig:per_relationship_numqueries} shows the results.
From this figure we conclude that there is no clear relationship and thus that NFCM is robust to the size of the training data for each relationship.

\begin{figure}[t]
	\centering
	\includegraphics[width=0.8\linewidth]{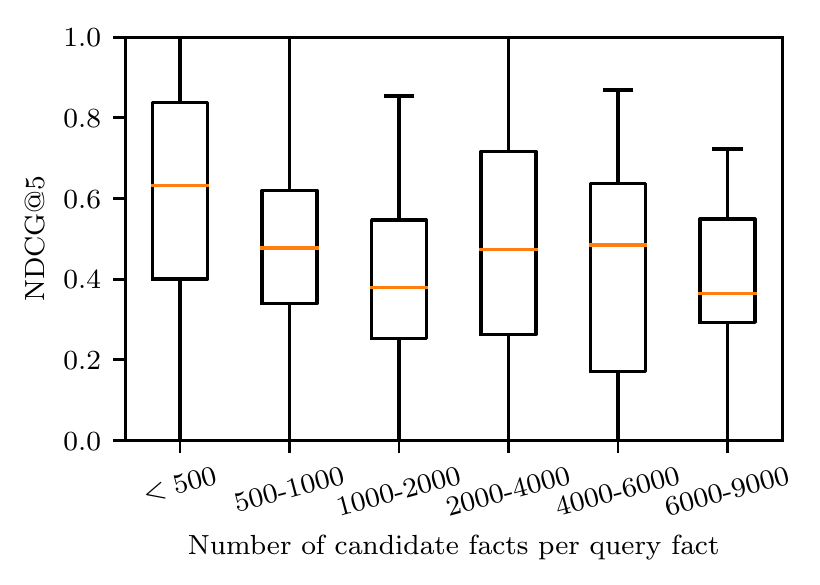}
	 \caption{
	 Box plot that shows NDCG@5 per number of candidate facts of each query fact (binned).
	 Each box shows the median score with an orange line and the upper and lower quartiles (maximum and lower values shown outside each box).
	 }
	 \label{fig:ndcg_5_pernumcandidates}
\end{figure}

Next, we analyse the performance of NFCM with respect to the number of candidates per query fact; Figure~\ref{fig:ndcg_5_pernumcandidates} shows the results.
We observe that the performance decreases when we have more candidate facts for a query, although not by a large margin, and that there does not seem to be a clear relationship between performance and the number of candidates to rank.

\section{Related Work}

The specific task we introduce in this chapter has not been addressed before, but there is related work in three main areas: entity relationship explanation, distant supervision, and fact ranking.

\subsection{Relationship explanation}

Explanations for relationships between pairs of entities can be provided in two ways: \emph{structurally}, i.e., by providing paths or sub-graphs in a KG containing the entities, or \emph{textually}, by ranking or generating text snippets that explain the connection.

\citet{fang2011rex} focus on explaining connections between entities by mining relationship explanation patterns from the KG. Their approach consists of two main components: explanation enumeration and explanation ranking. The first phase generates all patterns in the form of paths connecting the two entities in the KG, which are then combined to form explanations. In the final stage, the candidate explanations are ranked using notions of interestingness.
\citet{seufert2016espresso} propose a similar approach for entity sets. Their method focuses on explaining the connections between entity sets based on the concept of relatedness cores, i.e., dense subgraphs that have strong relations with both query sets. 
\citet{pirro2015explaining} also provide explanations of the relation between entities in terms of the top-k most informative paths between a query pair of entities; such paths are ranked and selected based on path informativeness and diversity, and pattern informativeness.

As to textual explanations for entity relationships, \citet{voskarides-learning-2015} focus on human-readable descriptions. They model the task as a learning to rank problem for sentences and employ a rich set of features.
\citet{huang2017learning} build on the aforementioned work and propose a pairwise ranking model that leverages clickthrough data and uses a convolutional neural network architecture.
While these approaches rank existing candidate explanations, \citet{voskarides-generating-2017} focus on generating explanations from scratch. They automatically identify the most common sentence templates for a particular relationship and, for each new relationship instance, these templates are ranked and instantiated using contextual information from the KG. 

The work described above focuses on explaining entity relationships in KGs; no previous work has focused on ranking additional KG facts for an input entity relationship as we do in this chapter.

\subsection{Distant supervision}

When obtaining labeled data is expensive, training data can be generated automatically.
\citet{mintz2009distant} introduce distant supervision for relation extraction; for a pair of entities that is connected by a KG relation, they treat all sentences that contain those entities in a text corpus as positive examples for that relation.
Follow-up work on relation extraction address the issue of noise related to distant supervision. \citet{riedel-2010-modeling, Surdeanu2012, alfonseca-2012-pattern} refine the model by relaxing the assumptions in the original method or by modeling noisy labels.

Beyond relation extraction, distant supervision has also been applied in other KG-related tasks.
\citet{ren-2015-clustype} introduce a joint approach entity recognition and classification based on distant supervision.
\citet{ling2012fine} used distant supervision to automatically label data for fine-grained entity recognition.

\subsection{Fact ranking}

In fact ranking, the goal is to rank a set of attributes with respect to an entity. \citet{hasibi2017dynamic} consider fact ranking as a component for entity summarization for entity cards. They approach fact ranking as a learning to rank problem. They learn a ranking model based on importance, relevance, and other features relating a query and the facts. \citet{meza2005ranking} explore a similar task, but rank facts with respect to a pair of entities to discover paths that contain informative facts between the pair.

Graph matching involves matching two graphs and discovering the patterns of relationships between them to infer their similarity \citep{cho2013}. Although our task can be considered as comparing a small query subgraph (i.e., query triples) and a knowledge graph, the goal is different from graph matching which mainly concerns aligning two graphs rather than enhancing one query graph.

\medskip\noindent%
Our work differs from the work discussed above in the following major ways. First, we enrich a query fact between two entities by providing relevant additional facts in the context of the query fact, taking into account both the entities and the relation of the query fact. Second, we rank whole facts from the KG instead of just entities. Last, we provide a distant supervision framework for generating the training data so as to make our approach scalable.

\section{Conclusion}
In this chapter, we introduced the knowledge graph fact contextualization task and proposed NFCM, a weakly-supervised method to address it.
NFCM first generates a candidate set for a query fact by looking at 1 or 2-hop neighbors and then ranks the candidate facts using supervised machine learning. 
NFCM combines handcrafted features with features that are automatically identified using deep learning.
We use distant supervision to boost the gathering of training data by using a large entity-tagged text corpus that has a high overlap with entities in the KG we use.
Our experimental results show that (i)~distant supervision is an effective means for gathering training data for this task, (ii)~NFCM significantly outperforms several heuristic baselines for this task, and (iii)~both the handcrafted and automatically-learned features contribute to the retrieval effectiveness of NFCM.

For future work, we aim to explore more sophisticated ways of combining handcrafted with automatically learned features for ranking.
Additionally, we want to explore other data sources for gathering training data, such as news articles and click logs.
Finally, we want to explore methods for combining and presenting the ranked facts in search engine result pages in a diversified fashion.

This chapter concludes our study on the first part of the thesis, which focuses on how to make structured knowledge more accessible to the user.
Next, in Chapter~\ref{chapter:sigir2020}, we address a different research theme, namely improving interactive knowledge gathering.

\part{Improving Interactive Knowledge Gathering}

\graphicspath{{05-sigir2020/figures/}}

\chapter{Query Resolution for Conversational Search with Limited Supervision}
\label{chapter:sigir2020}

\footnote[]{This chapter was published as~\citep{voskarides-2020-query}.}

In the second part of this thesis, we move to the research theme of improving interactive knowledge gathering and focus on conversational search.
In this Chapter, we aim to answer \textbf{\ref{rq:query-res}}: \acl{rq:query-res}
\section{Introduction}
Conversational AI deals with developing dialogue systems that enable interactive knowledge gathering~\cite{gao_neural_2018}.
A large portion of work in this area has focused on building dialogue systems that are capable of engaging with the user through chit-chat~\cite{li_diversity-promoting_2016} or helping the user complete small well-specified tasks~\cite{peng_deep_2018}.
In order to improve the capability of such systems to engage in complex information seeking conversations~\cite{qu_attentive_2019}, researchers have proposed information seeking tasks such as conversational question answering (QA) over simple contexts, such as a single-paragraph text~\cite{choi_quac:_2018,reddy_coqa_2019}. 
In contrast to conversational QA over simple contexts, in conversational search, a user aims to interactively find information stored in a large document collection~\cite{culpepper_research_2018}. 

In this chapter, we study multi-turn passage retrieval as an instance of conversational search: given the conversation history (the previous turns) and the current turn query, we aim to retrieve passage-length texts that satisfy the user's underlying information need~\cite{cast2019}.
Here, the current turn query may be under-specified and thus, we need to take into account context from the conversation history to arrive at a better expression of the current turn query.
Thus, we need to perform \emph{query resolution}, that is, add missing context from the conversation history to the current turn query, if needed.
An example of an under-specified query can be seen in Table~\ref{tab:quac_paragraph1}, turn \#4, for which the gold standard query resolution is: ``\emph{when was saosin 's first album released?}''.
In this example, context from all turns \#1 (``saosin''), \#2 (``band'') and \#3 (``first'') have to be taken into account to arrive to the query resolution.

Designing automatic query resolution systems is challenging because of phenomena such as zero anaphora, topic change and topic return, which are prominent in information seeking conversations~\cite{yatskar_qualitative_2019}.
These phenomena are not easy to capture with standard NLP tools (e.g., coreference resolution). 
Also, heuristics such as appending (part of) the conversation history to the current turn query are likely to lead to query drift~\cite{mitra_improving_1998}.
Recent work has modeled query resolution as a sequence generation task~\cite{raghu_statistical_2015,kumar_incomplete_2017,elgohary_can_2019}.
Another way of implicitly solving query resolution is by query modeling~\cite{guan2012effective,yang2015query,van2016lexical}, which has been studied and developed under the setup of session-based search~\cite{carterette2014overview,carterette2016evaluating}. 

\begin{table}[t]
	\centering
	\def\arraystretch{1}
	\caption{Excerpt from an example conversational dialog. Co-occurring terms in the conversation history and the relevant passage to the current turn (\#4) are shown in bold-face.
	}
	\begin{tabularx}{\linewidth}{l p{7cm}}
		\toprule
        \textbf{Turn} & \textbf{Query}
        \\\midrule
		1 & who formed \textbf{saosin}? \\
		2 & when was the band founded? \\
		3 & what was their \textbf{first} album?	\\ \hline
		4 & when was the album released? \\  		
		 & \emph{resolved:} when was saosin 's first album released? \\
		\midrule
		\multicolumn{2}{p{0.95\linewidth}}{\textit{Relevant passage to turn \#4}: The original lineup for \textbf{Saosin}, consisting of Burchell, Shekoski, Kennedy and Green, was formed in the summer of 2003. On June 17, the \textbf{band} released their \textbf{first} commercial production, the EP Translating the Name.}\\
        \bottomrule
	\end{tabularx}
	\label{tab:quac_paragraph1}
\end{table}

In this chapter, we propose to model query resolution for conversational search as a binary term classification task: for each term in the previous turns of the conversation decide whether to add it to the current turn query or not.
We propose QuReTeC (\textbf{Qu}ery \textbf{Re}solution by \textbf{Te}rm \textbf{C}lassification), a query resolution model based on bidirectional transformers~\cite{vaswani_attention_2017} -- more specifically BERT~\cite{devlin_bert:_2019}.
The model encodes the conversation history and the current turn query and uses a term classification layer to predict a binary label for each term in the conversation history.
We integrate QuReTeC in a standard two-step cascade architecture that consists of an initial retrieval step and a reranking step.
This is done by using the set of terms predicted as relevant by QuReTeC as query expansion terms.

Training QuReTeC requires binary labels for each term in the conversation history.
One way to obtain such labels is to use human-curated gold standard query resolutions~\cite{elgohary_can_2019}.
However, these labels might be cumbersome to obtain in practice.
On the other hand, researchers and practitioners have been collecting general-purpose passage relevance labels, either by the means of human annotations or by the means of weak signals, e.g., clicks or mouse movements~\cite{joachims_optimizing_2002}.
We propose a distant supervision method to automatically generate training data, on the basis of such passage relevance labels. 
The key assumption is that passages that are relevant to the current turn share context with the conversation history that is missing from the current turn query.
Table~\ref{tab:quac_paragraph1} illustrates this assumption: the relevant passage to turn \#4 shares terms with the conversation history. 
Thus, we label the terms that co-occur in the relevant passages\footnote{A relevance passage contains not only the answer to the question but also context and supporting facts that allow the algorithm or the human to reach to this answer.} and the conversation history as relevant for the current turn.

Our main contributions can be summarized as follows:
\begin{enumerate}[leftmargin=*,nosep]
	\item We model the task of query resolution as a binary term classification task and propose to address it with a neural model based on bidirectional transformers, QuReTeC.
	\item We propose a distant supervision approach that can use general-purpose passage relevance data to substantially reduce the amount of human-curated data required to train QuReTeC.
	\item We experimentally show that when integrating the QuReTeC model in a multi-stage ranking architecture we significantly outperform baseline models. Also, we conduct extensive ablation studies and analyses to shed light into the workings of our query resolution model and its impact on retrieval performance.
\end{enumerate}

\section{Related work}

\subsection{Conversational search}
Early studies on conversational search have focused on characterizing information seeking strategies and building interactive IR systems~\cite{oddy_information_1977,belkin1980anomalous,Croft:1987:IRN:35053.35054,belkin1995cases}.
\citet{Vtyurina:2017:ECS:3027063.3053175} investigated human behaviour in conversational systems through a user study and find that existing conversational assistants cannot be effectively used for conversational search with complex information needs.
\citet{Radlinski:2017:TFC:3020165.3020183} present a theoretical framework for conversational search, which highlights the need for multi-turn interactions.
\citet{cast2019} organize the Conversational Assistance Track (CAsT) at TREC 2019.
The goal of the track is to establish a concrete and standard collection of data with information needs to make systems directly comparable.
They release a multi-turn passage retrieval dataset annotated by experts, which we use to compate our method to the baseline methods.
\subsection{Query resolution}
Query resolution has been studied in the context of dialogue systems.
~\citet{raghu_statistical_2015} develop a pipeline model for query resolution in dialogues as text generation.
\citet{kumar_incomplete_2017} follow up on that work by using a sequence to sequence model combined with a retrieval model.
However, both these works rely on templates that are not available in our setting.
More related to our work, \citet{elgohary_can_2019} studied query resolution in the context of conversational QA over a single paragraph text.
They use a sequence to sequence model augmented with a copy and an attention mechanism and a coverage loss.
They annotate part of the QuAC dataset~\cite{choi_quac:_2018} with gold standard query resolutions on which they apply their model and obtain competitive performance.
In contrast to all the aforementioned works that model query resolution as text generation, we model query resolution as binary term classification in the conversation history.
\subsection{Query modeling}
\label{sec:session-search}
Query modeling has been used in session search, where the task is to retrieve documents for a given query by utilizing previous queries and user interactions with the retrieval system~\cite{carterette2014overview}.
~\citet{guan2012effective} extract substrings from the current and previous turn queries to construct a new query for the current turn.
~\citet{yang2015query} propose a query change model that models both edits between consecutive queries and the ranked list returned by the previous turn query.
\citet{van2016lexical} compare the lexical matching session search approaches and find that naive methods based on term frequency weighing perform on par with specialized session search models.
The methods described above are informed by studies of how users reformulate their queries and why~\cite{DBLP:journals/ir/SloanYW15}, which, in principle, is different in nature from conversational search. 
For instance, in session search users tend to add query terms more than removing query terms, which is not the case in (spoken) conversational search.
Another form of query modeling is query expansion. 
Pseudo-relevance feedback is a query expansion technique that first retrieves a set of documents that are assumed to be relevant to the query, and then selects terms from the retrieved documents that are used to expand the query~\cite{Lavrenko:2001:RBL:383952.383972,abdul2004umass,nogueira_task-oriented_2017}.
Note that pseudo-relevance feedback is fundamentally different from query resolution: in order to revise the query, the former relies on the top-ranked documents, while the latter only relies on the conversation history.
\paragraph{Distant supervision}
Distant supervision can be used to obtain large amounts of noisy training data.
One of its most successful  applications is relation extraction, first proposed by~\citet{mintz2009distant}.
They take as input two entities and a relation between them, gather sentences where the two entities co-occur from a large text corpus, and treat those as positive examples for training a relation extraction system.
Beyond relation extraction, distant supervision has also been used to automatically generate noisy training data for other tasks such as named entity recognition~\cite{yang_distantly_2018}, sentiment classification~\cite{ritter_named_2011}, knowledge graph fact contextualization~\cite{voskarides-weakly-supervised-2018} and dialogue response generation~\cite{ren-2020-thinking}.
In our work, we follow the distant supervision paradigm to automatically generate training data for query resolution in conversational search by using query-passage relevance labels.

\section{Multi-turn Passage Retrieval Pipeline}
In this section we provide formal definitions and describe our multi-turn passage retrieval pipeline. 
Table~\ref{tab:sigir2020-notation} lists notation used in this chapter.

 \begin{table}[t]
\centering
\caption{Notation used in the chapter.}
\begin{tabularx}{26em}{lp{7cm}}
\toprule
\bf Name  & \bf Description \\ 
\midrule
$terms(x)$  & Set of terms in term sequence $x$\\    %
$D$ & Passage collection \\ 
$q_i$ & Query at the current turn $i$\\
$q_{1:i-1}$ & Sequence of previous turn queries \\  
$q^*_i$ & Gold standard resolution of $q_i$\\
$E_{q_i}^*$   & Gold standard resolution terms for $q_i$, see Eq. \eqref{eq:gold-completion-terms} \\
$\hat{q}_i$ & Predicted resolution of $q_i$\\
$p^*_{q_i}$ & A relevant passage for $q_i$\\
\bottomrule 
\end{tabularx}
\label{tab:sigir2020-notation}
\end{table}

\subsection{Definitions}
\label{sec:definitions}

\paragraph{Multi-turn passage ranking} 
Let $[ q_1, \ldots, q_{i-1}, q_i ]$ be a sequence of conversational queries that share a common topic $T$.
Let $q_i$ be the current turn query and $q_{1:i-1}$ be the conversation history.
Given $q_i$ and $q_{1:i-1}$, the task is to retrieve a ranked list of passages $L$ from a passage collection $D$ that satisfy  the user's information need.\footnote{We follow the TREC CAsT setup and only take into account $q_{1:i-1}$ but not the passages retrieved for $q_{1:i-1}$.}

In the multi-turn passage ranking task, the current turn query $q_i$ is often underspecified due to phenomena such as zero anaphora, topic change, and topic return.
Thus, context from the conversation history $q_{1:i-1}$ must be taken into account to arrive at a better expression of the current turn query $q_i$.
This challenge can be addressed by query resolution.

\paragraph{Query resolution} 
Given the conversation history $q_{1:i-1}$ and the current turn query $q_i$, output a query $\hat{q}_i$ that includes both the existing information in $q_i$ and the missing context of $q_i$ that exists in the conversation history $q_{1:i-1}$.

\begin{figure}[t]
  \centering
   \includegraphics[scale=0.45]{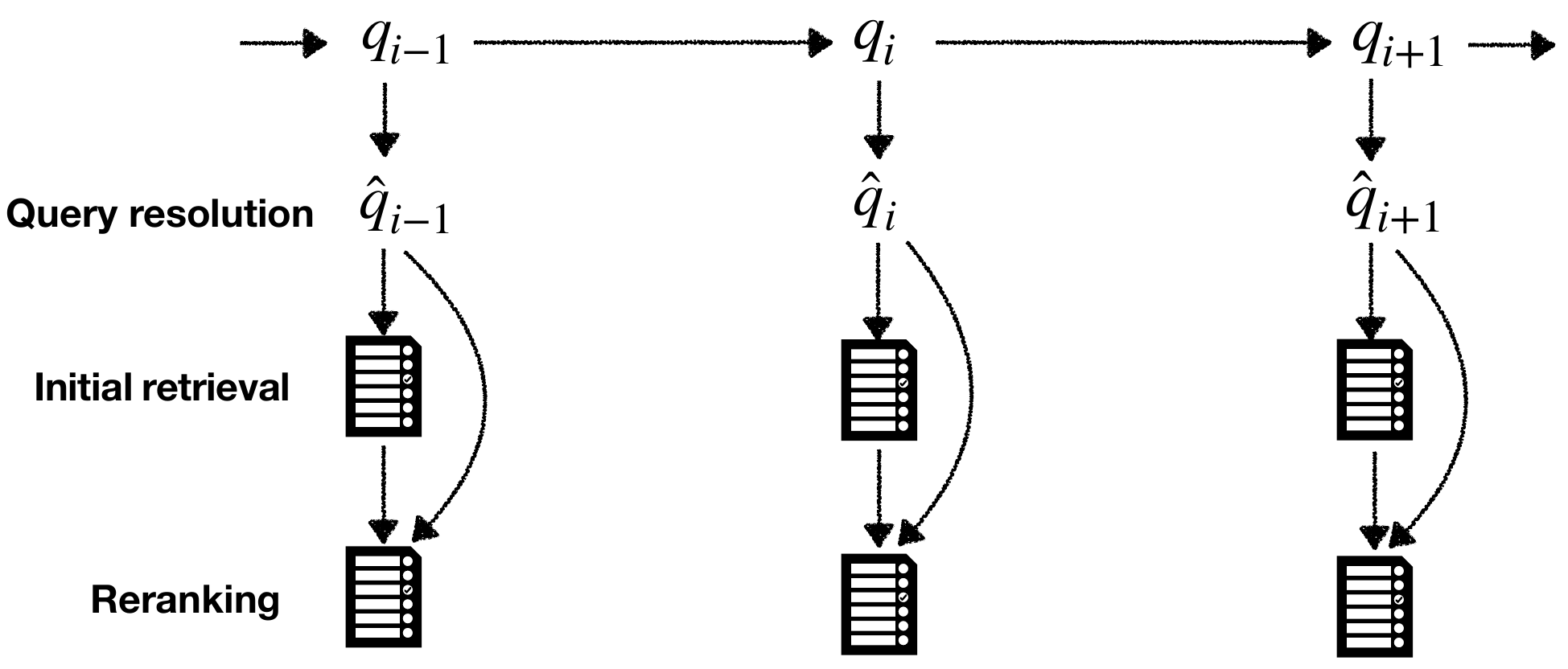}
   \caption{Illustration of our multi-turn passage retrieval pipeline for three turns.}
\label{fig:pipeline}
\end{figure}

\subsection{Multi-turn passage retrieval pipeline}

Figure~\ref{fig:pipeline} illustrates our multi-turn passage retrieval pipeline.
We use a two-step cascade ranking architecture~\cite{wang2011cascade}, which we augment with a query resolution module (Section~\ref{sec:query-resolution}). First, the unsupervised initial retrieval step outputs the initial ranked list $L_1$ (Section~\ref{sec:initial-retrieval}). Second, the re-ranking step outputs the final ranked list $L$ (Section~\ref{sec:reranking}).
Below we describe the two steps of the cascade ranking architecture.

\subsubsection{Initial retrieval step}
\label{sec:initial-retrieval}

In this step we obtain the initial ranked list $L_1$ by scoring each passage $p$ in the passage collection $D$ with respect to the resolved query $\hat{q}_i$ using a lexical matching ranking function $f_1$.
We use query likelihood (QL) with Dirichlet smoothing~\cite{zhai2004study} as $f_1$, since it 
 outperformed other ranking functions such as BM25 in preliminary experiments over the TREC CAsT dataset.

\subsubsection{Reranking step}
\label{sec:reranking}

In this step, we re-rank the list $L_1$ by scoring each passage $p \in L_1$ with a ranking function $f_2$ to obtain the final ranked list $L$.
To construct $f_2$, we use rank fusion and combine the scores obtained by $f_1$ (used in initial retrieval step) and a supervised neural ranker $f_n$.
Next, we describe the neural ranker $f_n$.

\paragraph{Supervised neural ranker}
We use BERT~\cite{devlin_bert:_2019} as the neural ranker $f_n$, as it has been shown to achieve state-of-the-art performance in ad-hoc retrieval~\cite{macavaney_cedr:_2019,qiao_understanding_2019,yang2019simple}.
Also, BERT has been shown to prefer semantic matches~\cite{qiao_understanding_2019}, and thereby can be complementary to $f_1$, which is a lexical matching method.
As is standard when using BERT for pairs of sequences, the input to the model is formatted as [ \texttt{<CLS>},  $\hat{q}_i$  \texttt{<SEP>}, $p$], where $\texttt{<CLS>}$ is a special token, $\hat{q}_i$ is the resolved current turn query, $p$ is the passage.
We add a dropout layer and a linear layer $l_a$ on top of the representation of the \texttt{<CLS>} token in the last layer, followed by a $\tanh$ function to obtain $f_n$~\cite{macavaney_cedr:_2019}.
We score each passage $p \in L_1$ using $f_n$ to obtain $L_n$  .
We fine-tune the pretrained BERT model using pairwise ranking loss on a large-scale single-turn passage ranking dataset~\cite{yang2019simple}.
During training we sample as many negative as positive passages per query.

\paragraph{Rank fusion}
We design $f_2$ such that it combines lexical matching and semantic matching~\cite{onal-neural-2018}.
We use Reciprocal Rank Fusion (RRF)~\cite{cormack_reciprocal_2009} to combine the score obtained by the lexical matching ranking function $f_1$, and the semantic matching supervised neural ranker $f_n$.
We choose RRF because of its effectiveness in combining individual rankers in ad-hoc retrieval and because of its simplicity (it has only one hyper-parameter).
We define $f_2$ as the RRF of $L_1$ and $L_n$~\cite{cormack_reciprocal_2009}:  
\begin{align}
	\label{eq:recip_rank_fusion}
	f_2(p) = \sum_{L' \in \{L_1, L_n\}} \frac{1}{k + rank(p, L')},
\end{align}
where $rank(p, L')$ is the rank of passage $p$ in a ranked list $L'$, and $k$ is a hyperparameter.\footnote{We set $k=60$ and do not tune it.} We score each passage $p$ in the initial ranked list $L_1$ with $f_2$ to obtain the final ranked list $L$. 

Since developing specialized re-rankers for the task at hand is not the focus of this work, we leave more sophisticated methods for choosing the neural ranker $f_n$ and for combining multiple rankers as future work. 
In the next section, we describe our query resolution model, QuReTeC, which is the focus of this work.

\section{Query Resolution}
\label{sec:query-resolution}

In this section we first describe how we model query resolution as term classification (Section~\ref{sec:query-resolution-as-tc}), then present our query resolution model, QuReTeC, (Section~\ref{sec:predicting-relevant-terms}), and finally describe how we generate distant supervision labels for the model (Section~\ref{sec:distant-supervision-for-query-resolution}).

\subsection{Query resolution as term classification in the conversation history}
\label{sec:query-resolution-as-tc}
Previous work has modeled query resolution as a sequence to sequence task~\cite{kumar_incomplete_2017,elgohary_can_2019}, 
where the source sequence is $q_{1:i}$ and the target sequence is $q^*_i$, where $q^*_i$ is a gold standard resolution of the current turn query $q_i$.
For instance, the gold standard resolution of turn \#4 in Table~\ref{tab:quac_paragraph1} is: ``When was Saosin's first album released?''

However, since (i) the initial retrieval step of our pipeline (Section~\ref{sec:initial-retrieval}) is a term-based model that treats queries as bag of words, and (ii) the supervised neural ranker we use in the re-ranking step (Section~\ref{sec:reranking}) is robust to queries that are not well-formed natural language texts~\cite{yang2019simple}, our query resolution model does not necessarily need to output a well-formed natural language query but rather a set of terms to expand the query.
Besides, sequence to sequence based models generally need a massive amount of data for training in order to get reasonable performance due to their generation objective~\cite{fadaee_data_2017}.
Therefore, we model query resolution as a term classification task: given the conversation history $q_{1:i-1}$ and the current turn query $q_i$, output a binary label (relevant or non-relevant) for each term in $q_{1:i-1}$.
Terms in the conversation history $q_{1:i-1}$ that are tagged as relevant are appended to the current turn query $q_i$ to form the predicted current turn query resolution $\hat{q}_i$.

We define the set of relevant resolution terms $E^*(q_i)$ as:
\begin{align}
	\label{eq:gold-completion-terms}
	 E_{q_i}^* = terms(q^*_i) \cap terms(q_{1:i-1}) \setminus terms(q_i),
\end{align}
where $q^*_i$ is a gold standard resolution of the current turn query $q_i$.
Under this formulation, the set of relevant terms $E_{q_i}^*$ represents the missing context from the conversation history $q_{1:i-1}$.
For instance, the set of gold standard resolution terms $E_{q_i}^*$ for turn \#4 in Table~\ref{tab:quac_paragraph1} is $\{  \text{Saosin}, \text{first} \}$.
Note that $E_{q_i}^*$ can be empty if $q_i = q^*_i$, i.e., the current turn query does not need to be resolved, or if $terms(q^*_i) \cap terms(q_{1:i-1})$ is empty. In our experiments $terms(q^*_i) \cap terms(q_{1:i-1}) \approx terms(q^*_i)$, and therefore almost all the gold standard resolution terms can be found in the conversation history.

\subsection{Query resolution model}
\label{sec:predicting-relevant-terms}
\begin{figure}[t]
    \centering
    \begin{subfigure}[t]{\textwidth}
        \centering
        \includegraphics[scale=0.55]{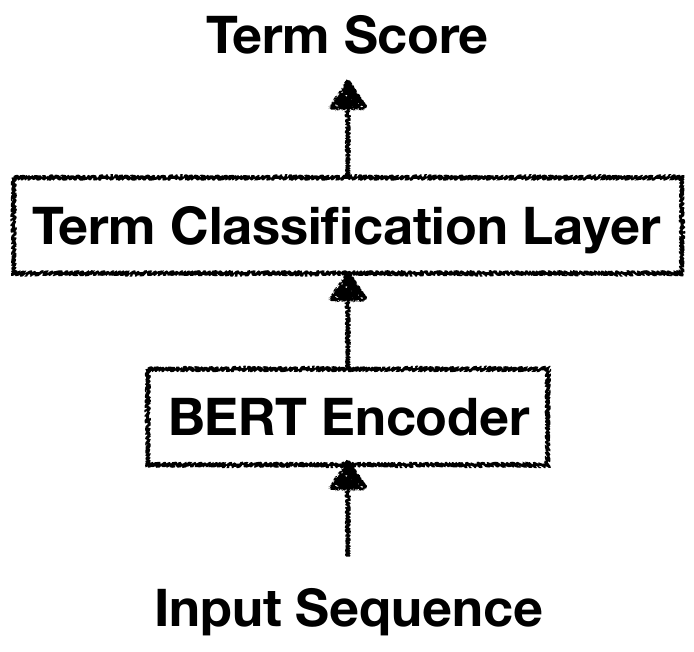}
    	\caption{QuReTeC model architecture.}
	\label{fig:model_architecture}
    \end{subfigure}\vspace{8mm}%
    \\ 
    \begin{subfigure}[t]{\textwidth}
        \centering
        \includegraphics[scale=0.35]{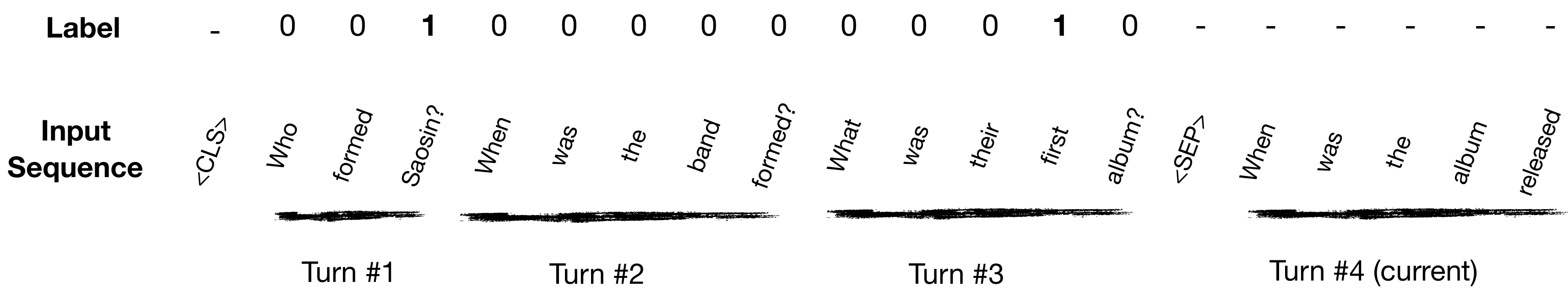}
    	\caption{Example input sequence and gold standard term labels (1: relevant, 0: non-relevant) for QuReTeC.}
	\label{fig:model_input_output}
    \end{subfigure}
    \caption{}
    \label{fig:model_architecture_and_input_output}
\end{figure}
In this section, we describe our query resolution model, QuReTeC.
Figure~\ref{fig:model_architecture} shows the model architecture of QuReTeC.
Each term in the input sequence is first encoded using bidirectional transformers~\cite{vaswani_attention_2017} -- more specifically BERT~\cite{devlin_bert:_2019}. 
Then, a term classification layer takes each encoded term as input and outputs a score for each term. 
We use BERT as the encoder since it has been successfully applied in tasks similar to ours, such as named entity recognition and coreference resolution~\cite{devlin_bert:_2019,joshi_bert_2019,lin2019open}.
Next we describe the main parts of QuReTeC in detail, i.e., input sequence, BERT encoder and Term classification layer.

\begin{enumerate}[leftmargin=*,nosep]
\item \textit{Input sequence.}
The input sequence consists of all the terms in the queries of the previous turns $q_{1:i-1}$ and the current turn $q_i$.
It is formed as: [\texttt{<CLS>}, $terms(q_1)$, \ldots, $terms(q_{i-1})$,  \texttt{<SEP>}, $terms(q_i)$], where \texttt{<CLS>} and \texttt{<SEP>} are special tokens.
We add a special separator token \texttt{<SEP>} between the previous turn $q_{i-1}$ and the current turn $q_i$ in order to inform the model where the current turn begins.
Figure~\ref{fig:model_input_output} shows an example input sequence and the gold standard term labels.

\item \textit{BERT encoder.}
BERT first represents the input terms with WordPiece embeddings using a 30K vocabulary.
After applying multiple transformer blocks, BERT outputs an encoding for each term.
We refer the interested reader to the original paper for a detailed description of BERT~\cite{devlin_bert:_2019}.

\item \textit{Term classification layer.}
The term classification layer is applied on top of the representation of the first sub-token of each term~\cite{devlin_bert:_2019}.
It consists of a dropout layer, a linear layer and a sigmoid function and outputs a scalar for each term.
We mask out the output of \texttt{<CLS>}
and the current turn terms, since we are not interested in predicting a label for those (see Equation~\eqref{eq:gold-completion-terms} for the definition and Figure~\ref{fig:model_input_output} for an example).
\end{enumerate}

In order to train QuReTeC we need a dataset containing gold standard resolution terms $E_{q_i}^*$ for each $q_i$. 
The terms in $E_{q_i}^*$ are labeled as relevant and the rest of the terms ($terms(q_{1:i-1}) \setminus E_{q_i}^*$) as non-relevant.
Assuming there exists a gold standard resolution $q^*_i$ for each $q_i$, we can derive $E_{q_i}^*$  using Equation~\eqref{eq:gold-completion-terms}.
We use standard binary cross entropy as the loss function. 

\subsection{Generating distant supervision for query resolution}
\label{sec:distant-supervision-for-query-resolution}
Recall that the gold standard resolution $q^*_i$ includes the information in $q_i$ and the missing context of $q_i$ that exists in the conversation history $q_{1:i-1}$.
As described above, we can train QuReTeC if we have a gold standard resolution $q^*_i$ for each $q_i$.
Obtaining such special-purpose gold standard resolutions is cumbersome compared to almost readily available general-purpose passage relevance labels for $q_i$.
We propose a distant supervision method to generate labels to train QuReTeC.
Specifically, we simply replace $q^*_i$ with a relevant passage $p^*_{q_i}$ in Equation~\eqref{eq:gold-completion-terms} to extract the set of relevant resolution terms $E_{q_i}^*$.
Table~\ref{tab:quac_paragraph1} illustrates this idea with an example dialogue and the relevant passage to the current turn query.
The gold standard resolution terms extracted with this distant supervision procedure for this example are $\{\text{Saosin}, \text{first}, \text{band}\}$.

Intuitively, the above procedure is noisy and can result in adding terms to $E_{q_i}^*$ that are non-relevant, or adding too few relevant terms to $E_{q_i}^*$.
Nevertheless, we experimentally show in Section~\ref{sec:distant-supervision-query-resolution} that this distant supervision signal can be used to substantially reduce the number of human-curated gold standard resolutions required for training QuReTeC.

The distant supervision method we describe here makes QuReTeC more generally applicable than other supervised methods such as the method in~\citet{elgohary_can_2019} that can only be trained with gold standard query resolutions.
This is because, apart from manual annotation, query-passage relevance labels can be potentially obtained at scale by using click logs~\cite{joachims_optimizing_2002}, or weak supervision~\cite{dehghani2017neural}.

\section{Experimental Setup}

\subsection{Research questions}
We aim to answer~\textbf{\ref{rq:query-res}}, which we break down to the following research sub-questions:
\begin{description}[nosep]%

\item[\textbf{RQ4.1}] How does the QuReTeC model perform compared to other state-of-the-art methods?

\item[\textbf{RQ4.2}] Can we use distant supervision to reduce the amount of human-curated training data required to train QuReTeC?

\item[\textbf{RQ4.3}] How does QuReTeC's performance vary depending on the turn of the conversation?

\end{description}
For all the research questions listed above we measure performance in both an intrinsic and an extrinsic sense.
\emph{Intrinsic} evaluation measures query resolution performance on term classification.  
\emph{Extrinsic} evaluation measures retrieval performance at both the initial retrieval and the reranking steps. 

\subsection{Datasets}

\begin{table*}[t]
\caption{TREC CAsT 2019 multi-turn passage retrieval dataset statistics.}
\begin{tabularx}{0.85\textwidth}{l c c r@{ $\pm$ }l r@{ $\pm$ }l r@{ $\pm$ }l r@{ $\pm$ }l }
\toprule
Split & \#Topics & \#Queries & 
 \multicolumn{2}{c}{\tabincell{l}{\#Labelled passages\\ per topic}}  & 
 \multicolumn{2}{c}{\tabincell{l}{\#Relevant passages \\ per topic}}   &  
 \multicolumn{2}{c}{\tabincell{l}{\#Labelled passages\\ per query}}  &  
 \multicolumn{2}{c}{\tabincell{l}{ \#Relevant passages\\ per query}} \\ %
\midrule
Test  & 20 & 173  &  1,467.50 & 252.86 & 406.00 & 190.18 & 169.65 & 36.69 & 46.94 & 31.53  \\
\bottomrule
\end{tabularx}
\label{tbl:passage-retrieval-data}
\end{table*}

\begin{table}[t]
\centering
\caption{Query resolution datasets statistics. In the Split column, we indicate the where the positive term labels originate from: either gold (gold standard resolutions) or distant (Section~\ref{sec:distant-supervision-for-query-resolution}).}
\setlength{\tabcolsep}{4pt}
\begin{tabularx}{28em}{ l l r r@{ $\pm$ }l r@{ $\pm$ }l}
\toprule
Dataset   & Split & \#Queries & \multicolumn{4}{c}{\#Terms (per query)}       \\ 
\cmidrule(l){4-7}
 & &  & \multicolumn{2}{c}{Total} & \multicolumn{2}{c}{Positive}   \\
\midrule
QuAC & Train (gold)   &   20,181 & 97.96 & 61.02 & 4.56 & 3.88     \\
& Train (distant)   &   31,538 & 99.78 & 62.36 & 6.90 & 5.59   \\
& Dev (gold)   &   2,196 & 95.49 & 58.79 & 4.49 & 3.90      \\
& Test (gold)   &    3,373 & 96.96 & 59.24 & 4.30 & 3.86      \\ 
\midrule
CAsT & Test  (gold)  & 153 & 39.97 & 17.97 & 1.89 & 1.62    \\
\bottomrule 
\end{tabularx}
\label{tbl:question-resolution-datasets}
\end{table}

\subsubsection{Extrinsic evaluation -- retrieval}
The TREC CAsT dataset is a multi-turn passage retrieval dataset~\cite{cast2019}.
It is the only such dataset that is publicly available.
Each topic consists of a sequence of queries.
The topics are open-domain and diverse in terms of their information need.
The topics are curated manually to reflect information seeking conversational structure patterns.
Later turn queries in a topic depend only on the previous turn queries, and not on the returned passages of the previous turns, which is a limitation of this dataset.
Nonetheless, the dataset is sufficiently challenging for comparing automatic systems, as we will show in Section~\ref{sec:query-resolution-reranking}.
Table~\ref{tbl:passage-retrieval-data} shows statistics of the dataset.
The original dataset consists of 30 training and 50 evaluation topics.
20 of 50 topics in the evaluation set were annotated  for relevance by NIST assessors on a 5-point relevance scale.
We use this set as the TREC CAsT test set.
The organizers also provided a small set of judgements for the training set, however we do not use it in our pipeline.
The passage collection is the union of two passage corpora, the MS MARCO~\cite{MSMARCO} (Bing), and the TREC CAR~\cite{treccar} (Wikipedia passages).\footnote{The Washington Post collection was also part of the original collection but it was excluded from the official TREC evaluation process and therefore we do not use it.}%

\subsubsection{Intrinsic evaluation -- query resolution}
\label{sec:datasets-intrinsic}
The original QuAC dataset~\cite{choi_quac:_2018}  contains dialogues on a single Wikipedia article section regarding people (e.g., early life of a singer).
Each dialogue contains up to 12 questions and their corresponding answer spans in the section.
It was constructed by asking two crowdworkers (a student and a teacher) to perform an interactive dialogue about a specific topic.
\citet{elgohary_can_2019} crowdsourced question resolutions for a subset of the original QuAC dataset~\cite{choi_quac:_2018}.
All the questions in the \emph{dev} and \emph{test} splits of~\cite{elgohary_can_2019} have gold standard resolutions. 
We use the \emph{dev} split for early stopping when training QuReTeC and evaluate on the \emph{test} set.
When training with gold supervision (gold standard query resolutions), we use the \emph{train} split from~\citep{elgohary_can_2019}, which is a subset of the train split of~\cite{choi_quac:_2018}; all the questions therein have gold standard resolutions.
Since QuAC is not a passage retrieval collection, in order to obtain distant supervision labels (Section~\ref{sec:distant-supervision-for-query-resolution}), we use a window of 50 characters around the answer span to extract passage-length texts, and we treat the extracted passage as the relevant passage.
When training with distant labels, we use the part of the \emph{train} split of~\cite{choi_quac:_2018} that does not have gold standard resolutions.

The TREC CAsT dataset~\cite{cast2019} also contains gold standard query resolutions for its test set. However, it is too small to train a supervised query resolution model, and we only use it as a complementary \emph{test} set. 

The two query resolution datasets described above have three main differences.
First, the conversations in QuAC are centered around a single Wikipedia article section about people whereas the conversations in CAsT are centered around an arbitrary topic. 
Second, the answers of the QuAC questions are spans in the Wiki\-pedia section whereas the CAsT queries have relevant passages that originate from different Web resources besides Wikipedia. 
Third, later turns in QuAC do depend on the answers in previous turns, while in CAsT they do not (Section~\ref{sec:definitions}).
Interestingly, in Section~\ref{sec:results-query-res-multi} we demonstrate that despite these differences, training QuReTeC on QuAC generalizes well to the CAsT dataset.

Table~\ref{tbl:question-resolution-datasets} provides statistics for the two datasets.\footnote{
Note that the first turn in each topic does not need query resolution because there is no conversation history at that point and thus the query resolution CAsT test has 20 (the number of topics) fewer queries than in Table~\ref{tbl:passage-retrieval-data}.}
First, we observe that the QuAC dataset is much larger than CAsT.
Also, QuAC has a larger number of terms on average than CAsT ($\sim$97 vs $\sim$40) and a larger negative-positive ratio ($\sim$20:1 vs $\sim$40:1).
This is because in QuAC the answers to the previous turns are included in the conversation history whereas in CAsT they are not. For this reason, we expect query resolution on QuAC to be more challenging than on CAsT.

\subsection{Evaluation metrics}
\subsubsection{Extrinsic evaluation -- retrieval}
We report NDCG@3 (the official TREC CAsT evaluation metric), Recall, MAP, and MRR at rank 1000.
We also provide performance metrics averaged per turn to show how retrieval performance varies across turns.

We report on statistical significance with a paired two-tailed t-test. We depict a significant increase for $p<0.01$ as $^\blacktriangle$.

\subsubsection{Intrinsic evaluation -- query resolution}
We report on Micro-Precision (P), Micro-Recall (R) and Micro-F1 (F1), i.e., metrics calculated per query and then averaged across all turns and topics. We ignore queries that are the first turn of the conversation when calculating the mean, since we do not predict term labels for those.

\subsection{Baselines}
We perform intrinsic and extrinsic evaluation by comparing against a number of query resolution baselines.
Next, we provide a detailed description of each baseline:
\begin{itemize}[leftmargin=*,nosep]
	\item \textbf{Original} This method uses the original form of the query. We explore different variations for constructing $\hat{q}_i$: (1) current turn only (cur), (2) current turn expanded by the previous turn (cur+prev), (3) current turn expanded by the first turn (cur+first), and (4) all turns.

	\item \textbf{RM3~\cite{abdul2004umass}} A state-of-the-art unsupervised pseudo-relevance feedback model.\footnote{Note that given the very small size of the TREC CAsT training set we do not compare to more sophisticated yet data-hungry pseudo-relevance feedback models such as~\cite{nogueira_task-oriented_2017}.} RM3 first performs retrieval and treats the top-$n$ ranked passages as relevant. Then, it estimates a query language model based on the top-$n$ results, and finally adds the top-$k$ terms to the original query. As with Original, we report on different variations for constructing the query: cur, cur+prev, cur+first and all turns. In order to apply RM3 for query resolution we append the top-$k$ terms to the original query $q_i$ to obtain $\hat{q}_i$.

	\item \textbf{NeuralCoref}\footnote{\url{https://medium.com/huggingface/state-of-the-art-neural-coreference-resolution-for-chatbots-3302365dcf30}}  A coreference resolution method designed for chatbots. It uses a rule-based system for mention detection and a feed-forward neural network that predicts coreference scores. We perform coreference resolution on the conversation history $q_{1:i-1}$ and the current turn query $q_i$. The output $\hat{q}_i$ consists of $q_i$ and the predicted terms in $q_{1:i-1}$ where terms in $q_i$ refer to.

	\item \textbf{BiLSTM-copy~\cite{elgohary_can_2019}} A neural sequence to sequence model for query resolution. It uses a BiLSTM encoder and decoder augmented with attention and copy mechanisms and also a coverage loss~\cite{see_get_2017}. It initializes the input embeddings with pretrained GloVe embeddings.\footnote{\url{https://nlp.stanford.edu/projects/glove/}} Given $q_{1:i-1}$ and  $q_i$, it outputs $\hat{q}_i$. It was optimized on the QuAC gold standard resolutions.

\end{itemize}

\subsubsection{Intrinsic evaluation -- query resolution}
In order to perform intrinsic evaluation on the aforementioned baselines, we take the query resolution they output ($\hat{q}_i$) and apply Equation \eqref{eq:gold-completion-terms} by replacing $q^*_i$ with $\hat{q}_i$ to obtain the set of predicted resolution terms.

\subsubsection{Extrinsic evaluation -- initial retrieval}
Here, apart from the aforementioned baselines, we also use the following baselines:
\begin{itemize}[leftmargin=*,nosep]
	\item \textbf{Nugget}~\cite{guan2012effective}. Extracts substrings from the current and previous turn queries to build a new query for the current turn.\footnote{We use the nugget version that does not depend on anchors text since they are not available in our setting.}
	\item \textbf{QCM}~\cite{yang2015query}. Models the edits between consecutive queries and the results list returned by the previous turn query to construct a new query for the current turn. 
	\item \textbf{Oracle} Performs initial retrieval using the gold standard resolution query. Released by the TREC CAsT organizers.
\end{itemize}

\subsubsection{Extrinsic evaluation -- reranking}
Since developing specialized rerankers for multi-turn passage retrieval is not the focus of this chapter, we evaluate the reranking step using ablation studies.
For reference, we also report on the performance of the top-ranked TREC CAsT 2019 systems~\cite{cast2019}:
\begin{itemize}[leftmargin=*,nosep]
	\item \textbf{TREC-top-auto} Uses an automatic system for query resolution and BERT-large for reranking.
	\item \textbf{TREC-top-manual} Uses the gold standard query resolution and BERT-large for reranking. 
\end{itemize}

\subsection{Implementation \& hyperparameters}
\paragraph{Multi-turn passage retrieval}
We index the TREC CAsT collections using Anserini with stopword removal and stemming.\footnote{\url{https://github.com/castorini/anserini}}
In the initial retrieval step (section~\ref{sec:initial-retrieval}) we retrieve the top 1000 passages using QL with Dirichlet smoothing (we set $\mu = 2500$).
We use the default value for the fusion parameter $k=60$~\cite{cormack_reciprocal_2009}  in Eq.~\eqref{eq:recip_rank_fusion}.
In the reranking step (section~\ref{sec:reranking}) we use a PyTorch implementation of BERT for retrieval~\cite{macavaney_cedr:_2019}.
We use the \texttt{bert-base-uncased} pretrained BERT model.
We fine-tune the BERT reranker with MSMARCO passage ranking dataset~\cite{bajaj_ms_2018}. We train on 100K randomly sampled training triples from its training set and evaluate on 100 randomly sampled queries of its development set.
We use the Adam optimizer with a learning rate of $0.001$ except for the BERT layers for which we use a learning rate of $3\mathrm{e}{-6}$. We apply dropout with a probability of $0.2$ on the output linear layer.
We apply early stopping on the development set with a patience of 2 epochs based on MRR.

\paragraph{Query resolution}
We use the \texttt{bert-large-uncased} model.
We implement QuReTeC on top of HuggingFace's PyTorch implementation of BERT.\footnote{\url{https://github.com/huggingface/transformers}} 
We use the Adam optimizer and tune the learning rate in the range $\{ 2\mathrm{e}{-5}$, $3\mathrm{e}{-5}$, $3\mathrm{e}{-6} \}$.
We use a batch size of 4 and do gradient clipping with the value of $1$.
We apply dropout on the term classification layer and the BERT layers in the range $\{0.1, 0.2, 0.3, 0.4\}$.
We optimize for F1 on the QuAC dev (gold) set.

\paragraph{Baselines}
\label{sec:exp-setup-baselines}
For RM3, we tune the following parameters: $n \in  \{3, 5, 10, 20, 30\}$ and $k \in \{5, 10\}$ and set the original query weight to the default value of $0.8$.
For Nugget, we set $k_{snippet}=10$ and tune $\theta \in \{0.95, 0.97,0.99\}$. 
For QCM, we tune $\alpha \in \{ 1.0, 2.2, 3.0\} $, $\beta \in \{1.6, 1.8, 2.0 \}$, $\epsilon \in \{0.06, 0.07, 0.08\}$ and $\delta \in \{0.2, 0.4, 0.6\}$.
For both Nugget and QCM we use \citet{van2016lexical}'s implementation.  For fair comparison, we retrieve over the whole collection rather than just reranking the top-1000 results.
The aformentioned methods are tuned on the small annotated training set of TREC CAsT.
For query resolution, we tune the greedyness parameter of NeuralCoref in the range $\{0.5, 0.75\}$.
We use the model of BiLSTM-copy released by ~\cite{elgohary_can_2019}, as it was optimized specifically for QuAC with gold standard resolutions.

\paragraph{Preprocessing}
We apply lowercase, lemmatization and stopword removal to $q^*_i$, $q_{1:i-1}$ and $q_i$ using Spacy\footnote{\url{http://spacy.io/}} before calculating term overlap in Equation~\ref{eq:gold-completion-terms}.

\section{Results \& Discussion}
\label{sec:results}
In this section we present and discuss our experimental results.
\subsection{Query resolution for multi-turn retrieval}
\label{sec:results-query-res-multi}
In this subsection we answer \textbf{RQ4.1}: we study how QuReTeC performs compared to other state-of-the-art methods when evaluated on term classification (Section~\ref{sec:query-resolution-classification}), when incorporated in the initial retrieval step (Section~\ref{sec:query-resolution-initial-retrieval}) and in the reranking step (Section~\ref{sec:query-resolution-reranking}).

\subsubsection{Intrinsic evaluation}
\label{sec:query-resolution-classification}
\begin{table}[t]
\centering
\caption{Intrinsic evaluation for query resolution on the QuAC test set. Cur, prev, first and all refer to using the current, previous, first or all turns respectively.}
\label{tab:query-resolution-quac}
\begin{tabular}{llll}
\toprule
\textbf{Method}  & \textbf{P} & \textbf{R} & \textbf{F1} \\ \midrule
Original (cur+prev) & 22.3 & 46.4 & 30.1 \\
 Original (cur+first) & 41.1 & 49.5 & 44.9 \\
 Original (all) & 12.3 & \bf 100.0 & 21.9 \\  \midrule
NeuralCoref  & 65.5 & 30.0 & 41.2 \\
BiLSTM-copy  &  67.0 & 53.2 & 59.3 \\ \midrule
 
QuReTeC & \bf 71.5 & 66.1 & \bf 68.7 \\ \bottomrule
\end{tabular}

\end{table}
\begin{table}[t]
\centering
\caption{Intrinsic evaluation for query resolution on the TREC CAsT test set. Cur, prev, first and all refer to using the current, previous, first, or all turns respectively.}
\label{tab:query-resolution-cast}
\begin{tabular}{llll}
\toprule
\textbf{Method}  & \textbf{P} & \textbf{R} & \textbf{F1} \\ \midrule
Original (cur+prev) & 32.5 & 43.9 & 37.4 \\
Original (cur+first) & 43.0 & 74.0 & 54.4 \\
Original (all) & 18.6 & \bf 100.0 & 31.4 \\ \midrule
RM3 (cur) & 35.8 & 8.3 & 13.5 \\
 RM3 (cur+prev) & 34.6 & 32.5 & 33.5 \\
 RM3 (cur+first) & 40.9 & 32.9 & 36.5 \\
 RM3 (all) & 41.5 & 38.8 & 40.1 \\  \midrule
NeuralCoref & \bf 83.0 & 28.7 & 42.7 \\
BiLSTM-copy  & 51.5 & 36.0 & 42.4 \\ \midrule
QuReTeC  & 77.2 & 79.9 & \bf 78.5 \\ 
\bottomrule
\end{tabular}
\end{table}
In this experiment we evaluate query resolution as a term classification task.\footnote{Note that the performance of Original (cur) is zero by definition when using the current turn only  (see Eq.~\ref{eq:gold-completion-terms}). Thus, we do not include it in Tables~\ref{tab:query-resolution-quac} and ~\ref{tab:query-resolution-cast}. Also, RM3 is not applicable in Table~\ref{tab:query-resolution-quac} since QuAC is not a retrieval dataset.}
Table~\ref{tab:query-resolution-quac} shows the query resolution results on the QuAC dataset.
We observe that QuReTeC outperforms all the variations of Original and the NeuralCoref by a large margin in terms of F1, precision and recall -- except for Original (all) that has perfect recall but at the cost of very poor precision.
Also, QuReTeC substantially outperforms BiLSTM-copy on all metrics. Note that BiLSTM-copy was optimized on the same training set as QuReTeC (see Section~\ref{sec:exp-setup-baselines}).
This shows that QuReTeC is more effective in finding missing contextual information from previous turns.

Table~\ref{tab:query-resolution-cast} shows the query resolution results on the CAsT dataset. 
Generally, we observe similar patterns in terms of overall performance as in Table~\ref{tab:query-resolution-quac}.
Interestingly, we observe that QuReTeC generalizes very well to the CAsT dataset (even though it was only trained on QuAC) and outperforms all the baselines in terms of F1 by a large margin.
In contrast, BiLSTM-copy fails to generalize and performs worse than Original (cur+first) in terms of F1.
NeuralCoref has higher precision but much lower recall compared to QuReTeC.
Finally, RM3 has relatively poor query resolution performance. This indicates that pseudo-relevance feedback is not suitable for the task of query resolution.

\subsubsection{Query resolution for initial retrieval}
\label{sec:query-resolution-initial-retrieval}
\begin{table}[t]
\centering
\caption{Initial retrieval performance on the TREC CAsT test set for different query resolution methods. The retrieval model is fixed (same as in Section~\ref{sec:initial-retrieval}). 
Significance is tested against RM3 (cur+first) since it has the best NDCG@3 among the baselines.}
\label{tab:initial-retrieval}
\begin{tabular}{lllll}
\toprule
\textbf{Method} & \textbf{Recall} & \textbf{MAP} & \textbf{MRR} & \textbf{NDCG@3} \\ \midrule
Original (cur) & 0.438 & 0.129 & 0.310 & 0.155 \\
 Original (cur+prev) & 0.572 & 0.181 & 0.475 & 0.235 \\
Original (cur+first) & 0.655 & 0.214 & 0.561 & 0.282 \\
 Original (all) & 0.694 & 0.190 & 0.552 & 0.256 \\ \midrule
RM3  (cur) & 0.440 & 0.140 & 0.320 & 0.158 \\
 RM3 (cur+prev) & 0.575 & 0.200 & 0.482 & 0.254 \\
 RM3 (cur+first) & 0.656 & 0.225 & 0.551 & 0.300 \\
 RM3 (all) & 0.666 & 0.195 & 0.544 & 0.266 \\ \midrule
Nugget & 0.426 & 0.101 & 0.334 & 0.145 \\
QCM  & 0.392 & 0.091 & 0.317 & 0.127 \\ \midrule
NeuralCoref & 0.565 & 0.176 & 0.423 & 0.212 \\
BiLSTM-copy & 0.552 & 0.171 & 0.403 & 0.205 \\ \midrule
QuReTeC  & \textbf{0.754}$^\blacktriangle$ & \textbf{0.272}$^\blacktriangle$ & \textbf{0.637}$^\blacktriangle$ & \textbf{0.341}$^\blacktriangle$ \\ \midrule
Oracle & 0.785 & 0.309 & 0.660 & 0.361  \\ \bottomrule

\end{tabular}

\end{table}

In this experiment, we evaluate query resolution when incorporated in the initial retrieval step (Section~\ref{sec:initial-retrieval}).
We compare QuReTeC to the baseline methods in terms of initial retrieval performance.
Table~\ref{tab:initial-retrieval} shows the results.
First, we observe that QuReTeC outperforms all the baselines by a large margin on all metrics.  
Also, interestingly, QuReTeC achieves performance close to the one achieved by the Oracle performance (gold standard resolutions).
Note that there is still plenty of room for improvement even when using Oracle, which indicates that exploring other ranking functions for initial retrieval is a promising direction for future work.
QuReTeC outperforms all Original and RM3 variations, which perform similarly.
The session search methods (Nugget and QCM) perform poorly even compared to the Original variations, which indicates that session search is different in nature than conversational search. 
BiLSTM-copy performs poorly compared to QuReTeC but also compared to the Original variations, which means that it does not generalize well to CAsT.

\subsubsection{Query resolution for reranking}
\label{sec:query-resolution-reranking}
\begin{table}[t]
\centering
\caption{Reranking performance on the TREC CAsT test set. 
All the methods in the first group use QuReTeC for query resolution. 
Significance is tested against BERT-base.}
\label{tab:reranking}
\begin{tabular}{llll}
\toprule
\textbf{Method} & \textbf{MAP} & \textbf{MRR} & \textbf{NDCG@3} \\ \midrule
Initial & 0.272 & 0.637 & 0.341 \\
BERT-base & 0.272 & 0.693 & 0.408 \\
RRF (Initial + BERT-base)& \textbf{0.355}$^\blacktriangle$ & \bf 0.787$^\blacktriangle$ & \textbf{0.476}$^\blacktriangle$ \\ 
Oracle  & 0.754 & 0.956  & 0.926 \\ 
\midrule
TREC-top-auto & 0.267 & 0.715 & 0.436 \\
TREC-top-manual & 0.405 & 0.879 & 0.589 \\ \bottomrule 
\end{tabular}
\end{table}

In this experiment, we study the effect of QuReTeC when incorporated in the reranking step (Section~\ref{sec:reranking}).
We keep the initial ranker fixed for all QuReTeC models.
Table~\ref{tab:reranking} shows the results.
First, we see that BERT-base improves over the initial retrieval model that uses QuReTeC for query resolution on the top positions (second line).
Second, when we fuse the ranked listed retrieved by BERT-base and the ranked list retrieval by the initial retrieval ranker using RRF, we significantly outperform BERT-base on all metrics (third line).
This shows that the two rankers can be effectively combined with RRF, which is a very simple fusion method that only has one parameter which we do not tune.
We also see that our best model outperforms TREC-top-auto on all metrics.
Furthermore, by comparing RRF (line 3) to Oracle (line 4) we see that there is still plenty of room for improvement for reranking, which is a clear direction for future work.
This also shows that the TREC CAsT dataset is sufficiently challenging for comparing automatic systems.
Note that TREC-top-manual uses the gold standard query resolutions and is thereby not directly comparable with the rest of the methods.

\subsection{Distant supervision for query resolution}
\label{sec:distant-supervision-query-resolution}
In this section we answer \textbf{RQ4.2}: Can we use distant supervision to reduce the amount of human-curated query resolution data required to train QuReTeC?
\begin{figure}[th]
    \centering
  \centering
  \includegraphics[width=0.8\linewidth]{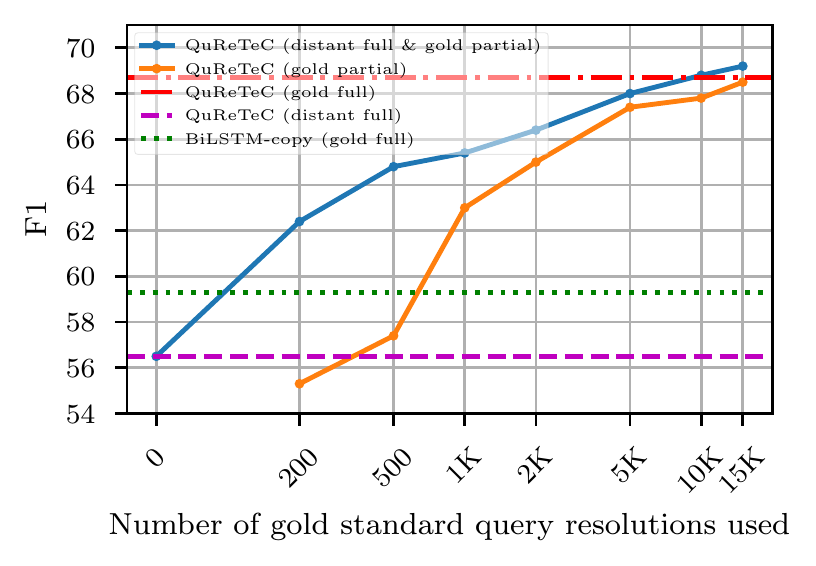}  
\caption{Query resolution performance (intrinsic) on the QuAC test set on different supervision settings. 
Gold refers to the QuAC train (gold) dataset and distant refers to the QuAC train (distant) dataset. 
Full refers to the whole and partial refers to a part of the corresponding dataset (gold or distant).  
The x-axis is plotted in log-scale.}
\label{fig:dist_sup_percentage_f1}

\end{figure}
Figure~\ref{fig:dist_sup_percentage_f1} shows the query resolution performance when training QuReTeC under different settings (see figure caption for a more detailed description).
For QuReTeC (distant full \& gold partial) we first pretrain QuReTeC on distant and then resume training with different fractions of gold. 
First, we see that QuReTeC performs competitively with BiLSTM-copy even when it does not use any gold resolutions (distant full).\footnote{Also, when trained with distant full, QuReTeC performs better than an artificial method that uses the label of the distant supervision signal as the prediction in terms of F1 (56.5 vs 41.6). This is in line with previous work that successfully uses noisy supervision signals for retrieval tasks~\cite{dehghani2017neural,voskarides-weakly-supervised-2018}.}
More importantly, when only trained on distant, QuReTeC performs remarkably well in the low data regime.
In fact, it outperforms BiLSTM-copy (trained on gold) even when using a surprisingly low number of gold standard query resolutions (200, which is $\sim$1\% of gold).
Last, we see that as we add more labelled data, the effect of distant supervision becomes smaller. This is expected and is also the case for the model trained on QuAC train (gold).\footnote{In fact (not shown in Figure~\ref{fig:dist_sup_percentage_f1}), performance stabilizes after 15K query resolutions ($\sim$75\% of gold full).}

In order to test whether our distant supervision method can be applied on different encoders, we performed an additional experiment where we replaced BERT with a simple BiLSTM as the encoder in QuReTeC.
Similarly to the previous experiment, we observed a substantial increase in F1 when retraining with 2K gold standard resolutions (+12 F1) over when only using gold resolutions.

In conclusion, our distant supervision method can be used to substantially decrease the amount of human-curated training data required to train QuReTeC.
This is especially important in low resource scenarios (e.g. new domains or languages), where large-scale human-curated training data might not be readily available.

\subsection{Analysis}
In this section we perform analysis on QuReTeC when trained with gold standard supervision.
\subsubsection{Query resolution performance per turn}
\begin{figure}[t]
  \centering
\begin{subfigure}{0.45\textwidth}
  \centering
  \includegraphics[width=\linewidth]{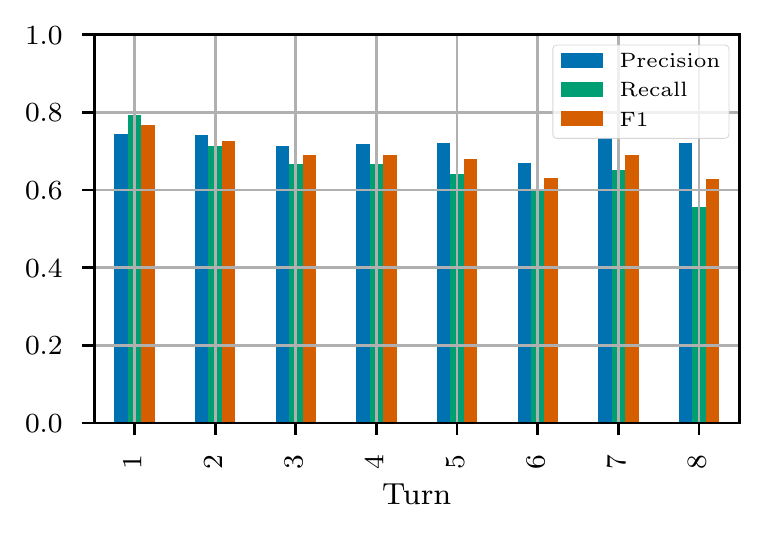}  
  \caption{QuAC \emph{test}}
  \label{fig:tc_per_turn_quac}
\end{subfigure}
\begin{subfigure}{0.45\textwidth}
  \centering
  \includegraphics[width=\linewidth]{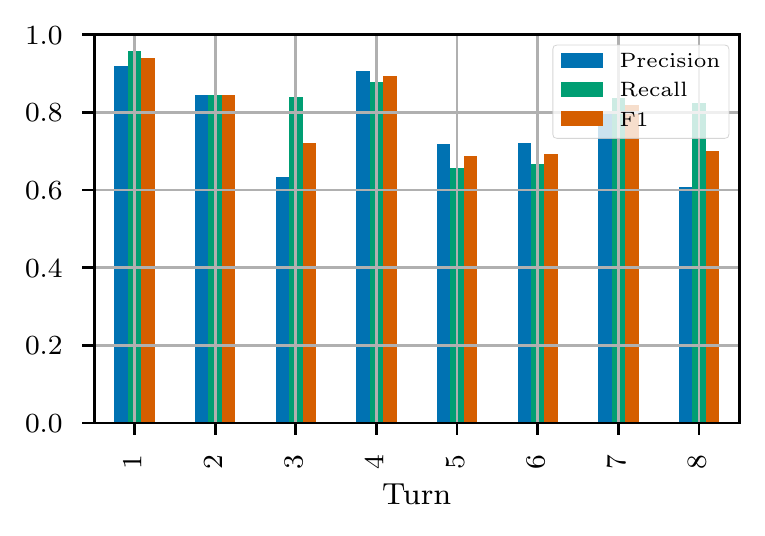}  
  \caption{CAsT \emph{test}}
  \label{fig:tc_per_turn_cast}
\end{subfigure}
\caption{Intrinsic query resolution evaluation (term classification performance) for QuReTeC, averaged per turn.}
\label{fig:tc_per_turn}
\end{figure}
Here we answer \textbf{RQ4.3} by analyzing the robustness of QuReTeC at later conversation turns.
We expect query resolution to become more challenging as the conversation history becomes larger (later in the conversation).

\textbf{Intrinsic} Figure~\ref{fig:tc_per_turn} shows the QuReTeC performance averaged per turn on the QuAC and CAsT datasets.
Even though performance decreases towards later turns as expected, we observe that it decreases very gradually, and thus we can conclude that QuReTeC is relatively robust across turns.

\textbf{Extrinsic  -- initial retrieval} 
\begin{figure}[t]
  \centering
  \begin{subfigure}{.45\textwidth}
  \centering
  \includegraphics[width=\linewidth]{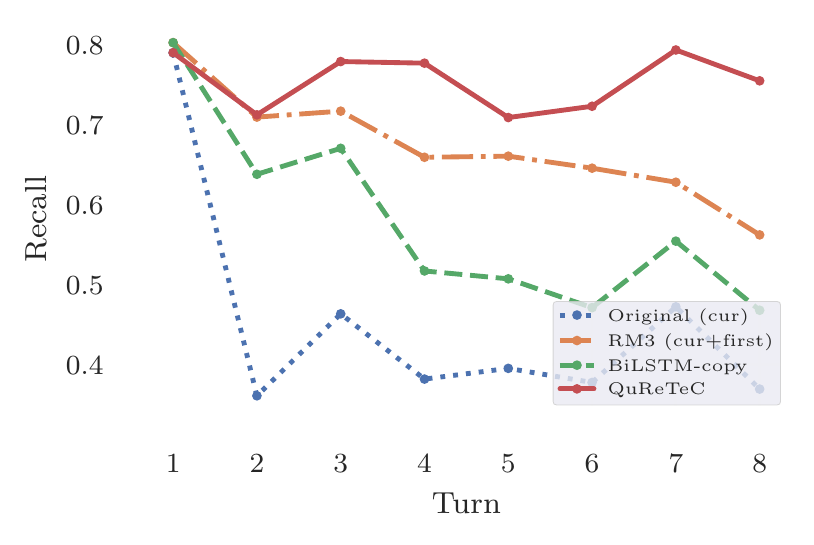}  
  \caption{Recall}
  \label{fig:initial_retrieval_per_turn_recall}
\end{subfigure}
    \begin{subfigure}{.45\textwidth}
  \centering
  \includegraphics[width=\linewidth]{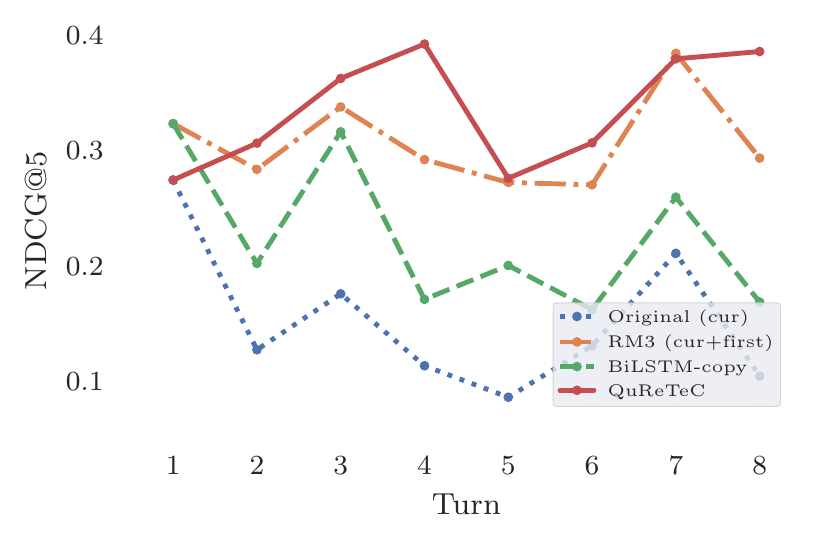}  
  \caption{NDCG@5}
  \label{fig:initial_retrieval_per_turn_recall_ndcg_5}
\end{subfigure}
\caption{Initial retrieval performance per turn for different query resolution methods CAsT \emph{test}
}
\label{fig:initial_retrieval_per_turn}
\end{figure}
Figure~\ref{fig:initial_retrieval_per_turn} shows the performance of different query resolution methods when incorporated in the initial retrieval step.
We observe that QuReTeC is robust to later turns in the conversation, whereas the performance of all the baseline models decreases faster (especially in terms of recall).
For reranking, we observe similar patterns as with initial retrieval; we do not include those results for brevity.

\subsubsection{Qualitative analysis}
Here we perform qualitative analysis by sampling specific instances from the data.
\begin{table}[t]
\centering
\caption{Qualitative analysis for QuReTeC on query resolution (intrinsic). 
We denote true positive terms with underline and false negative terms in italics. The examples are sampled from the QuAC dev set.}
\label{tab:qualitative-intrinsic}
\begin{tabularx}{\linewidth}{p{0.97\linewidth}}
\toprule
\textbf{Success case} -- no mistakes \\ 
\midrule
Q1: What was \underline{Bipasha} \underline{Basu}'s debut?\\
A1: In 2001, Basu finally made her debut opposite Akshay Kumar in Vijay Galani 's \underline{Ajnabee}.\\
Q2: Did this help her become well known?\\
A2: It was a moderate box-office success and attracted unfavorable reviews from critics.\\
Q3 (current): Why did she receive unfavorable reviews? \\
\midrule
\textbf{Failure case} -- misses two relevant terms: \emph{dehusking}, \emph{machine}\\
\midrule
Q1: How old was \underline{Alexander} \underline{Graham} \underline{Bell} when he made his first invention?\\
A1: The age of 12.\\
Q2: What did he invent?\\
A2: Bell built a homemade device that combined rotating paddles with sets of nail brushes.\\
Q3: What was it for?\\
A3: A simple \emph{dehusking} \emph{machine}.\\
Q4 (current): By inventing this, what happened to allow him to continue inventing things? \\
\bottomrule
\end{tabularx}
\end{table}
\begin{table}[t]
\centering
\caption{Qualitative analysis for initial retrieval (extrinsic) when using QuReTeC or RM3 (cur+first) for query resolution. The example is sampled from the TREC CAsT dataset.}
\label{tab:qualitative-extrinsic}
\begin{tabularx}{\linewidth}{p{0.97\linewidth}}
\toprule
Q1: What is a real-time database?\\
Q2: How does it differ from traditional ones?\\
Q3: What are the advantages of real-time processing?\\
Q4: What are examples of important ones?\\
Q5: What are important applications?\\
Q6: What are important cloud options?\\
Q7: Tell me about the Firebase DB? \\
Q8 (current): How is it used in mobile apps?\\
 \hline
\textbf{Predicted terms -- QuReTeC}: \{``database'', ``firebase'', ``db'' \} \\
\textbf{Top-ranked passage -- QuReTeC} \\
Firebase is a mobile and web application platform \ldots Firebase's initial product was a realtime database, \ldots Over time, it has expanded its product line to become a full suite for app development \ldots \\ \midrule
\textbf{Predicted terms -- RM3 (cur+first)}: \{``real'', ``time'', ``database''\} \\ 
\textbf{Top-ranked passage -- RM3 (cur+first)} \\
There are two options in Jedox to access the central OLAP database and software functionality on mobile devices: 
Users can access reports through the touch-optimized Jedox Web Server \ldots
 on their smart phones and tablets.\\
\bottomrule
\end{tabularx}

\end{table}
\textbf{Intrinsic}
Table~\ref{tab:qualitative-intrinsic} shows one success and one failure case for QuReTeC from the QuAC dev set.
In the success case (top) we observe that QuReTeC succeeds in resolving ``she'' $\rightarrow$ \{``Bipasha'', ``Basu''\} and  ``reviews'' $\rightarrow$ ``Anjabee''. Note that ``Anjabee'' is a movie in which Basu acted but is not mentioned explicitly in the current turn.
In the failure case (bottom) we observe that QuReTeC succeeds in resolving ``him'' $\rightarrow$ \{``Alexander'', ``Graham'' ``Bell''\} but misses the connection between ``this'' and ``dehusking machine''.

\textbf{Extrinsic -- initial retrieval}
Table~\ref{tab:qualitative-extrinsic} shows an example from the CAsT test set where QuReTeC succeeds and RM3 (cur+first), the best performing baseline for initial retrieval, fails.  
First, note that a topic change happens at Q7 (the topic changes from general real-time databases to Firebase DB). 
We observe that QuReTeC predicts the correct terms, and a relevant passage is retrieved at the top position.
In contrast, RM3 (cur+first) fails to detect this topic change and therefore an irrelevant passage is retrieved at the top position that is about real-time databases on mobile apps but not about Firebase DB.

\section{Conclusion}

In this chapter, we studied the task of query resolution for conversational search.
We proposed to model query resolution as a binary term classification task: whether to add terms from the conversation history to the current turn query.
We proposed QuReTeC, a neural query resolution model based on bidirectional transformers.
We proposed a distant supervision method to gather training data for QuReTeC.
We found that QuReTeC significantly outperforms multiple baselines of different nature and is robust across conversation turns.
Also, we found that our distant supervision method can substantially reduce the required amount of gold standard query resolutions required for training QuReTeC, using only query-passage relevance labels. 
This result is especially important in low resource scenarios, where gold standard query resolutions might not be readily available.

As for future work, we aim to develop specialized rankers for both the initial retrieval and the reranking steps that incorporate QuReTeC in a more sophisticated way.
Also, we want to study how to effectively combine QuReTeC with text generation query resolution methods as well as pseudo-relevance feedback methods.
Finally, we aim to explore weak supervision signals for training QuReTeC~\cite{dehghani2017neural}.

In this chapter, we focused on how to improve interactive knowledge gathering and studied multi-turn passage retrieval as an instance of conversational search. 
In Chapter~\ref{chapter:chapter6}, we focus on a different research theme, namely supporting knowledge exploration for narrative creation.

\part{Supporting Knowledge Exploration for Narrative Creation}

\graphicspath{{06-chapter6/figures/}}

\chapter{News Article Retrieval in Context for~Event-centric~Narrative~Creation}
\label{chapter:chapter6}

\footnote[]{This chapter was published as~\citep{voskarides-2020-event}.}

In the third and final part of this thesis we study the research theme of supporting knowledge exploration for narrative creation. 
In this chapter, we address \textbf{\ref{rq:narrativecreation}}: \acl{rq:narrativecreation}

\section{Introduction}
Professional writers such as journalists generate narratives centered around specific events or topics.
As shown in recent studies, such writers envision automatic systems that suggest material relevant to the narrative they are creating~\cite{huurnink_search_2010,diakopoulos_automating_2019}. %
This material may provide background information or connections that can help writers generate new angles on the narrative and thus help engage the reader~\cite{kirkpatrick_putting_2015}. %

Previous work has focused on developing automatic systems to support writers explore content relevant to the narrative they are writing about.
Such systems use content originating from various sources such as as social media~\cite{diakopoulos_finding_2012,zubiaga_curating_2013,cucchiarelli_topic_2019}, political speeches and conference transcripts~\cite{maclaughlin_context-based_2020}, %
or news articles~\cite{maiden_inject_2020}. %

Writers in the news domain often develop narratives around a single main event, and refer to other, related events that can serve different functions in relation to the narrative~\cite{dijk_news_1988}.
These include explaining the cause or the context of the main event or providing supporting information, among others~\cite{choubey_discourse_2020}.
Recent work has focused on automatically profiling news article content (i.e., paragraphs or sentences) in relation to their discourse function~\cite{yarlott_identifying_2018,choubey_discourse_2020}.

\begin{table}[t]
\centering
\caption{Example incomplete narrative $q$ (consisting of a main event $e$ and context $c$), and a news article $d^*$ that is relevant to $q$ because it is relevant to both the main event $e$ and to the narrative context $c$ in the sense explained in the main text.}
\label{tab:examples-wapo}
\begin{tabularx}{\linewidth}{@{}p{1\textwidth}}
\toprule
\textbf{Incomplete narrative $q$} \\
\textbf{-- Main event} ($e$)\\
(\#1) Malta’s armed forces storm merchant ship taken over by rescued migrants.
\\
(\#2) Maltese armed forces on Thursday stormed a merchant vessel taken over by rescued migrants who were allegedly demanding to be transported to Europe, rather than back to Libya.
\\
\textbf{-- Narrative context} ($c$)\\
(\#3) In earlier years of Europe's migration crisis---when flows from the Middle East and North Africa were much higher---the Mediterranean was patrolled by Italian and European vessels, as well as by humanitarian groups, 
which would rescue migrants from flimsy dinghies and transport them to safety, typically to Italy.
\\
\midrule
\textbf{Relevant news article} ($d^*$)\\
(\#4) Italy's new government sends immigration message by rejecting rescue ship	
\\
(\#5) Italy's new populist government has delivered a jolt to European migration politics, prompting a diplomatic standoff with its refusal to accept a rescue vessel overloaded with migrants.
\\
\bottomrule
\end{tabularx}
\end{table}
In this chapter, instead of profiling existing narratives, we consider a scenario where a writer has generated an incomplete narrative about a specific event up to a certain point, and aims to explore other news articles that discuss relevant events to include in their narrative.
A news article that discusses a different event from the past is relevant to the writer's incomplete narrative if it relates to the narrative's main event and to the \emph{narrative's context}.
Relevance to the narrative's main event is topical in nature but, importantly, relevance to the narrative's context is not only topical: to be relevant to the narrative's context, a news article should enable the continuation of the narrative by expanding the narrative discourse~\cite{caswell_automated_2018}.
Table~\ref{tab:examples-wapo} shows an example of an incomplete narrative and a news article relevant to it.
The relevant article discusses an event about a subject mentioned in the narrative context (\emph{Italy}).
Here, the relevant news article is relevant to the topic of the incomplete narrative (\emph{migration crisis}) and also relevant to the narrative context in the sense that it is used by the writer to expand the narrative by making a comparison: the previous government of Italy was more welcoming to immigrants than the current.
To avoid confusion, in the remainder of this chapter \emph{relevance} without further restriction or scope is taken to mean both \emph{topical relevance} and \emph{relevance to the narrative context}.

We model the problem of finding a relevant news article given an incomplete narrative as a retrieval task where the query is an incomplete narrative and the unit of retrieval is a news article.
We automatically generate retrieval datasets for this task by harvesting links from existing narratives manually created by journalists.
Using the generated datasets, we analyze the characteristics of this task and study the performance of different rankers on this task.
We find that state-of-the-art lexical and semantic rankers are not sufficient for this task and that combining those with a ranker that ranks articles by their reverse chronological order outperforms those rankers alone.

Our main contributions are: 
\begin{enumerate*}[label=(\roman*)]
\item we propose the task of news article retrieval in context for event-centric narrative creation; 
\item we propose an automatic retrieval dataset construction procedure for this task; and
\item we empirically evaluate  the performance of different rankers on this task and perform an in-depth analysis of the results to better understand the characteristics of this task.
\end{enumerate*}

\section{Problem Statement}

\subsection{Preliminaries}

A \emph{news article} $d$ published at time $t$ consists of its headline $H$---which introduces the topic of the article~\cite{dijk_news_1988}---and a sequence of paragraphs $p_1, p_2, \ldots$. Each paragraph $p_i$ consists of a sequence of sentences $a_{i,1}, a_{i,2}, \ldots$. 

The \emph{lead paragraph} $L$ of a news article $d$ is its first paragraph $p_1$, which summarizes the main topic of the article~\cite{dijk_news_1988}.

An \emph{event} $e$ is characterized by interactions between entities such as countries, organizations, or individuals---that deviate from typical interaction patterns~\cite{chaney_detecting_2016}.
We assume that each news article $d$ is associated with a single main event $e$.

A \emph{link sentence} $a_{i,j}$ in article $d$ is a sentence that contains a hyperlink to a news article $d^*$.

A \emph{context} is a sequence of sentences already generated by the writer that introduces a new idea or subtopic in a narrative.%

A \emph{query} $q=(e,c,t)$ is an incomplete narrative at time $t$ that consists of an event $e$ and a context $c$.

\subsection{Task definition}
\label{sec:task-def}

The task of \emph{news article retrieval in context for event-centric narrative creation} is defined as follows.
Given a query $q=(e,c,t)$ and a collection of news articles $D$ published before time $t$, we need to rank articles in $D$ w.r.t.\ their relevance to $q=(e, c, t)$. 
Here, ``relevance to $e$'' is to be interpreted as topical, whereas ``relevance to $c$'' is not only topical, but it should also enable the continuation of the narrative by expanding the narrative discourse~\cite{caswell_automated_2018}. 
``Relevance to $q$'' is taken to mean the same as ``relevance to $e$ and to $c$''.
An article relevant to $q$ can thus be used by the writer to create the next sentence in the yet incomplete narrative.
Table~\ref{tab:examples-wapo} shows an example query $q$ and a relevant news article $d^*$ published at time $t^* < t$.

\section{Retrieval Dataset Construction}

\subsection{Dataset construction procedure}
\label{sec:dataset-construction}
In order to construct a retrieval dataset for our news article retrieval task, we rely on existing news articles to simulate incomplete narratives as well as relevant documents.
We capitalize on the fact that (complete) news articles often contain links to other news articles manually inserted by journalists in the form of hyperlinks.

The automatic retrieval dataset construction procedure that we propose takes as input a news article $d$ and outputs a set of ($q, d^*$) pairs, where $q=(e, c, t)$ is a query and $d^*$ is the (unique) relevant news article to $q$. 
We assume that the event $e$ associated with $d$ is described by the headline $H$ and the lead paragraph $L$ of $d$~\cite{choubey_discourse_2020}.

In order to construct the context $c$ of $q$, we iteratively look for link sentences $a_{i,j}$ in $d$ that contain a hyperlink to another news article $d^*$.
We enforce $i > 1$ so that the paragraph where the link sentence appears is after the lead paragraph.
We also enforce $j > 1$ motivated by the fact that links after the first sentence of a paragraph are tightly related to the main idea of the paragraph, therefore the sentences preceding the link sentence can be considered as context~\cite{hacker2011writer}. %
If such a link sentence $a_{i,j}$ exists, we consider the sentences $a_{i,1}, \ldots, a_{i,j-1}$ as the narrative context $c$ and the article $d^*$ as the relevant article for  $q$.

\paragraph{Example}
To illustrate the procedure described above, consider the example in Table~\ref{tab:examples-wapo}.
Sentences \#1 and \#2 in Table~\ref{tab:examples-wapo} are the headline and lead paragraph of a news article $d$ respectively.
Sentence \#3 in Table~\ref{tab:examples-wapo} is the first sentence $a_{i,j-1}$ of a paragraph $p_i$, $i > 1$ in $d$, which constitutes the narrative context $c$. 
The link sentence $a_{i,j}$ (not shown in the table) is:

\begin{quote}
 But over the past year, \emph{Italy has closed its ports} to migrants rescued by humanitarian boats.
 \end{quote}
where the part in italics is (the anchor text of) a hyperlink to the relevant news article $d^*$ shown in Table~\ref{tab:examples-wapo}, where sentences \#4 and \#5 are the headline and lead of $d^*$, respectively.

\subsection{Retrieval dataset description}
\begin{table}[t]
\centering
\caption{Statistics of the retrieval datasets derived from the WaPo and Guardian newspaper collections. Because of the way we construct the retrieval datasets (see Section~\ref{sec:dataset-construction}), each query has a single relevant news article.}
\label{tab:datasets-stats}
\begin{tabular}{llrrrrr}
\toprule
\textbf{Dataset} & \textbf{Split} & \textbf{\# $q$} & \textbf{\# uniq. $d$} & \textbf{\# uniq. $d^*$} & \multicolumn{2}{c}{\textbf{Link sentence} ($a_{i,j}$)} \\
& & & & & $i$ mean/  & $j$ mean/ \\ 
& & & & & median & median \\
\midrule
WaPo & Train & 32,963 & 23,537 & 24,279 & 7.9/7 & 2.5/2 \\
 & Dev. & 1,831 & 1,286 & 1,585 & 8.4/8 & 2.4/2 \\
 & Test & 1,832 & 1,216 & 1,555 & 9.1/9 & 2.4/2 \\ \hline
Guardian & Train & 31,329 & 21,730 & 22,935 & 7.3/6 & 2.4/2\\
 & Dev. & 1,740 & 1,128 & 1,526 & 8.0/7 & 2.4/2 \\
 & Test & 1,742 & 1,064 & 1,532 & 7.3/7 & 2.5/2 \\ 
 \bottomrule
\end{tabular}
\end{table}
We consider two collections of news articles written in English and published by major newspapers.
The first is a set of news articles published by The Washington Post (WaPo), released by the TREC News Track~\cite{soboroff2018trec}.
It contains 671,947 news articles and blog posts published from January 2012 to December 2019.
The second is a set of news articles published by The Guardian, between November 2013 to June 2017, which we crawl ourselves. 
We also crawl the out-links of each article in this set; the final set contains 572,639 news articles published between January 2000 and March 2018. 

The articles in both newspapers cover multiple genres and domains.
In order to ensure that the news articles describe real-world events, we filter out blog posts and opinion news articles, and only keep articles in the following domains: \emph{news}, \emph{world}, \emph{business}, \emph{environment}, \emph{technology}, \emph{society}, \emph{science}, \emph{culture}, \emph{education}, \emph{global}, \emph{healthcare}, \emph{media}, \emph{money}, \emph{teacher}, \emph{local}, \emph{national}.
After filtering for genre and domain, we are left with 386,196 articles in WaPo and 185,034 in The Guardian.

\begin{figure}[t]
    \centering
    \begin{subfigure}[t]{0.45\linewidth}
        \centering
        \includegraphics[scale=0.4]{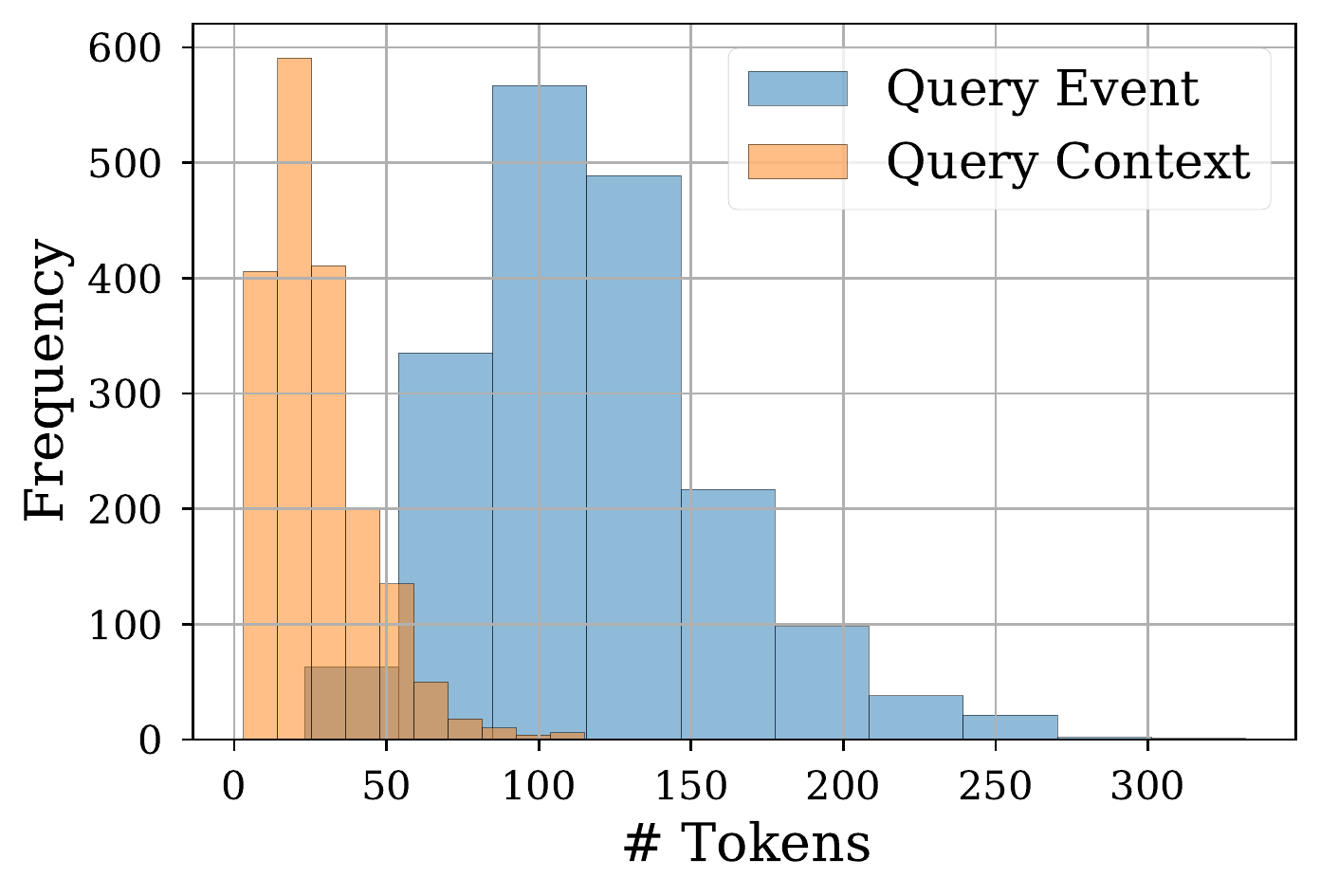}
        \caption{WaPo dev.}
    \end{subfigure}
    ~
    \begin{subfigure}[t]{0.45\linewidth}
        \centering
        \includegraphics[scale=0.4]{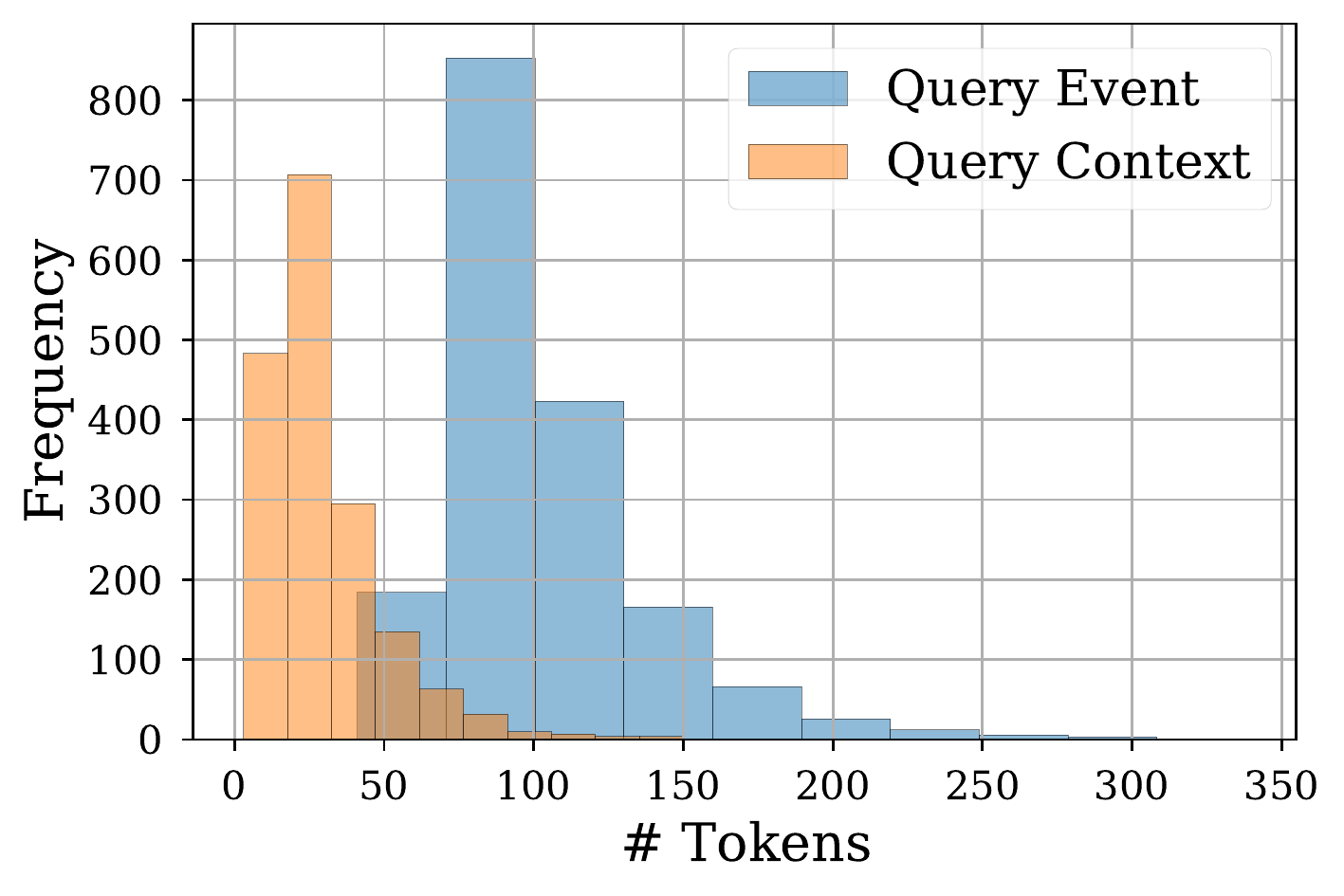}
        \caption{Guardian dev.}
    \end{subfigure}
    \caption{Histogram of the number of tokens in the query event $e$ and the query context $e$.}
    \label{fig:dev-query-len}
\end{figure}
We then apply the dataset construction procedure described in Section~\ref{sec:dataset-construction} to construct a retrieval dataset for both collections.
We split the retrieval datasets chronologically and keep the first 90\% for training, the next 5\% for development, and the last 5\% for testing.
Table~\ref{tab:datasets-stats} shows basic statistics for both retrieval datasets.
Figure~\ref{fig:dev-query-len} shows a histogram of the number of tokens in the query event $e$ and the query context $c$.
We observe that the query context is shorter than the query event in both datasets. 
Also, the query event is longer in WaPo than in Guardian because the way those newspapers perform paragraph splitting is different.

\begin{figure}[t]
    \centering
    \begin{subfigure}[t]{0.45\linewidth}
        \centering
        \includegraphics[scale=0.38]{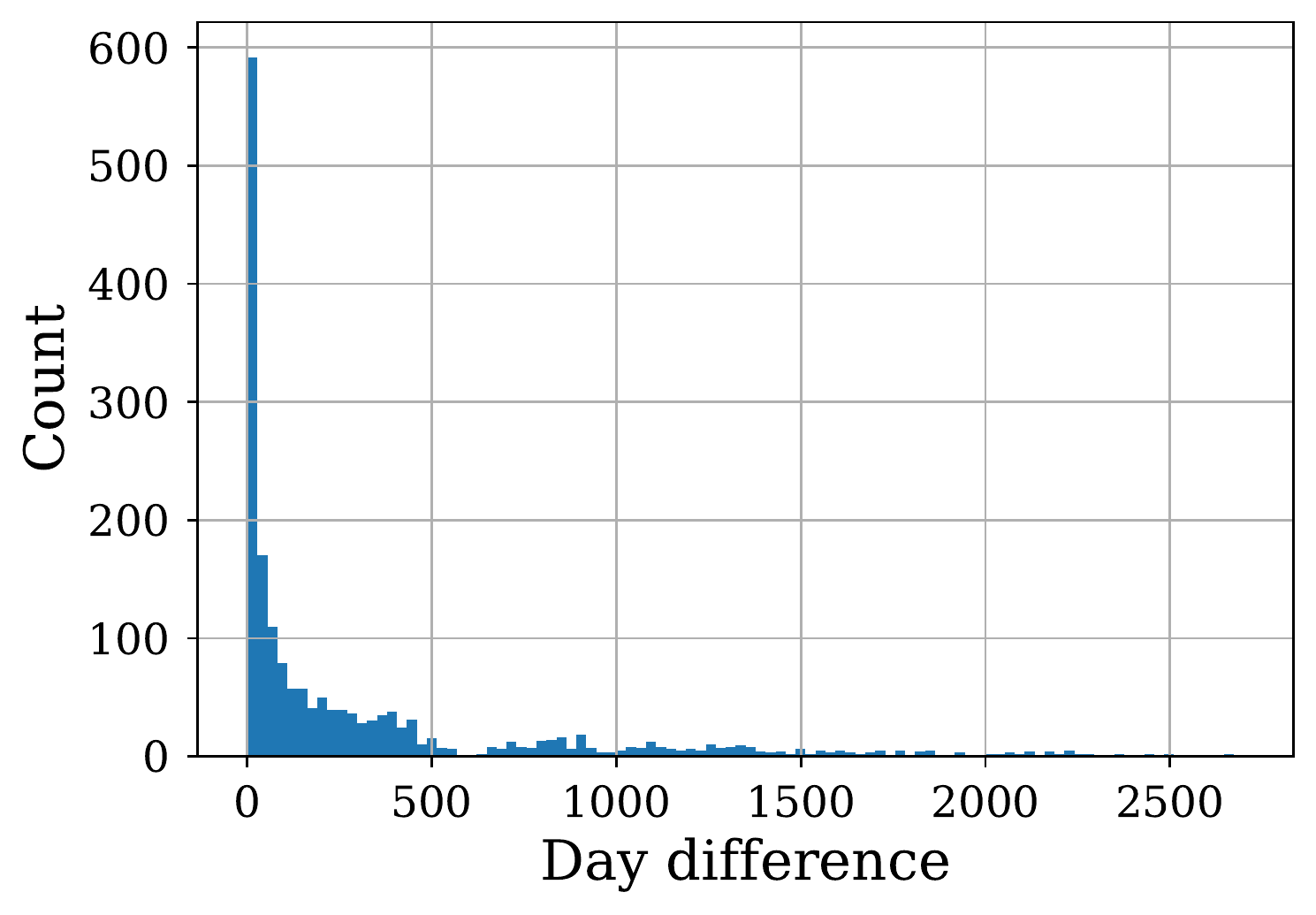}
        \caption{WaPo dev.}
    \end{subfigure}
    ~
    \begin{subfigure}[t]{0.45\linewidth}
        \centering
        \includegraphics[scale=0.38]{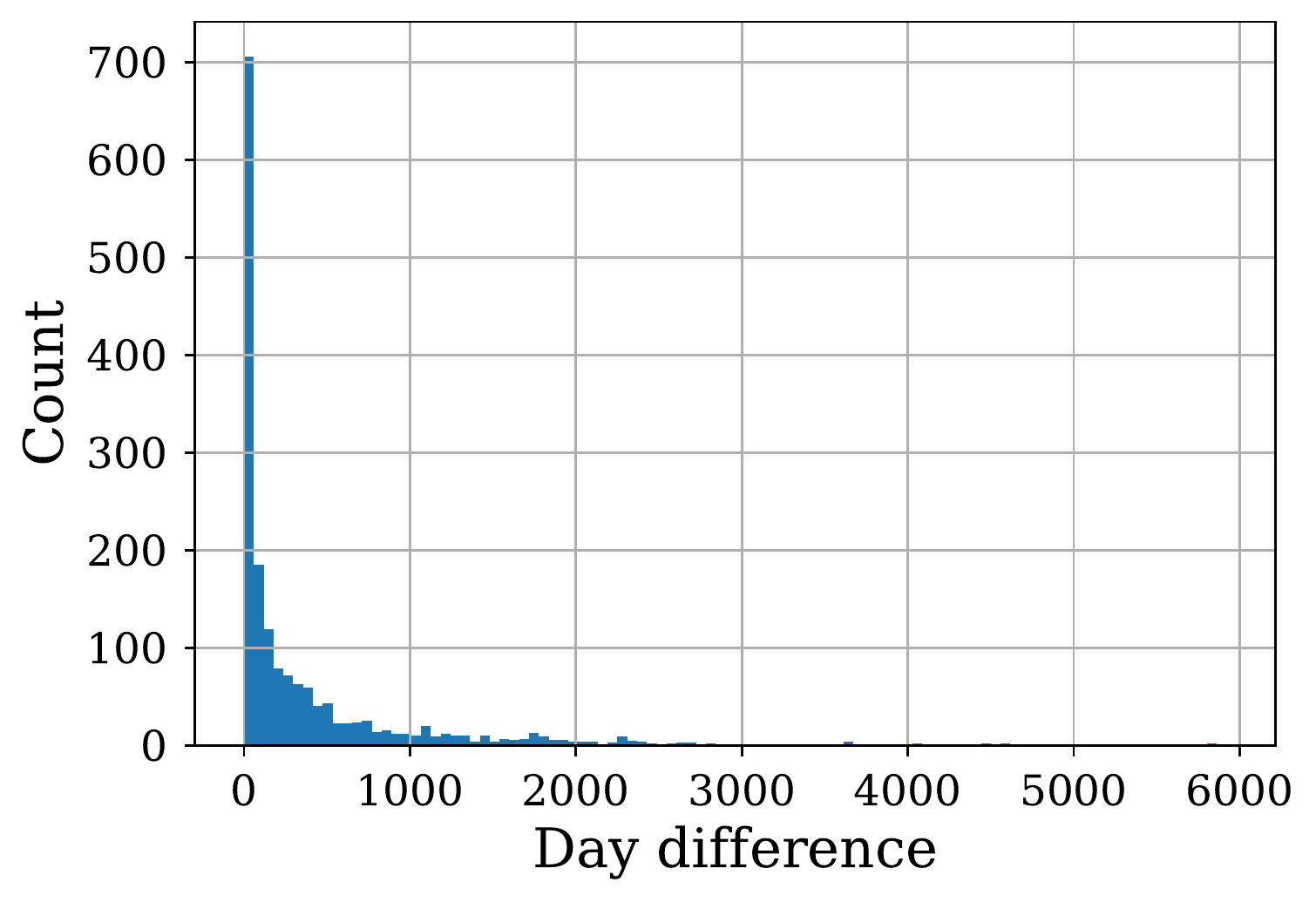}
        \caption{Guardian dev.}
    \end{subfigure}
    \caption{Histogram of day difference between the query and its relevant news article.}
    \label{fig:dev-day-diff}
\end{figure}
Figure~\ref{fig:dev-day-diff} shows a histogram of the difference in number of days between the publication date of the query and the publication date of the relevant news article on the development sets of the two datasets.
The retrieval datasets have a strong recency bias, which is in line with studies on content generation in the news domain~\cite{niculae_quotus_2015}. 
Typical examples of recent, relevant articles are those discussing a previous development of a query event or of an event mentioned in the narrative context.
And a typical example of a less recent, relevant article can be found when discussing an event that is similar to one mentioned in the query (e.g., an earthquake) but involving different entities (e.g., a person, location, or organization).

\subsection{Retrieval dataset quality}
The dataset construction procedure we described in Section~\ref{sec:dataset-construction} assumes that an article $d^*$ is relevant to $q$ since the writer has chosen to link to it in a particular context, which is a fair assumption to make.
Nevertheless, we further assess the quality of the automatically constructed retrieval datasets with respect to our task definition (Section~\ref{sec:task-def}) by performing two annotation tasks.
In the first task, we show $e$ and $d^*$ to a human annotator and ask whether they understand their connection (binary).  %
In the second task, which is done after the completion of the first task, we additionally show the context $c$ and ask whether it enhances their understanding of the connection of $e$ and $d^*$ (binary).
The two tasks can help us validate whether $d^*$ is topically relevant to $e$, and relevant to $c$ in a way that enables the continuation of the narrative (Section~\ref{sec:task-def}).

One annotator annotated 100 examples from the development set of each dataset (i.e., 200 examples in total).
In order to assess the quality of the annotations, a second assessor annotated a subset of 50 examples from each dataset (100 examples in total). 
The Cohen's $\kappa$ \cite{cohen_multiple_1968} score is 0.61 for Task 1 and 0.50 for Task 2, both of which are considered moderate agreement.

\begin{table}[t]
\centering
\caption{Results of the annotation exercise: assessing relevance of document $d^*$ w.r.t. $e$ only (Task 1), and then $c$ (Task 2). We show the fraction of times the annotator labeled a sample as positive for the task.}
\label{tab:dataset-quality}
\begin{tabular}{l @{~~} l @{~~} l @{~~} l}
\toprule
  Dataset& Task 1 & Task 2  & Either \\ %
 \midrule
WaPo & 0.90 & 0.77 & 0.91 \\ 
Guardian & 0.85 & 0.83 & 0.92 \\ 
\bottomrule
\end{tabular}
\end{table}
The results can be seen in Table~\ref{tab:dataset-quality}.
We see that, for both datasets, the context $c$ enhances the understanding of the connection to $d^*$ for more than 3/4 of the cases (Task 2).
Also, for the vast majority of the cases, either the event $e$ or the context $c$ is sufficient to understand the connection (third column). 
We conclude that the automatic dataset construction procedure we proposed in Section~\ref{sec:dataset-construction} can produce reliable datasets for this task.

\section{Retrieval Method}

We follow a standard two-step retrieval pipeline that consists of (1) an unsupervised initial retrieval step and (2) a re-ranking step~\cite{wang2011cascade}.
Note that we do not focus on proposing new methods but rather on studying existing ones on this novel task.

\subsection{Initial retrieval}
\label{sec:NC-initial-retrieval}
In this step, we score each news article $d$ in $D$ w.r.t. $q=(e, c, t)$ to obtain the initial ranked list $L_1$. 
Here, we are interested in achieving high recall at lower depths in the ranking, since this step is followed by a more sophisticated reranking step.
We use BM25~\cite{robertson_probabilistic_2009}, an unsupervised lexical matching function, which is effective for ad-hoc retrieval and other tasks, such as question answering~\cite{yang_anserini_2017}.
In order to construct the lexical query, we simply concatenate $e$ and $c$.

\subsection{Reranking}
\label{sec:NC-reranking}
Here we rerank the initial ranked list $L_1$ obtained in the previous step by combining the results of multiple rankers using Reciprocal Rank Fusion (RRF), an unsupervised ranking fusion function~\cite{cormack_reciprocal_2009}:
\begin{align}
	\sum_{L \in \cal{L}} \frac{1}{k + rank(d, L)},
\end{align}
where $\cal{L}$ is a set of ranked lists, $rank(d, L)$ is the rank of article $d$ in the ranked list $L$, and $k$ is a parameter, set to its default value (60). 

We use the following rankers:
\paragraph{BM25} The initial retrieval step ranker (Section~\ref{sec:NC-initial-retrieval}), often used in combination with more sophisticated ranking models~\cite{macavaney_opennir_2020}. 
\paragraph{BERT} 
BERT~\cite{devlin_bert:_2019} has recently achieved state-of-the-art performance for retrieval and recommendation tasks in the news domain~\cite{yang2019simple,wu_mind_2020}.
BERT has been shown to prefer semantic matches and it is often used in combination with lexical matching ranking functions~\cite{qiao_understanding_2019}.
Given the query $q$ and a candidate news article $d$, we follow~\cite{maclaughlin_context-based_2020} and construct the input to BERT as follows: [\texttt{<CLS>} $e$ \texttt{<unused>} $c$ \texttt{<SEP>} $d$], 
 where $\texttt{<CLS>}$ is a special token,  \texttt{<unused>} is a special token that informs the model where the context begins and \texttt{<SEP>} is a special token that informs the model where the document $d$ begins.
We add a dropout layer on top of the $\texttt{<CLS>}$ token, and a linear layer with a scalar output to obtain the final matching score, which is used to rank the articles in $L_1$.
Note that, because of the limit of BERT in the number of tokens, we only take into account the headline and lead of $d$.

\paragraph{Recency}
This ranker simply sorts the candidate articles in $L_1$ by their reversed chronological order. %

\medskip\noindent%
Note that we have also experimented with combining the scores of the above rankers as features in supervised learning to rank models but they only gave minor improvements over RRF. Thus we do not discuss them in this chapter.

\section{Experimental setup}
\label{sec:NC-exp-setup}

\subsection{Evaluation metrics}
We use standard IR metrics: Mean Reciprocal Rank (MRR) and recall at different cut-offs (R@20, R@1000).
Note that because of the way we construct our dataset (Section~\ref{sec:dataset-construction}), we only have one relevant news article per query and thus MRR is equivalent to MAP.
We use a cut-off of 20 at recall since we expect writers to be willing to navigate the ranked list to lower positions~\cite{kim_automatic_2011}.
We report on statistical significance with a paired two-tailed t-test. %

\subsection{Implementation and hyperparameters}

We use the BM25 implementation of Anserini~\cite{yang_anserini_2017} with default parameters and retrieve the top-1000 articles (Section~\ref{sec:NC-initial-retrieval}). 

We use the OpenNIR implementation of BERT for retrieval~\cite{macavaney_opennir_2020}. We fine-tune the \emph{bert-base} pre-trained model on the training set of each of our datasets separately.
We assign a maximum 300 tokens for the query $q$ and 200 for the article $d$.
We use a batch size of 16 with gradient accumulation of 2, we apply max grad norm of 1 and tune the following hyperparameters for MRR on the development set of each dataset separately: number of negatives $\{1,2,3\}$ and learning rate $\{5e-6, 1e-5, 2e-5\}$.
During training we sample one negative example from the initial ranked list obtained in Section~\ref{sec:NC-initial-retrieval}, and train the model with pairwise ranking loss.

\paragraph{Preprocessing and word vectors}
We use Spacy\footnote{\url{http://spacy.io/}} for sentence splitting, POS tagging and Named Entity Recognition.
We use the \emph{en\_core\_web\_lg} model to obtain word vectors.

\section{Results}
\label{sec:NC-results}

In this section we present our experimental results.
\subsection{Initial retrieval}
\begin{table}[t]
\centering
\caption{Initial retrieval performance of BM25 on the test sets for different variations of the query $q = (e, c, t)$, or the link sentence (LS).}
\label{tab:initial-retrieval-query-variation}
\begin{tabular}{lrr|rr}
\toprule
 & \multicolumn{2}{c}{\textbf{WaPo}} & \multicolumn{2}{c}{\textbf{Guardian}} \\ 
 \midrule
Query & \small{MRR} & \small{R@1000} & \small{MRR} & \small{R@1000} \\ 
  \midrule
$e$ & 0.117 & 0.745 & 0.104 & 0.723 \\
$c$ & 0.167 & 0.737 & \bf 0.154 & 0.714 \\
$e$ \& $c$ & \bf 0.172 & \bf 0.832 & 0.149 & \bf \bf 0.806 \\
 \midrule
LS &  0.459 &  0.944 &  0.427 &  0.929 \\
\bottomrule
\end{tabular}
\end{table}

We examine the performance of the initial retrieval step when different variations of the query $q$ are used.
Table~\ref{tab:initial-retrieval-query-variation} shows the results.
We observe that, for both datasets, when using both the event $e$ and the context $c$ we get better results than when using either of the two alone, especially in terms of R@1000.
This shows that both the event $e$ and the context $c$ are important for our task.

In Table~\ref{tab:initial-retrieval-query-variation} (bottom row) we also show ranking performance when using the link sentence as the query (see Section~\ref{sec:dataset-construction}).
Even though we do not use the link sentence as part of the query in our task definition (Section~\ref{sec:task-def}),  this can give us a reference point for the ``upper bound'' performance in this step, since the link sentence has a high lexical overlap with the relevant article $d^*$~\cite{peng_news_2016}.  We observe that, indeed, when using the link sentence as the query, ranking performance is much higher than when using $q$, achieving close to perfect R@1000.
Nevertheless, R@1000 when using $e$ \& $c$ is relatively close to when using LS, which is an encouraging result given that in this step we are more interested in recall.

\subsection{Reranking}
Here, we report results on the individual rankers described in Section~\ref{sec:NC-reranking} and their combinations with RRF. 
\begin{table}[t]
\centering
\caption{Retrieval performance when reranking the ranked list obtained by BM25 (first row).}
\label{tab:NC-reranking}
\begin{tabular}{lrr|rr}
\toprule
 & \multicolumn{2}{c}{\textbf{WaPo}} & \multicolumn{2}{c}{\textbf{Guardian}} \\ 
 \midrule
\bf Method & \small{MRR} & \small{R@20} & \small{MRR} & \small{R@20} \\ 
  \midrule

BM25 & 0.172 & 0.433  & 0.149 & 0.382  \\\midrule
Recency & 0.086 & 0.284 & 0.065 & 0.065 \\ 

BERT & 0.182 & 0.451  & 0.173 & 0.447  \\  \midrule
RRF-recency & 0.206 & 0.509 &  0.195 & 0.477    \\ 
RRF & \bf 0.236 & \bf 0.588 & \bf 0.212 &  \bf 0.533  \\ %

\bottomrule
\end{tabular}
\end{table}
Table~\ref{tab:NC-reranking} shows the results.
First, we see that the performance of the Recency ranker is poor.
Also, we see that BERT outperforms BM25 on both datasets, while only using the headline and the lead of the candidate news article.
RRF-recency combines BERT and BM25 achieves an increase over BERT.
Finally, when also adding the Recency ranker in RRF, we observe a significant ($p<0.01$) increase on all metrics.
We conclude that RRF, albeit simple, is effective in combining the three rankers and that all three rankers are useful for this task.

\if
Note that the effectiveness of BERT in reranking the list obtained by BM25 is less critical than the one reported in a similar task to ours~\cite{maclaughlin_context-based_2020}. 
We hypothesize that this is because (i) we re-rank full-length articles (even though we use only their title and lead) instead of just short paragraphs as in their case, and (ii) that our task is more 
\fi

\section{Analysis}
\label{sec:analysis}

In this section we analyze our results along different dimensions to gain further insights into this task.
For our analysis we use the development set of the WaPo and Guardian datasets.%

\subsection{Vocabulary gap}
The vocabulary gap is a well known challenge in information retrieval~\cite{li_semantic_2014}.
Here, we analyze the performance of the rankers under comparison for this task based on the vocabulary gap between the query $q$ and the relevant article $d^*$.

\begin{figure}[t]
    \centering
    \begin{subfigure}[t]{0.48\linewidth}
        \centering
        \includegraphics[scale=0.4]{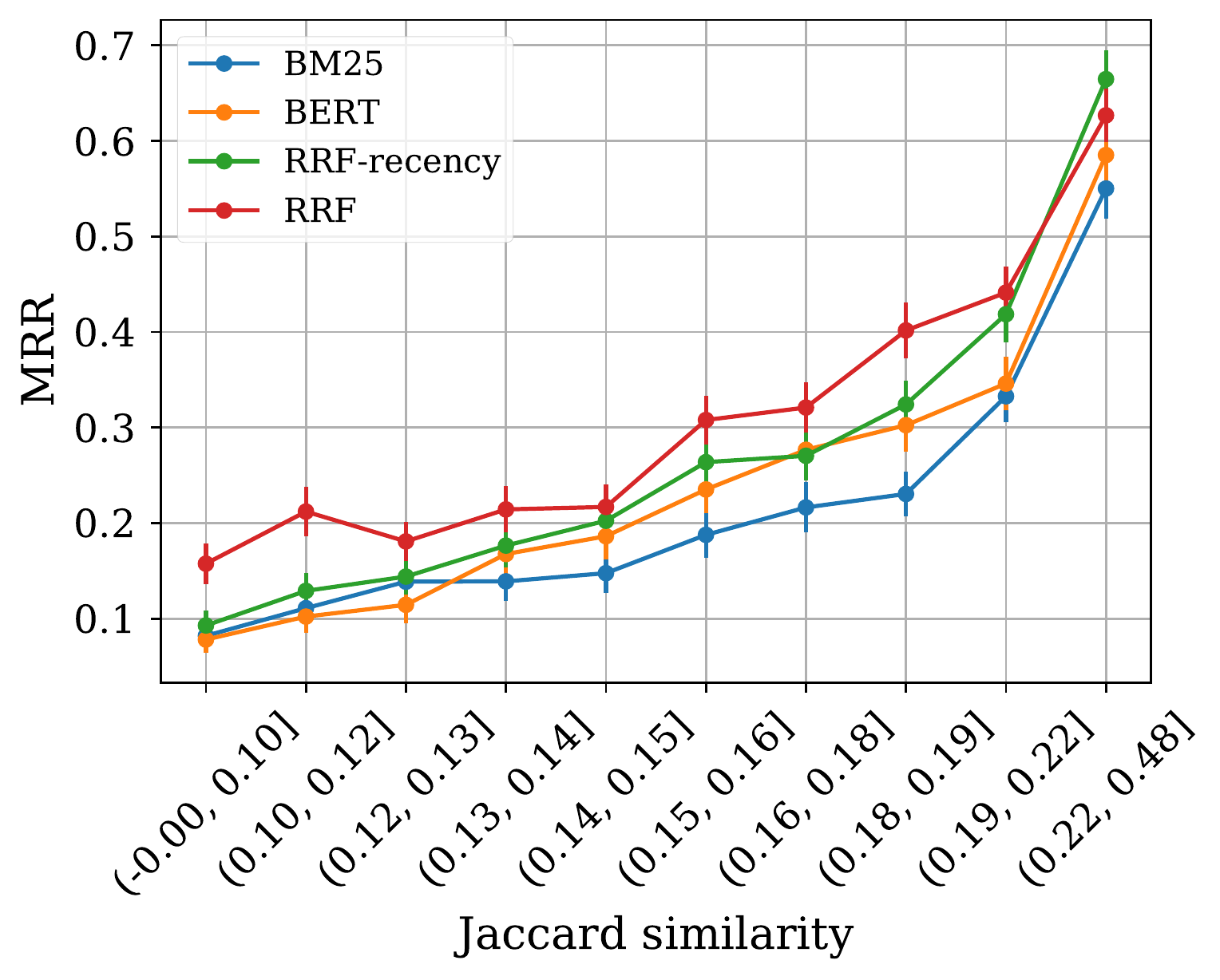}
        \caption{WaPo}
    \end{subfigure}
    ~
    \begin{subfigure}[t]{0.48\linewidth}
        \centering
        \includegraphics[scale=0.4]{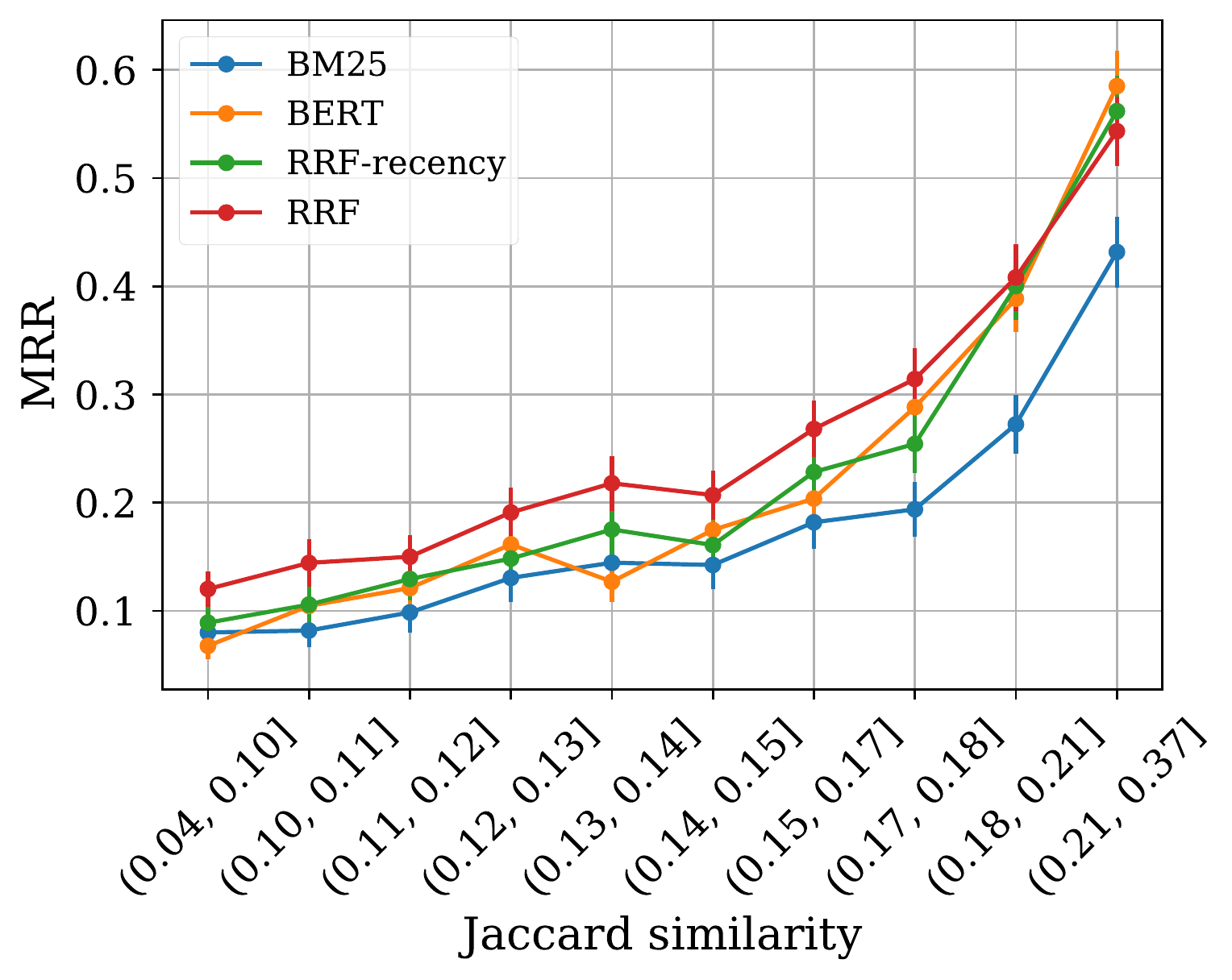}
        \caption{Guardian}
    \end{subfigure}
    \caption{MRR vs Jaccard similarity between query $q$ and $d^*$.}
    \label{fig:runs-jaccard-query-target-term-entity}
\end{figure}
\begin{figure}[t]
    \centering
    \begin{subfigure}[t]{0.48\linewidth}
        \centering
        \includegraphics[scale=0.4]{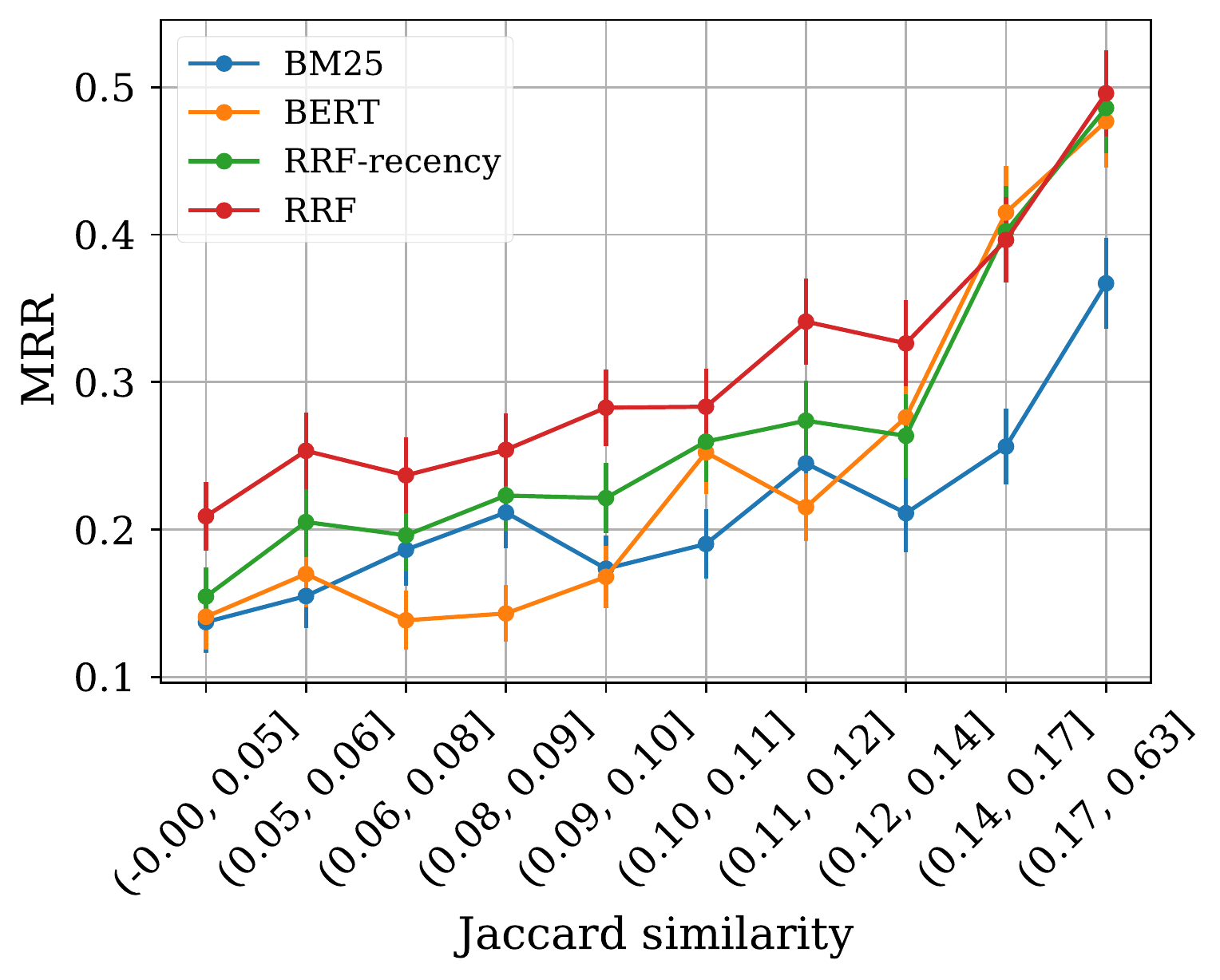}
        \caption{WaPo}
    \end{subfigure}
    ~
    \begin{subfigure}[t]{0.48\linewidth}
        \centering
        \includegraphics[scale=0.4]{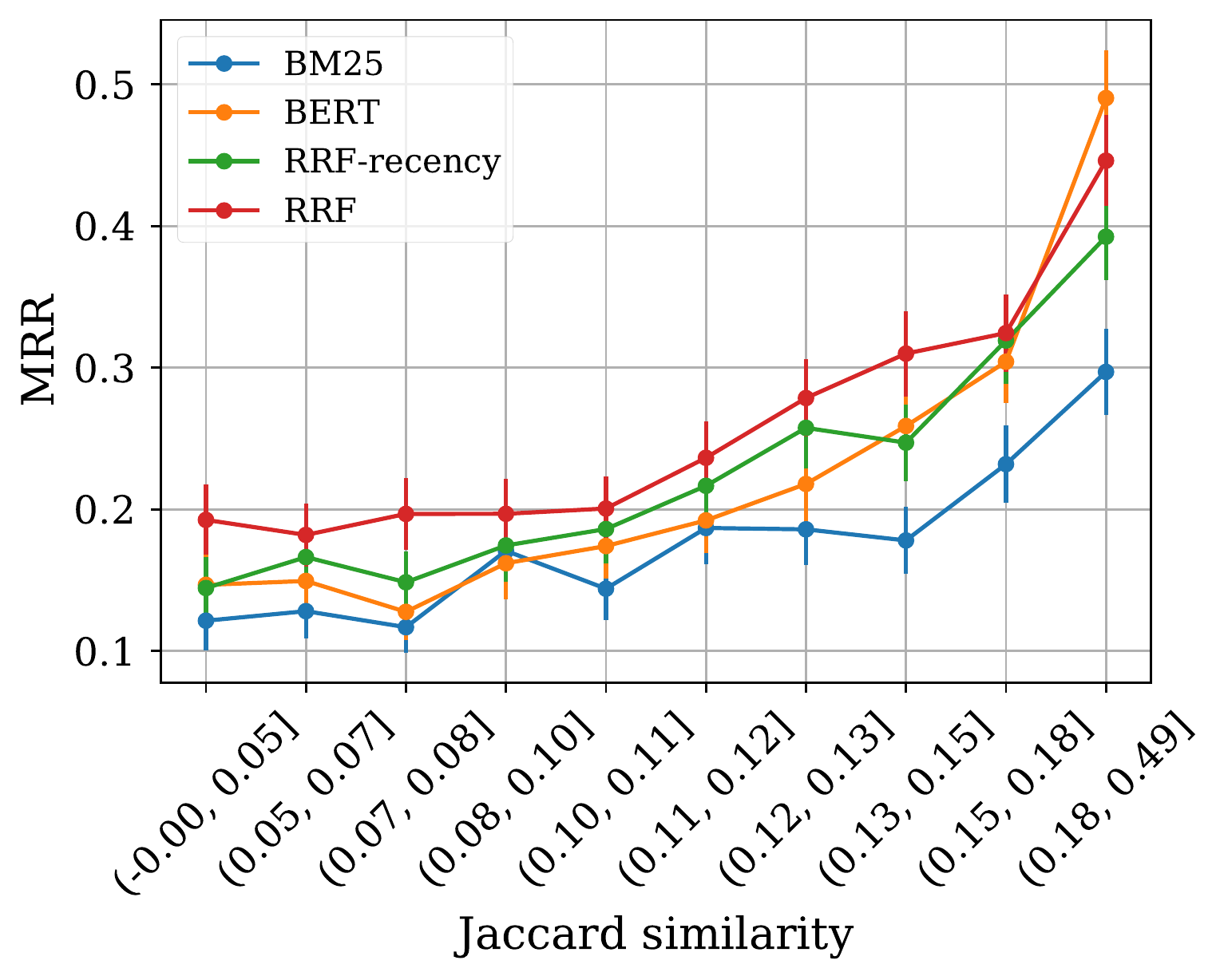}
        \caption{Guardian}
    \end{subfigure}
    \caption{MRR vs Jaccard similarity between narrative's context $c$ and $d^*$.}
    \label{fig:runs-jaccard-ps-target-term-entity}
\end{figure}

In Figure~\ref{fig:runs-jaccard-query-target-term-entity} we observe that the higher the lexical overlap between $q$ and $d^*$ (small vocabulary gap) the higher the performance for all rankers, for both datasets.
Also, we see that when the lexical overlap is low (large vocabulary gap), all rankers fail to bring the relevant article at the top positions of the ranking.
This shows that more sophisticated methods are needed to handle the large vocabulary gap in this task.
In Figure~\ref{fig:runs-jaccard-ps-target-term-entity} we show the lexical overlap between the narrative's context $c$ only and the relevant $d^*$.
Even though it follows the same trend as in Figure~\ref{fig:runs-jaccard-query-target-term-entity}, we see that BERT is consistently better than BM25 as the term overlap between the narrative's context $c$ and $d^*$ increases, for both datasets.
This shows that BERT is able to better take into account the narrative's context $c$ than BM25.

We next show examples of high/low lexical overlap between $q$ and $d^*$ in Table~\ref{tab:low_jaccard_query_target}.
In the first example (high lexical overlap), we see that because of high term overlap, all rankers are able to rank $d^*$ at the top 1--2 positions.
In the second example (low lexical overlap), the relevant article $d^*$ discusses the execution of Alfredo Prieto: this is a case in which Morrogh, a prosecutor in Virginia, was involved in (Morrogh is mentioned in the narrative's context $c$).
However, the fact that Morrogh is involved in the case is not mentioned explicitly in $d^*$ and thus all rankers fail to rank the relevant article at the top positions.
Incorporating the fact that Morrogh is related to Prieto in the ranking model could potentially be achieved by exploiting knowledge graphs that store event information~\cite{rospocher_building_2016,gottschalk_eventkg_2018}. We leave the exploration towards this direction for future work.

\begin{table*}
\caption{Examples from the WaPo dev. set with high/low lexical overlap between $q$ and $d^*$ (top/bottom).}
\tiny
\begin{tabularx}{\linewidth}{p{2cm}p{2cm} | p{2cm}  | p{2cm}X | p{2cm} X | X X X X}
\toprule
\multicolumn{2}{c}{\textbf{Query $q$}} & \textbf{Link sentence}
& \multicolumn{2}{c}{\textbf{Relevant article $d^*$}}
& \multicolumn{2}{c}{\textbf{Top-ranked article RRF}}
& \multicolumn{4}{c}{\textbf{Rank of $d^*$}}
\\
\midrule
\textbf{Query event $e$} & \textbf{narrative's context $c$} &    &  
Headline \& Lead & Day diff. &
Headline \& Lead & Day diff. &
 BM25 & BERT & RRF-recency & RRF
\\ \midrule
What {\textquoteleft}arrest{\textquoteright} means for the Canadians detained in China {\textemdash} and the epic battle over Huawei~. BEIJING {\textemdash} Over the past five months, as Beijing and Washington have exchanged fire on trade and technology, two Canadian men have been held in near-isolation in Chinese detention facilities. 
  & Last week, a Chinese court scheduled Schellenberg{\textquoteright}s appeal hearing to begin hours after Meng faced an extradition hearing in Vancouver.
  & After a Canadian court pushed back a decision in Meng{\textquoteright}s case, the Chinese court announced~it would delay a ruling on whether Schellenberg would be put to death.
  & Chinese court delays ruling on Canadian{\textquoteright}s death sentence appeal. BEIJING {\textemdash} A Chinese court has delayed ruling on a Canadian man{\textquoteright}s appeal against his death~sentence for drug smuggling, just hours after a Canadian court set a September date for the next hearing in an extradition case against a top Chinese executive.
  & 7
  &  \multicolumn{2}{c}{$d^*$}
  & 1
  & 2
  & 1
  & 1 \\ \hline 
Fairfax race for prosecutor puts focus on pace of criminal justice reform. Political races are usually about striking contrasts, but in the first Democratic primary for prosecutor in Virginia{\textquoteright}s largest county in 55 years, both candidates give themselves the same title: progressive. %
  & Morrogh, who was first elected commonwealth{\textquoteright}s attorney in 2007, has spent nearly all of his career in the prosecutor{\textquoteright}s office, where he has won some high-profile cases and avoided major scandal.
  & Morrogh helped secure the convictions of D.C. sniper Lee Boyd Malvo, serial killer Alfred Prieto and more recently the MS-13 gang members who killed a 15-year-old girl.
  & The execution of Alfredo Prieto: Witnessing a serial killer{\textquoteright}s final moments. JARRATT, Va. {\textemdash} It is undeniably disturbing to drive to the scheduled killing of another. A hurricane brewing in the distance, slicing steady rain through the gray day. %
  & 1339
  & Money from PAC funded by George Soros shakes up prosecutor races in Northern Virginia. A political action committee funded by Democratic mega\-donor and billionaire George Soros has made large contributions to two upstart progressive candidates attempting to unseat Democratic prosecutors in Northern Virginia primary races.%
  & 38
  & 371
  & 98
  & 151
  & 252 \\ \hline 
\end{tabularx}
\label{tab:low_jaccard_query_target}
\end{table*}

\subsection{Temporal aspects}
\label{sec:temporal-aspect}
As discussed in Section~\ref{sec:dataset-construction}, the retrieval datasets we derived for this task have a strong recency bias.
Here, we analyze the performance of the rankers under comparison based on the temporal aspect, i.e., how recent the relevant article is.
\begin{figure}[t]
    \centering
    \begin{subfigure}[t]{0.48\linewidth}
        \centering
        \includegraphics[scale=0.4]{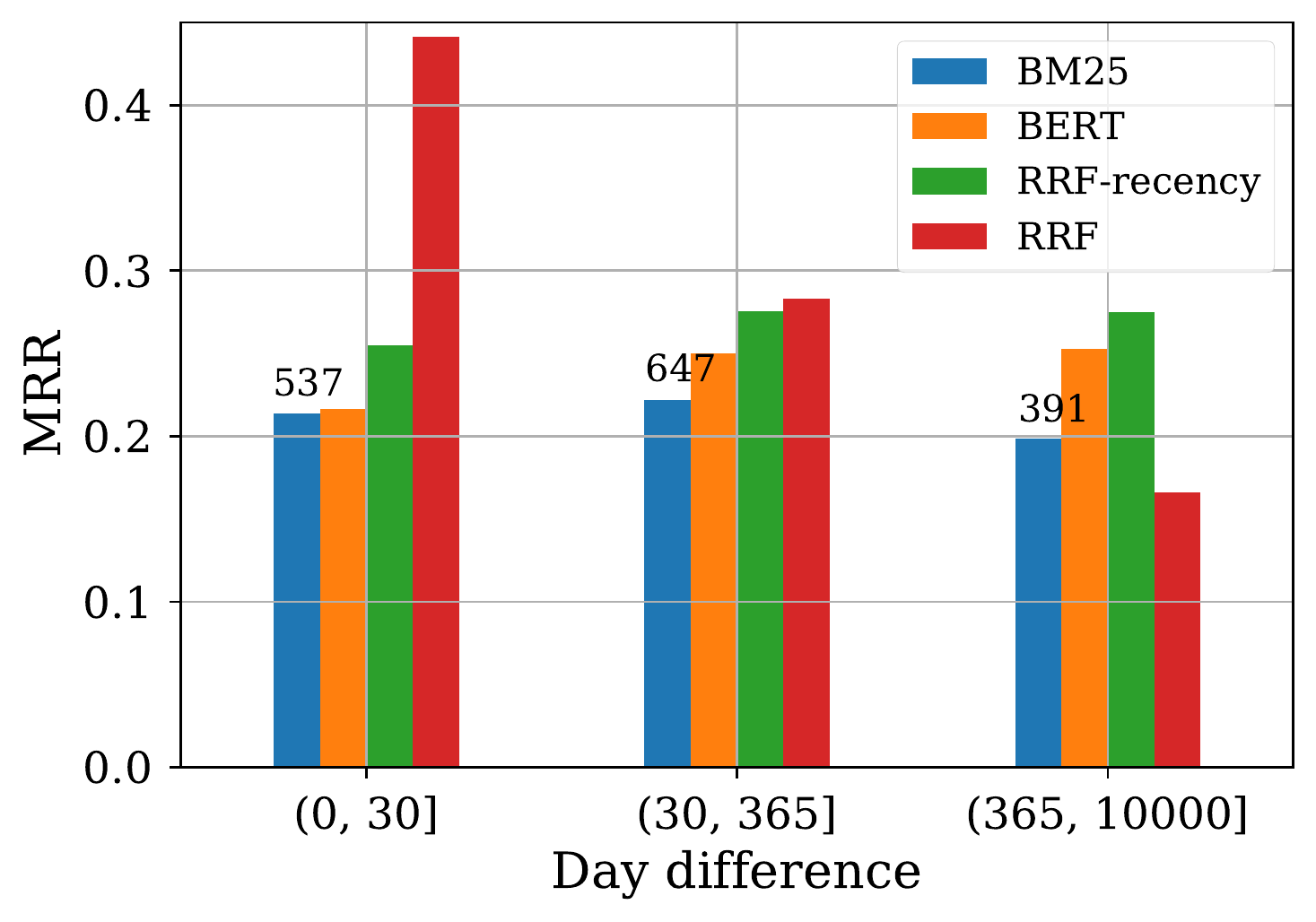}
        \caption{WaPo}
    \end{subfigure}
    ~
    \begin{subfigure}[t]{0.48\linewidth}
        \centering
        \includegraphics[scale=0.4]{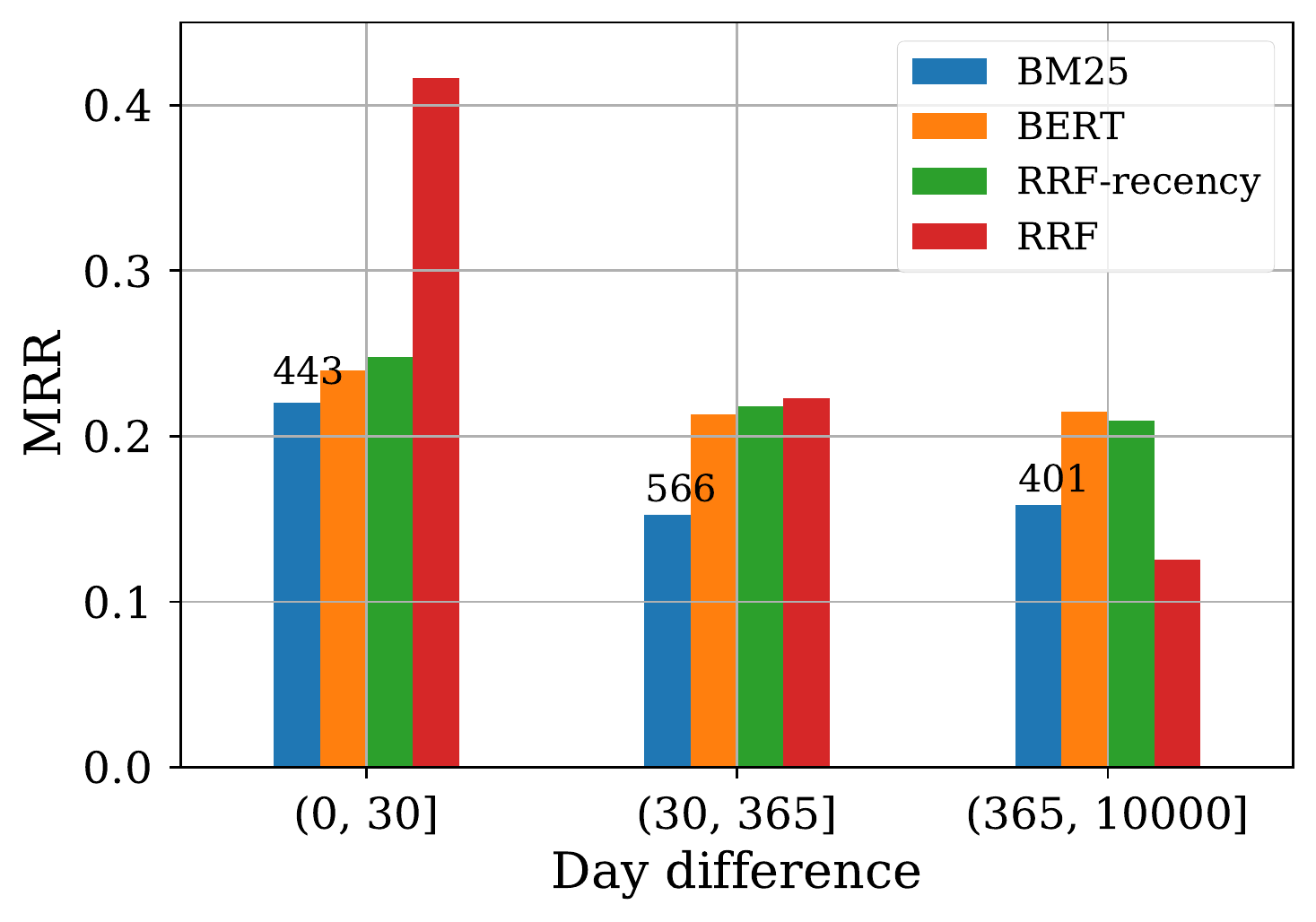}
        \caption{Guardian}
    \end{subfigure}
    \caption{MRR for retrieval methods grouped per day difference of the query and the relevant article.}
    \label{fig:runs-dev-day-diff}
\end{figure}

In Figure~\ref{fig:runs-dev-day-diff} we show the performance of the retrieval methods for different day differences between the query $q$ and the relevant article $d^*$.
As expected, we observe that for RRF, which uses the recency signal, the performance increases substantially on average when the relevant article is recent, and decreases when it is older.

\begin{table*}
\caption{Examples from the WaPo dev. set with a recent relevant article where RRF ranks the relevant article at the top, while RRF-Recency ranks it lower.}
\tiny

\begin{tabularx}{\linewidth}{p{2cm}p{2cm} | p{2cm}  | p{2cm}X | p{2cm} X | X X X X}
\toprule
\multicolumn{2}{c}{\textbf{Query $q$}} & \textbf{Link sentence}
& \multicolumn{2}{c}{\textbf{Relevant article $d^*$}}
& \multicolumn{2}{c}{\textbf{Top-ranked article RRF-recency}}
& \multicolumn{4}{c}{\textbf{Rank of $d^*$}}
\\
\midrule
\textbf{Query event $e$} & \textbf{narrative's context $c$} &    &  
Headline \& Lead & Day diff. &
Headline \& Lead & Day diff. &
 BM25 & BERT & RRF-recency & RRF
\\ \midrule

China{\textquoteright}s influence on campus chills free speech in Australia, New Zealand. SYDNEY {\textemdash} Chinese students poured into Australia and New Zealand in the hundreds of thousands over the past 20 years, paying sticker prices for university degrees that made higher education among both countries{\textquoteright} top export earners. 
  & After years of feeling fortunate about their economic relationship with China, Australians are starting to worry about the cost.
  & On Thursday, a ruling party lawmaker, Andrew Hastie, compared China{\textquoteright}s expansion to the rise of Germany before World War II, suggesting it posed a direct military threat.
  & Threat from China recalls that of Nazi Germany, Australian lawmaker says. The West{\textquoteright}s approach to containing China is akin to its failure to prevent Nazi Germany{\textquoteright}s aggression, an influential Australian lawmaker warned, earning a rebuke from Beijing while highlighting the difficulty the U.S. ally faces in weighing its security needs against economic interests. 
  & 3
  & China{\textquoteright}s meddling in Australia {\textemdash} and what the U.S. should learn from it. While American attention remains focused on Russia{\textquoteright}s interference in the 2016 presidential election, Australia {\textemdash} perhaps the United States{\textquoteright} closest ally {\textemdash} is debating the designs that a different country altogether has on its political system, economy and public opinion. That country is China. 
  & 788
  & 32
  & 38
  & 14
  & 1 \\ \hline 
Despite national security concerns, GOP leader McCarthy blocked bipartisan bid to limit China{\textquoteright}s role in U.S. transit. House Minority Leader Kevin McCarthy (R-Calif.) blocked a bipartisan attempt to limit Chinese companies from contracting with U.S. transit systems, a move that benefited a Chinese government-backed manufacturer with a plant in his district, according to multiple people familiar with the matter. %
  & Lawmakers frequently take a stance on legislation that could affect campaign contributors or hometown companies.  But McCarthy{\textquoteright}s intervention was striking because the close ally of President Trump sought to protect Chinese interests at a time when Trump and many lawmakers on Capitol Hill are attempting to curb Beijing{\textquoteright}s access to U.S. markets, particularly in industries deemed vital to national security.
  & Just last week, Trump put Chinese telecom giant Huawei on a trade {\textquotedblleft}blacklist{\textquotedblright} that severely restricts its access to U.S. technology.
  & Trump administration cracks down on giant Chinese tech firm, escalating clash with Beijing. The Trump administration on Wednesday slapped a major Chinese firm with an extreme penalty that makes it very difficult for it to do business with any U.S. company, a dramatic escalation of the economic clash between the two nations. %
  & 5
  & Trump says he{\textquoteright}ll spare Chinese telecom firm ZTE from collapse, defying lawmakers. President Trump said late Friday he had allowed embattled Chinese telecommunications giant ZTE Corp. to remain open despite fierce bipartisan opposition on Capitol Hill, defying lawmakers who have warned that the huge technology company should be severely punished for breaking U.S. law. %
  & 360
  & 98
  & 9
  & 3
  & 1 \\ \hline

\end{tabularx}
\label{tab:temporal-examples-rrf-better}
\end{table*}

\begin{table*}
\caption{Examples from the WaPo dev. set with an old relevant article where RRF-recency ranks the relevant article at the top, while RRF ranks it lower.}
\tiny
\begin{tabularx}{\linewidth}{p{2cm}p{2cm} | p{2cm}  | p{2cm}X | p{2cm} X | X X X X}
\toprule
\multicolumn{2}{c}{\textbf{Query $q$}} & \textbf{Link sentence}
& \multicolumn{2}{c}{\textbf{Relevant article $d^*$}}
& \multicolumn{2}{c}{\textbf{Top-ranked article RRF}}
& \multicolumn{4}{c}{\textbf{Rank of $d^*$}}
\\
\midrule
\textbf{Query event $e$} & \textbf{narrative's context $c$} &    &  
Headline \& Lead & Day diff. &
Headline \& Lead & Day diff. &
 BM25 & BERT & RRF-recency & RRF
\\ \midrule
Max Scherzer{\textquoteright}s knuckle injury might keep him from being ready for Opening Day. The knuckle at the base of Max Scherzer{\textquoteright}s right ring finger became the most analyzed joint in the Washington Nationals{\textquoteleft} clubhouse on Thursday, knocking Stephen Strasburg{\textquoteright}s right elbow out of its familiar spotlight, and delivering an unexpected blow to the early-season stability of the Nationals{\textquoteright} rotation. %
  & Scherzer expected the sprain to heal with regular rest in the offseason.  But the symptoms did not improve by December, when another MRI exam revealed the fracture.
  & A month later, the fracture still had not healed, so he told Team USA Manager Jim Leyland he would not be able to pitch in the World Baseball Classic.
  & Max Scherzer won{\textquoteright}t pitch in WBC because of stress fracture in finger. Nationals ace~Max Scherzer, one of the first and highest-profile players to commit to play~for the United States in the upcoming World Baseball Classic, will not participate in the tournament because of {\textquotedblleft}the ongoing rehabilitation stress fracture in the knuckle of his right ring finger,{\textquotedblright} the club announced Monday afternoon in a statement.%
  & 927
  & Nationals place Stephen Strasburg on disabled list (again) with pinched nerve. MIAMI {\textemdash} Nothing appeared amiss for Stephen Strasburg on Wednesday. He played catch at Miller Park in Milwaukee as scheduled, a day before he was to take the mound for the Washington Nationals against the Miami Marlins on Thursday night. But the throwing session didn{\textquoteright}t go well. %
  & 363
  & 12
  & 1
  & 1
  & 3 \\ \hline 
A mysterious sickness has killed nearly 100 children in India. Could litchi fruit be the cause? NEW DELHI {\textemdash} The children go to sleep as best they can in the sweltering heat. Early in the morning, the fever spikes and the seizures begin. 
  & In August 2017, India witnessed a notorious outbreak of encephalitis in the city of Gorakhpur in the neighboring state of Uttar Pradesh.
  & More than 30 children died over two days at one hospital~after its oxygen ran out.
  & {\textquoteleft}It{\textquoteright}s a massacre{\textquoteright}: At least 30 children die in Indian hospital after oxygen is cut off. NEW DELHI {\textemdash} One by one, the infants and children slipped away Thursday night, their parents watching helplessly as oxygen supplies at the government hospital ran dangerously low.%
  & 674
  & {\textquoteleft}It is horrid{\textquoteright}: India roasts under heat wave with temperatures above 120 degrees. NEW DELHI {\textemdash} When the temperature topped 120 degrees (49 Celsius), residents of the northern Indian city of Churu stopped going outside and authorities started hosing down the baking streets with water.
  & 11
  & 1
  & 1
  & 1
  & 3 \\ \hline 

\end{tabularx}
\label{tab:temporal-examples-rrf-worse}
\end{table*}
We next look at specific examples to better understand the results.
Table~\ref{tab:temporal-examples-rrf-better} shows examples where the relevant article is recent and RRF ranks it at the top of the ranking, while RRF-recency ranks it lower.
In both examples, RRF-recency's top-ranked article seems to also be relevant to $q$, however the writer chose to refer to a more recent event~\cite{niculae_quotus_2015}. 
Note that the fact that only one article is relevant to each query is an artifact of our dataset and not of the task itself. 
Table~\ref{tab:temporal-examples-rrf-worse} shows examples where the relevant article is old and RRF-recency ranks it at the top of the ranking, while RRF ranks it lower.
In the first example, the relevant article discusses a development on the injury of Scherzer, a player of the Washington Nationals team, and RRF-recency correctly brings that at the top position. 
However, RRF ranks a more recent event at the top position that discusses an injury of a different player of the same team.
In the second example, RRF brings at the top position an article that discusses an event about India that is more recent than the one that the relevant article discusses, however the article is off-topic. 

The above phenomena suggest that more sophisticated methods that model recency should be explored for this task.
For instance, it would be interesting to try to predict which queries are of temporal nature based on the characteristics of the underlying collection~\cite{kanhabua_temporal_2015}.
However, methods that build on features derived from user interactions are not applicable to our setting~\cite{dong_towards_2010}.

\subsection{Entity popularity}
Entities play a central role in event-centric narratives, especially in the news domain~\cite{rospocher_building_2016}.
We examine whether entity popularity affects retrieval performance in our task by measuring the IDF (Inverse Document Frequency) of entities in the query~\cite{manning2008introduction}. 
An entity with a high IDF in the collection is less popular than an entity with a low IDF.
\begin{figure}[t]
    \centering
    \begin{subfigure}[t]{0.48\linewidth}
        \centering
        \includegraphics[scale=0.4]{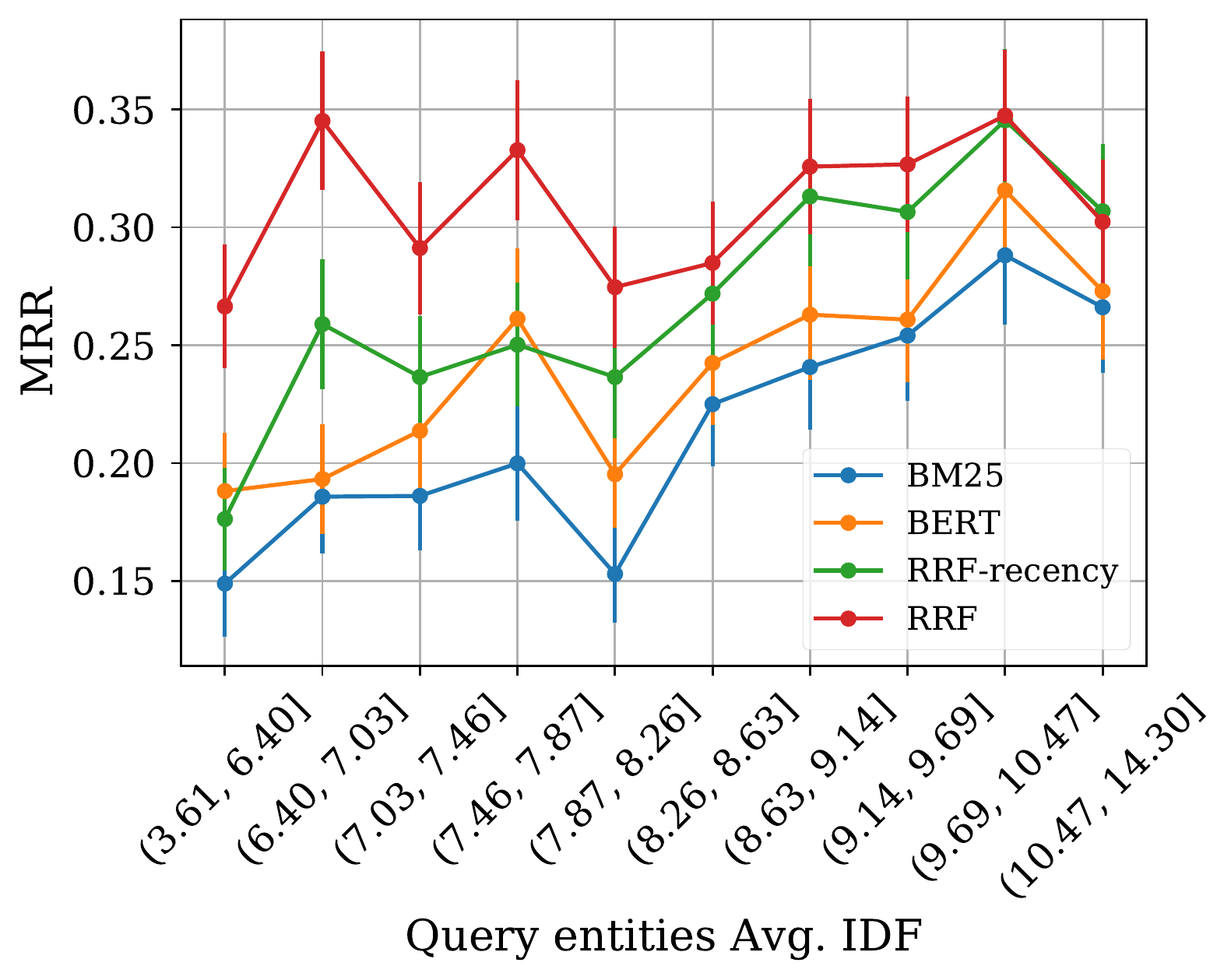}
        \caption{WaPo}
    \end{subfigure}
    ~
    \begin{subfigure}[t]{0.48\linewidth}
        \centering
        \includegraphics[scale=0.4]{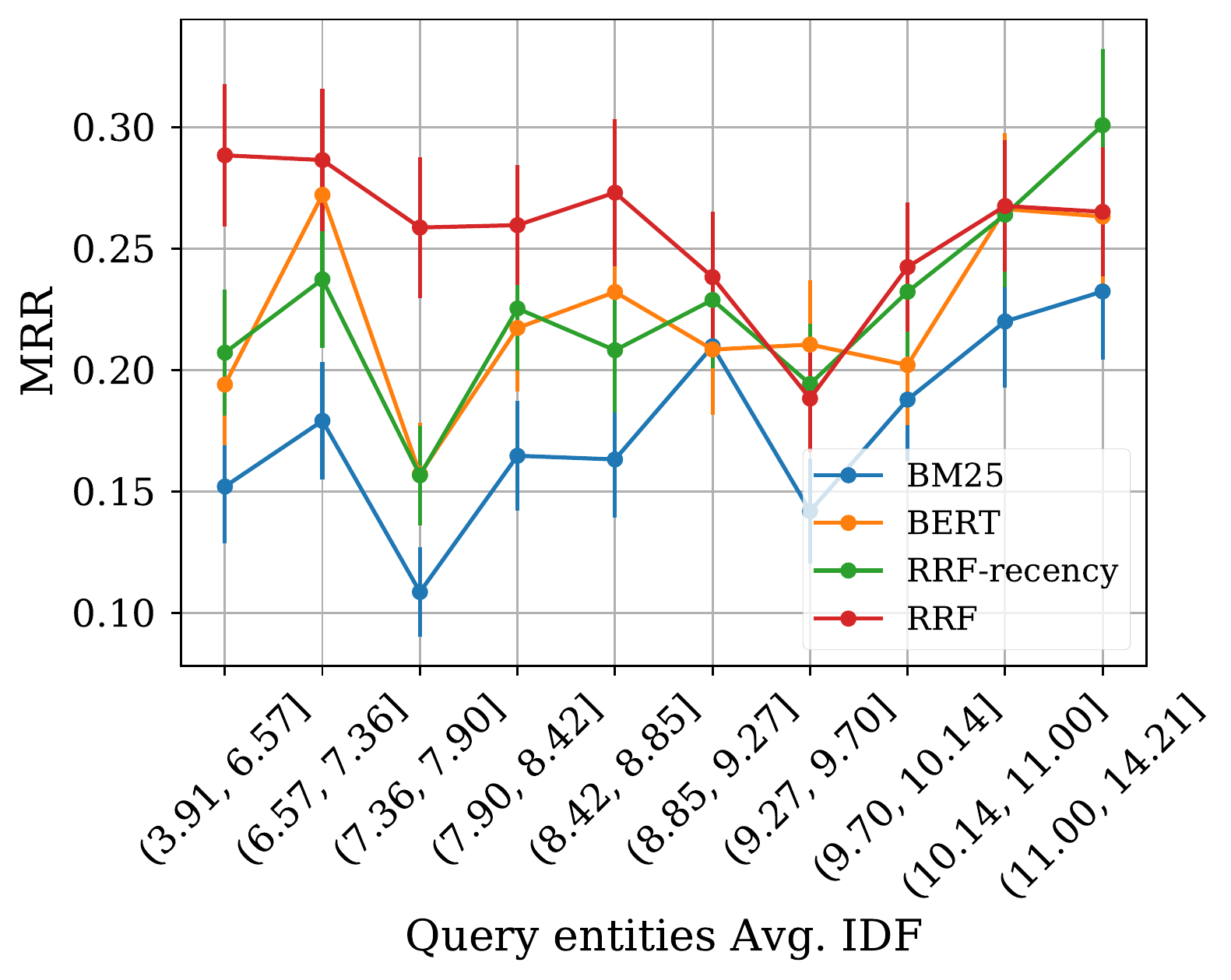}
        \caption{Guardian}
    \end{subfigure}
    \caption{MRR vs avg. IDF of the entities in the query $q$.}
    \label{fig:runs_query_entity_temporal_features_idf_mean}
\end{figure}

In Figure~\ref{fig:runs_query_entity_temporal_features_idf_mean} we show the performance depending on the average IDF of the entities in the query in the underlying collection.
We observe that the rankers that use the query and article text (BM25, BERT, RRF-recency) perform worse for queries with more popular entities (low IDF) than for queries with less popular entities.
This is because popular entities appear in multiple events, and thus there are many potentially relevant articles for a query.
We also see that RRF, which takes recency into account, is more robust to entity popularity.
This might also be related to the fact that a recent event that involves a popular entity is more likely to be relevant in general than a less recent event that involves the same entity (also see examples in Section~\ref{sec:temporal-aspect}, Table~\ref{tab:temporal-examples-rrf-better}).

\subsection{Link sentence}

\begin{table*}
\caption{Examples from the WaPo dev. set where the link sentence contains crucial information for the connection of the complete narrative and the relevant article.}
\tiny
\begin{tabularx}{\linewidth}{p{2cm}p{2cm} | p{2cm}  | p{2cm}X | p{2cm} X | X X X X}
\toprule
\multicolumn{2}{c}{\textbf{Query $q$}} & \textbf{Link sentence}
& \multicolumn{2}{c}{\textbf{Relevant article $d^*$}}
& \multicolumn{2}{c}{\textbf{Top-ranked article RRF}}
& \multicolumn{4}{c}{\textbf{Rank of $d^*$}}
\\
\midrule
\textbf{Query event $e$} & \textbf{narrative's context $c$} &    &  
Headline \& Lead & Day diff. &
Headline \& Lead & Day diff. &
 BM25 & BERT & RRF-recency & RRF
\\ \midrule
  Americans are drinking more ‘gourmet’ coffee. This doesn’t mean they’re drinking great coffee. The National Coffee Association USA recently dropped its annual survey results, and, as usual, there’s a wealth of information to sift through to better understand the state of coffee drinking in America. %
  & According to this year’s finding, coffee remains the No.  1 drink: Sixty-three percent of the respondents said they drank a coffee beverage (drip coffee, espresso, latte, cold brew, Unicorn Frappuccino, etc.) the previous day, a click down from 64 percent in 2018. 
  & By the way, the second-most consumed beverage was unflavored bottled water, which might help explain the Great Pacific Garbage Patch.

  & Plastic within the Great Pacific Garbage Patch is ‘increasing exponentially,’ scientists find. Seventy-nine thousand tons of plastic debris, in the form of 1.8 trillion pieces, now occupy an area three times the size of France in the Pacific Ocean between California and Hawaii, a scientific team reported on Thursday. %
    & 371
    & N/A

    & N/A
    & N/A
    & N/A
    & N/A
    & N/A

  \\ \hline
Turkey{\textquoteright}s elections show the limits of Erdogan{\textquoteright}s nationalism. Ahead of local elections throughout his country last weekend, Turkish President Recep Tayyip Erdogan resorted to his usual tactics. He cast some of his ruling party{\textquoteright}s opponents as traitors in league with terrorists. %
  & There{\textquoteright}s a broader story to be told, as well.
  & Well before President Trump, Indian Prime Minister Narendra Modi or Hungarian Prime Minister Viktor Orban, Erdogan arrived at the politics of the zeitgeist.
  & Trump{\textquoteright}s populism is about creating division, not unity. President Trump~begins his third week in office with the worst approval ratings~of any new American president since polls began tracking such results. %
  & 786
  & Stunning setbacks in Turkey{\textquoteright}s elections dent Erdogan{\textquoteright}s aura of invincibility. ISTANBUL {\textemdash} Turkish President Recep Tayyip Erdogan faced the prospect Monday of a stinging electoral defeat in Istanbul, the city whose politics he dominated for a quarter of a century, with vote results showing what appeared to be an opposition victory in the race for the city{\textquoteright}s mayor.%
  & 1
  & 447
  & 471
  & 487
  & 535 \\ \hline

\end{tabularx}
\label{tab:wapo_dev_query_underdefined_rrf}
\end{table*}

Recall that we do not use the link sentence as part of the query (see Section~\ref{sec:dataset-construction}). Thus, our rankers are not aware of its content.
However, we found that in some cases the link sentence contains information that is crucial for the connection of the \emph{complete} narrative and the relevant news article. 
Thus, in such cases, the query event $e$ and the narrative's context $c$ are not sufficient.
Table~\ref{tab:wapo_dev_query_underdefined_rrf} shows examples of such cases.
Note that in the first example, the relevant article was not even retrieved in the top-1000 of the initial retrieval step (see~Section~\ref{sec:NC-initial-retrieval}).
In the second example, the relevant article is ranked very low by all rankers.

One direction for future work would be to detect parts of the link sentence that contain such crucial information and add them to the narrative's context $c$. 
This could be performed as a manual annotation task or modeled as a prediction task~\cite{jha_nlp-driven_2017}.

\section{Related work}

\subsection{Supporting narrative creation}
Recent work on developing automatic applications to support writers has focused on designing tools that track and filter information from social media to support journalists~\cite{diakopoulos_finding_2012,zubiaga_curating_2013}. 
\citet{cucchiarelli_topic_2019} track the Twitter stream and Wikipedia edits to suggest potentially interesting topics that relate to a new event that a writer can include in their narrative when reporting on the event. 
In contrast, instead of relying on external sources, we aim to retrieve news articles that describe events from the past that can help the writer expand the incomplete narrative about a specific event.

Perhaps the closest to our task are the works by \citet{maiden_inject_2020} and \citet{maclaughlin_context-based_2020}. 
\citet{maiden_inject_2020} focus on suggesting articles that would help journalists discover new, creative angles on a current incomplete narrative.
The difference with our work is that they aim to suggest creative angles on articles and retrieve articles depending on the angle the writer selects.
In addition, they evaluate their system in a living lab scenario, whereas we create static retrieval datasets from historical data and use them to train ranking functions.
Evaluating our system in a living lab scenario would be a promising direction for future work.

\citet{maclaughlin_context-based_2020} retrieve paragraphs that contain quotes from political speeches and conference transcripts, so that writers can use them in their incomplete narratives.
Even though their retrieval task definition is similar to ours, our task differs in that our unit of retrieval is a news article from a large news article collection instead of a paragraphs from a single document (e.g., a political speech).
Moreover, our unit of retrieval (article) is timestamped, which makes the temporal aspect prominent in our task.
\subsection{Context-aware citation recommendation}
The task of context-aware citation recommendation is to find articles that are relevant to a specific piece of text a writer has generated~\cite{he_context-aware_2010}.
It has mainly been studied in the scientific domain~\cite{huang_neural_2015,ebesu_neural_2017,jeong_context-aware_2019,saier_semantic_2020}, but also in the news domain~\cite{li_related_2017}.
The main difference of the aforementioned works and our task is that we aim to retrieve articles to expand existing incomplete narratives instead of finding citations for complete narratives.

\subsection{Event extraction \& retrieval}
Events are the starting points of narrative news items. 
Recent work has focused on extracting and characterizing events from large streams of documents~\cite{chaney_detecting_2016} and extracting the most dominant events from news articles~\citep{choubey_identifying_2018}.
In our work, we assume that a news article is associated with a single main event, which is described by the article's headline and lead paragraph~\cite{choubey_discourse_2020}.

More related to our task is work focused on retrieving events given a query event~\cite{li_related_2017,setty_event2vec_2018}. However, this work does not consider additional context in the query as we do and thus it is not directly comparable to ours.

\section{Conclusion and Future Work}
In this chapter, we proposed and studied the task of news article retrieval in context for event-centric narrative creation.
We proposed an automatic dataset construction procedure and showed that it can generate reliable evaluation sets for this task.
Using the generated datasets, we compared lexical and semantic rankers and found that they are insufficient.
We found that combining those rankers with one that ranks articles by their reverse chronological order significantly improves retrieval performance over those rankers alone.

Our analysis showed that the vocabulary gap for this task is large, and therefore more advanced methods for semantic matching are needed. 
This could be achieved by exploiting external knowledge about events stored in knowledge graphs~\cite{gottschalk_eventkg_2018}.
To this end we aim to build on insights gained from our studies in Chapters~\ref{chapter:acl2015},~\ref{chapter:ecir2017} and~\ref{chapter:sigir2018} to improve semantic matching.
For instance, we could first detect KG facts in the query and the articles and then use the method we proposed in Chapter~\ref{chapter:acl2015} to retrieve descriptions of the detected KG facts. The retrieved descriptions can then be used to provide additional knowledge to the BERT ranker and thus improve semantic matching~\cite{petroni2020how}.

Moreover, our analysis showed that the temporal aspect is prominent in this retrieval task, which was not the case for the tasks we studied in the previous chapters of this thesis. Therefore, future work would aim to find more robust ways to incorporate the temporal aspect in the ranking function~\cite{kanhabua_temporal_2015}.

Furthermore, we found that this task is more challenging when the query event involves entities that appear more frequently in the collection, which we plan to further study in the future.
Another direction for future work is to categorize queries in relation to their discourse function in the narrative~\cite{teufel_annotation_1999,teufel_automatic_2006}, for example in relation to their function with respect to the main event of the narrative~\cite{choubey_discourse_2020}, and develop specialized rankers for each category.

We found that in some cases the link sentence contains crucial information for the connection between the \emph{complete} narrative and the relevant news article.
However, since the link sentence is not part of the query according to our dataset construction procedure, the constructed query may miss this key piece of information to capture the connection.
For future work, we aim to address this limitation by detecting such information in the link sentence and adding it to the query, or by using natural language generation techniques to fill in these blanks automatically. 

Finally, it is important to note that even though our dataset construction procedure can generate reliable retrieval datasets, the fact that we only have a single relevant article for each query may be limiting as more than one article may be relevant.
Thus, some of our findings might be an artifact of that procedure and not the task itself.
We plan to overcome this limitation in future work by asking journalists to qualitatively assess the output of different rankers to enrich the automatically constructed datasets with more relevant articles per query~\cite{maclaughlin_context-based_2020,maiden_inject_2020}.

\bookmarksetup{startatroot} %
\addtocontents{toc}{\bigskip} %

\chapter{Conclusions}
\label{chapter:conclusions}
In this thesis, we studied three research themes aimed at supporting search engines with knowledge and context: 
\begin{enumerate*}[label=(\arabic*)]
\item making structured knowledge more accessible to the user, 
\item improving interactive knowledge gathering, and
\item supporting knowledge exploration for narrative creation.
\end{enumerate*}
We studied several algorithmic tasks within these themes and proposed solutions to address them.

In this concluding chapter, we first revisit the research questions that we introduced in Chapter~\ref{chapter:introduction} and describe our main findings in Section~\ref{sec:conclusion-main-findings}.
In Section~\ref{sec:conclusion-future-work}, we discuss limitations and future directions.

\section{Main Findings} %
\label{sec:conclusion-main-findings}

\subsection{Making structured knowledge more accessible to the user}

Within this research theme we asked and answered three research questions motivated by the need of presenting \acf{KG} facts to users in a natural way.
In Chapter~\ref{chapter:acl2015}, we asked the following question:
\begin{description}\item[\acs{rq:retrieve-explanations}]\acl{rq:retrieve-explanations}\end{description}
To answer this question, we formalized the task of retrieving textual descriptions of KG facts from a corpus of sentences.
We developed a method for this task that consists of two steps.
First, we extract and enrich candidate sentences from the corpus and then rank them by how well they describe the KG fact.
In the first step, we detect sentences in the corpus that contain surface forms of any of the two entities in the KG fact and apply coreference resolution and entity linking to enrich them.
In the second step, we rank the extracted sentences using learning to rank, that combines a rich set of features of different types.
To evaluate our method, we construct a manually annotated dataset that contains descriptions of KG facts that involve people.
We found that our method improves performance over state-of-the-art sentence retrieval methods and that all groups of features contribute to retrieval performance, with relation-based features being the most important.
Moreover, we found that training relationship-dependent rankers is beneficial to improving retrieval performance.
Importantly, we also found that almost one third of the facts in our dataset did not correspond to any relevant sentence in the corpus.
This is usually the case for facts of which the entities are less popular.
Not being able to provide a meaningful description of certain facts limits the applicability of our method in real world scenarios.
This finding led us to the following research question in Chapter~\ref{chapter:ecir2017}:
\begin{description}\item[\acs{rq:generate-explanations}]\acl{rq:generate-explanations}\end{description}
To answer this question, we formalized the task of generating textual descriptions of KG facts.
We proposed a method that first generates sentence templates for a specific relationship and then, given a specific KG fact selects the most relevant template and fills it with information from the KG to create a novel sentence.
In order to create sentence templates, we designed a graph-based algorithm that combines information contained in existing sentences and the KG.
In order to select the most relevant template for a KG fact, we designed a supervised feature-based scoring function.
To evaluate our method, we automatically extracted a dataset for KG fact description generation and performed both automatic and manual evaluation.
We found that our method can generate grammatically correct and generally informative descriptions, and that a supervised scoring function outperforms an unsupervised one for selecting templates.
In addition, our error analysis showed that generating KG fact descriptions that are valid under the KG closed-world assumption is challenging and needs to receive more attention.

Next, in Chapter~\ref{chapter:sigir2018} we turned to a closely related problem and asked the following question:
\begin{description}\item[\acs{rq:contextualize-facts}]\acl{rq:contextualize-facts}\end{description}
To answer this question, we formalized the problem of contextualizing KG facts as a retrieval task.
We designed NFCM, a neural fact contextualization method that first generates a set of candidate facts that are part of the immediate neighborhood of the query fact in the KG, and subsequently ranks the candidate facts by how relevant they are to the query fact. 
We designed a neural network ranking model that combines information from multiple paths connecting the query and the candidate facts in the KG using recurrent neural networks to learn automatic features.
We further augmented the representation power of this model by using existing and novel hand-crafted features.
Since it is expensive to manually obtain human-curated training data to train this model, we turned to distant supervision to automatically generate training data for this task.
We evaluated NFCM using a human-curated dataset separate from the one used for distant supervision.
We found that when trained on distant supervision, NFCM significantly outperforms several heuristic baselines on this task.
Additionally, we found that NFCM benefits from both automatically learned and hand-crafted features.
Finally, we found that NFCM is relatively robust to the number of training data for each relationship.
\subsection{Improving interactive knowledge gathering}
We then moved to the theme of improving interactive knowledge gathering and studied multi-turn passage retrieval as an instance of conversational search.
In Chapter~\ref{chapter:sigir2020}, we asked the following question:
\begin{description}\item[\acs{rq:query-res}]\acl{rq:query-res}\end{description}
To answer this question, we formulated the task of query resolution for conversational search as a term classification task.
We proposed QuReTeC, a neural term classification model based on bidirectional transformers, more specifically BERT.
QuReTeC encodes the conversation history and the current turn query and predicts which terms from the history are relevant to the current turn.
We integrated QuReTeC in a standard, two-step retrieval pipeline by appending the terms predicted as relevant to the current turn query.
We performed evaluation both in terms of term classification and retrieval performance using a recently constructed multi-turn passage retrieval dataset.
We found that QuReTeC significantly outperforms state-of-the-art methods on this task when trained on gold standard query resolutions.
Furthermore, we found that QuReTeC is robust across conversation turns.
Since collecting such gold standard query resolutions for training QuReTeC might be cumbersome, we designed a distant supervision method that automatically generates training data for query resolution using query-passage relevance labels.
We found that this distant supervision method can substantially reduce the number of gold standard query resolutions required for training QuReTeC, a result especially important in low resource scenarios.

\subsection{Supporting knowledge exploration for narrative creation}
Our next study was in the theme of supporting knowledge exploration for narrative creation.
In Chapter~\ref{chapter:chapter6}, we asked the following question:
\begin{description}\item[\acs{rq:narrativecreation}]\acl{rq:narrativecreation}\end{description}
To answer this question, we formalized the task of event-centric news article retrieval in context.
We proposed an automatic retrieval dataset construction procedure that can produce reliable datasets for this task.
We generated two retrieval datasets using this procedure and used the generated datasets to evaluate automatic methods for this task.
We found that an unsupervised combination of state-of-the-art lexical and semantic rankers and a ranker that ranks articles by reverse chronological order outperforms those rankers alone.
We performed an in-depth quantitative and qualitative analysis to acquire insights into the characteristics of this task.
We found that this task has a large vocabulary gap, which highlights the need for semantic matching that takes into account structured knowledge about events.
In addition, we found that the temporal aspect is prominent in this task and thus more advanced temporal query and collection characteristics need to be explored.
Moreover, we found that this task is more challenging for queries that contain entities that appear more frequently in the underlying news article collection.
Last, we found that our dataset construction procedure is sometimes prone to generating queries that are not sufficiently defined, which is a clear future work direction.

\medskip\medskip\noindent
We now reflect on the main question we asked in Chapter~\ref{chapter:introduction}, namely how to support search engines in leveraging knowledge while accounting for different types of context.
In the first part of this thesis (Chapters~\ref{chapter:acl2015}, \ref{chapter:ecir2017} and \ref{chapter:sigir2018}), we proposed tasks and methods that make structured knowledge more accessible to the user (when the search engine proactively provides context to enrich search results) by retrieving existing or generating novel descriptions of KG facts, and also by contextualizing KG facts with other, related facts.
In the second part of this thesis (Chapter~\ref{chapter:sigir2020}), we proposed a method for query resolution that improves interactive knowledge gathering in conversational search by adding missing context from the conversation history to the current turn query.
In the third part of this thesis (Chapter~\ref{chapter:chapter6}), we proposed and studied the task of retrieving news articles that are relevant to the user's broad query (the query event) and a context that further specifies the query, thereby supporting knowledge exploration for narrative creation.

\section{Future Directions}
\label{sec:conclusion-future-work}
In this section, we discuss limitations of our study and directions for future work that would overcome those limitations and further expand our work.

\subsection{Making structured knowledge more accessible to the user}
\paragraph{Validity of KG fact descriptions}
Ensuring the validity of automatically generated KG fact descriptions is crucial when presenting such descriptions to the user~\cite{gatt_survey_2018}.
In Chapter~\ref{chapter:ecir2017} we found that generating valid KG fact descriptions is a challenging task.
This is a challenge not only for template-based generation methods such as ours but also for neural sequence to sequence generation methods~\cite{bahdanau2014neural,xu_fact-based_2020,maynez_faithfulness_2020}.
A possible direction towards overcoming this challenge is to learn discrete templates jointly with learning how to generate~\cite{wiseman_learning_2018}.
Another possible direction is to learn to edit existing descriptions instead of generating  descriptions from scratch~\cite{guu_generating_2018}.
Moreover, it would be interesting to assess the ability of recently developed large-scale pretrained language models for generating valid KG fact descriptions~\cite{raffel_exploring_2019}.

\paragraph{Richness of KG fact descriptions}
In Chapter~\ref{chapter:sigir2018}, we proposed NFCM, a neural fact contextualization method.
Relevant facts retrieved by NFCM can be used to  improve KG fact description retrieval by better modeling the relevance of existing descriptions (Chapter~\ref{chapter:acl2015}).
In addition, they can be used to select more informative templates in KG fact description generation (Chapter~\ref{chapter:ecir2017}).

\paragraph{Source of KG fact descriptions}
In Chapters~\ref{chapter:acl2015}, ~\ref{chapter:ecir2017} and ~\ref{chapter:sigir2018} we used Wikipedia as the source of existing descriptions of KG facts.
Using other sources of such descriptions could widen the applicability of our proposed tasks and methods to less popular entities.
~\citet{huang2017learning} performed an initial exploration towards this direction by using web pages as the source of descriptions, with an application to KG fact description retrieval.
Their results showed that using the web as the source of descriptions poses further challenges that would be interesting to explore even further.
\paragraph{Query-dependent KG fact information}
Deciding what information about a KG fact to present in a SERP may depend on the user's query~\cite{hasibi2017dynamic}.
Future work could develop query-dependent methods for all three tasks we considered: KG fact description retrieval and generation, and KG fact contextualization.
A study with real search engine users interacting with KG fact information on SERPs could provide further insights into this direction.

\subsection{Improving interactive knowledge gathering}

\paragraph{Incorporating the system's response} 
In Chapter~\ref{chapter:sigir2020}, we followed the TREC CAsT 2019 setup and only took into account the previous turn queries (the ones that preceded the current turn query) but not the passages retrieved by the system for those queries (the system's response). 
In future work, we will evaluate QuReTeC on a more realistic scenario where the passages retrieved for the previous turn queries are also taken into account.
\paragraph{Distant supervision for query resolution}
In Chapter~\ref{chapter:sigir2020}, we proposed a distant supervision method for reducing the amount of query resolution training data required to train QuReTeC.
Our distant supervision method relies on query-passage relevance labels.
Future work could address how to combine our distant supervision method with methods that generate relevance labels with weak supervision~\cite{dehghani2017neural}, pseudo-relevance feedback~\cite{Lavrenko:2001:RBL:383952.383972} or user signals~\cite{joachims_optimizing_2002}.
Also, we would like to explore noise reduction methods to improve the quality of the distant supervision signal~\cite{roth_survey_2013}.

\paragraph{Term classification and rewriting for query resolution}
In Chapter~\ref{chapter:sigir2020}, we formulated query resolution for conversational search as a term classification task.
This gave us flexibility not only in terms of modeling but also in terms of where we can get the supervision signal from.
In two studies contemporaneous to ours, query resolution was formulated as a sequence generation task~\cite{yu_few-shot_2020,vakulenko_question_2020}.
Combining the strengths of both formulations of the query resolution task could result in developing more powerful models.
\paragraph{Specialized rankers in low resource settings}
In Chapter~\ref{chapter:sigir2020}, we focused on query resolution for conversational search and used existing rankers for both the initial retrieval and reranking steps.
State-of-the-art neural ranking models rely on large-scale annotated ranking datasets that are not yet available in conversational search~\cite{yang2019simple,cast2019}.
Therefore, future work could develop specialized rankers for conversational search in low resource settings, possibly by learning to perform query resolution and ranking in a joint manner.
\subsection{Supporting knowledge exploration for narrative creation}

\paragraph{Incorporate structured knowledge about events}
In Chapter~\ref{chapter:chapter6}, we found that the vocabulary gap for the retrieval task we studied is large, and that the retrieval methods we considered are not able to effectively account for that.
One possible direction for future work is to incorporate structured knowledge about events (and the entities involved in them) in the retrieval methods~\cite{rospocher_building_2016,gottschalk_eventkg_2018}.
Such knowledge includes relationships between entities (which we studied in Chapters~\ref{chapter:acl2015},~\ref{chapter:ecir2017} and~\ref{chapter:sigir2018}), or sub-event relations~\cite{araki_interoperable_2018,gottschalk_happening_2019}.

\paragraph{Temporal aspect}
In Chapter~\ref{chapter:chapter6}, we found that a simple combination of lexical and semantic rankers with a ranker that ranks articles by reverse chronological order improves performance when the relevant article is recent but harms performance otherwise, as expected.
In future work, we aim to incorporate the temporal aspect in the ranking function in a more robust way.
A possible way to achieve that is to identify temporal phenomena such as trending terms or entities in the underlying news article collection~\cite{kanhabua_temporal_2015} or in external sources such as social media~\cite{cucchiarelli_topic_2019}.

\paragraph{Dataset construction} 
In Chapter~\ref{chapter:chapter6}, we proposed a dataset construction procedure that can produce reliable datasets for this task.
The main limitation of this procedure is that only one article is relevant for each query, even though more than one article may be relevant.
In future work, we aim to ask experts (journalists) to qualitatively assess the output of different rankers to enrich the automatically constructed datasets with more relevant articles per query~\cite{maclaughlin_context-based_2020,maiden_inject_2020}.
Another limitation of our dataset construction procedure is that in some cases the query (usually the narrative context) misses crucial information for the connection of the complete narrative and the relevant article that is contained in the link sentence.
In future work we aim to ask experts to manually add such missing information to the narrative context.
Since this task can be cumbersome, we will try to semi-automate this procedure by casting this as a prediction task~\cite{jha_nlp-driven_2017}.

\paragraph{Specialized rankers per discourse function}
Previous work has studied how to categorize parts of existing narratives according to their discourse function with respect to the main event of the narrative~\cite{teufel_automatic_2006,choubey_discourse_2020}.
We hypothesize that, in the narrative creation task we studied in Chapter~\ref{chapter:chapter6}, queries that serve a certain discourse function would have relevant news articles of specific characteristics.
In future work, we aim to categorize queries to different discourse functions and perform manual analysis to validate this hypothesis.
We will then design specialized rankers for each category to improve retrieval effectiveness.

\backmatter

\renewcommand{\bibsection}{\chapter{Bibliography}}
\renewcommand{\bibname}{Bibliography}
\markboth{Bibliography}{Bibliography}
\renewcommand{\bibfont}{\footnotesize}
\setlength{\bibsep}{0pt}

\bibliographystyle{abbrvnat}
\bibliography{thesis,02-acl2015/acl2015-ere,03-ecir2017/references,04-sigir2018/bibliography,05-sigir2020/bibliography,06-chapter6/bibliography}

\chapter{Summary}
Search engines leverage knowledge to improve information access.
Such knowledge comes in different forms: unstructured knowledge (e.g., textual documents) and structured knowledge (e.g., relationships between real-world objects and topics).
In order to effectively leverage knowledge, search engines should account for context, i.e., additional information about the user and the query.
In this thesis, we aim to support search engines in leveraging knowledge while accounting for different types of context.

In the first part of this thesis, we study how to make structured knowledge more accessible to the user when the search engine proactively provides such knowledge as context to enrich search results. %
We focus on knowledge graphs (KGs), which store world knowledge in the form of facts, i.e., relationships between entities (e.g., persons, locations, organizations).
Since \Ac{KG} facts are stored in a formal form, they are not suitable for presentation to end users.
As a first task, we study how to retrieve natural language descriptions of \ac{KG} facts from a text corpus.
We propose a method that successfully extracts and then ranks descriptions of \ac{KG} facts.
The method breaks down when a description for a certain \ac{KG} fact does not exist.
This leads us to our second task, where we study how to automatically generate \ac{KG} fact descriptions.
We propose a method that first creates sentence templates and then fills them with relevant information from the \ac{KG}.
\Ac{KG} fact descriptions often contain mentions to other related facts that can increase the understanding of the fact as a whole.
As a third task, we study how to contextualize \ac{KG} facts, that is, automatically find facts related to a query fact.
We propose a method that enumerates \ac{KG} facts in the neighborhood of the query fact and then ranks them with respect to their relevance to the query fact.

In the second part of this thesis, we move to a different research theme and study how to improve interactive knowledge gathering.
We focus on conversational search, where the user interacts with the search engine to gather knowledge over large unstructured knowledge repositories.
Here, the search engine should account for context that stems from interactions between the user and the search engine in a conversational search session.
We focus on multi-turn passage retrieval as an instance of conversational search.
A prominent challenge is that the current turn query may be underspecified.
Thus, we need to perform query resolution, that is, add missing context from the conversation history to the current turn.
We propose to model query resolution as a term classification task and propose a method to address it.

In the third and final part of this thesis, we focus on a specific type of search engine users, professional writers in the news domain.
We study how to support such writers create event-narratives by exploring knowledge from a corpus of news articles. 
We focus on a scenario where the writer has already generated an incomplete narrative that consists of a main event and a context, and aims to retrieve news articles that discuss relevant events from the past.
We formally define the task of news article retrieval in context for event-centric narrative creation.
We propose a retrieval dataset construction procedure for this task that relies on existing news articles to simulate incomplete narratives and relevant articles.
We study the performance of multiple rankers, lexical and semantic, and perform an in-depth quantitative and qualitative analysis to acquire insights into the characteristics of this task.

\chapter{Samenvatting}
Zoekmachines maken gebruik van kennis om de toegang tot informatie te verbeteren.
Dergelijke kennis komt in verschillende vormen voor: ongestructureerde kennis (bijv. tekstdocumenten) en gestructureerde kennis (bijv. relaties tussen objecten in de echte wereld en onderwerpen).
Om kennis effectief te benutten, moeten zoekmachines rekening houden met de context, in dit geval, aanvullende informatie over de gebruiker en de zoekopdracht.
In dit proefschrift willen we zoekmachines ondersteunen bij het benutten van kennis en tegelijkertijd rekening houden met verschillende soorten context.

In het eerste deel van dit proefschrift bestuderen we hoe gestructureerde kennis toegankelijker  voor de gebruiker kan worden gemaakt wanneer de zoekmachine proactief kennis zoals context verschaft om zoekresultaten te verrijken.
We richten ons op kennisgrafen (KG's), die wereldkennis opslaan in de vorm van feiten, d.w.z. relaties tussen entiteiten (bijv. personen, locaties, organisaties).
Omdat KG-feiten in een formele vorm worden opgeslagen, zijn ze niet geschikt voor presentatie aan eindgebruikers.
Als eerste taak bestuderen we hoe we beschrijvingen van KG-feiten in natuurlijke taal uit een tekstcorpus kunnen halen.
We stellen een methode voor die met succes de beschrijving van KG-feiten extraheert en rangschikt.
De methode werkt echter niet als er geen beschrijving voor een bepaald KG-feit bestaat.
Dit leidt ons naar onze tweede taak, waar we bestuderen hoe we automatisch KG-feitbeschrijvingen kunnen genereren.
We stellen een methode voor die eerst zinssjablonen maakt en deze vervolgens vult met relevante informatie uit de KG.
KG-feitbeschrijvingen bevatten vaak vermeldingen van andere gerelateerde feiten die het begrip van het feit als geheel kunnen vergroten.
Als derde taak bestuderen we hoe we KG-feiten kunnen contextualiseren, dat wil zeggen, automatisch feiten vinden die verband houden met een vraagfeit.
We stellen een methode voor die KG-feiten opsomt in de buurt van het vraagfeit en deze vervolgens rangschikt met betrekking tot hun relevantie voor het vraagfeit.

In het tweede deel van dit proefschrift stappen we over op een ander onderzoeksthema en bestuderen we hoe we interactieve kennisvergaring kunnen verbeteren.
We richten ons op conversational search, waarbij de gebruiker interactie heeft met de zoekmachine om kennis te vergaren over grote ongestructureerde kennisverzamelingen.
Hier moet de zoekmachine rekening houden met de context die voortkomt uit interacties tussen de gebruiker en de zoekmachine in een conversatiezoeksessie.
We richten ons op het ophalen van passages in meerdere beurten als een voorbeeld van \emph{conversational search}.
Een prominente uitdaging is dat de vraag voor een enkele beurt mogelijk te weinig is gespecificeerd.
We moeten dus een zoekopdrachtresolutie uitvoeren, dat wil zeggen: ontbrekende context uit de gespreksgeschiedenis toevoegen aan de huidige beurt.
We stellen voor om zoekopdrachtresolutie te modelleren als een termclassificatietaak en stellen een methode voor om deze aan te pakken.

In het derde en laatste deel van dit proefschrift richten we ons op een specifiek type gebruikers van zoekmachines, namelijk professionele schrijvers in het nieuwsdomein.
We bestuderen hoe we dergelijke schrijvers kunnen ondersteunen bij het creëren van verhalen over gebeurtenissen door kennis uit een corpus van nieuwsartikelen te verkennen.
We richten ons op een scenario waarin de schrijver al een onvolledig verhaal heeft geschreven dat bestaat uit een hoofdgebeurtenis en een context, en beoogt nieuwsartikelen op te halen die relevante gebeurtenissen uit het verleden bespreken.
We definiëren de taak van het ophalen van nieuwsartikelen formeel in de context voor het creëren van op gebeurtenissen gerichte verhalen.
We stellen voor deze taak een procedure voor het construeren van een dataset voor, die gebaseerd is op bestaande nieuwsartikelen om onvolledige verhalen en relevante artikelen te simuleren.
We bestuderen de prestaties van meerdere rangschikmodellen, lexicaal en semantisch, en voeren een diepgaande kwantitatieve en kwalitatieve analyse uit om inzicht te verwerven in de kenmerken van deze taak.

\end{document}